\documentclass[aps,showpacs,preprintnumbers,amsmath,notitlepage,nofootinbib,amssymb]{revtex4-1}

\usepackage{amsmath,amssymb}
\usepackage[utf8]{inputenc}
\usepackage{graphicx}
\usepackage{mathrsfs}
\usepackage{bm}
\usepackage{indentfirst}
\usepackage{epstopdf}
\usepackage{color}
\usepackage[colorlinks,citecolor=blue,urlcolor=blue,linkcolor=black]{hyperref}
\usepackage{braket}
\usepackage{placeins}
\usepackage{multirow}
\usepackage{hyperref}
\usepackage{lineno}
\newcommand{\pbp}{\langle\bar{\psi}\psi\rangle}
\newcommand{\suthree}{SU(3)}
\newcommand{\uone}{U(1)}
\newcommand{\av}[1]{\left\langle #1 \right\rangle} 
\newcommand{\e}{\text{e}} 
\newcommand{\tr}{\mathrm{tr}\,}  
\newcommand{\fsla}[1]{{\ooalign{\hfil/\hfil\crcr$#1$}}} 
\newcommand{\massmat}{M }
\newcommand{\diag}{\mathrm{diag}\,}  
\newcommand{\im}{\text{i}} 

\begin{document}
\title{Chiral properties of (2+1)-flavor QCD in strong magnetic fields at zero temperature}

\author{
	H.-T. Ding$^{\rm a}$, S.-T. Li$^{\rm b,a}$, A. Tomiya$^{\rm c}$, X.-D. Wang$^{\rm a}$ and Y. Zhang$^{\rm a}$}

\affiliation{
	$^{\rm a}$Key Laboratory of Quark and Lepton Physics (MOE) and Institute of Particle Physics, \\
	Central China Normal University, Wuhan 430079, China\\
	$^{\rm b}$Institute of Modern Physics, Chinese Academy of Sciences, Lanzhou 730000, China \\
	$^{\rm c}$RIKEN BNL Research center, Brookhaven National Laboratory, Upton, New York, 11973, USA 
}


\begin{abstract}
	We present lattice QCD results for masses and magnetic polarizabilities of light and strange pseudoscalar mesons, chiral condensates, decay constants of neutral pion, and neutral kaon in the presence of background magnetic fields with $eB$ ranging up to around 3.35 GeV$^2$ ($\sim70~M_\pi^2$) in the vacuum. The computations were carried out in (2+1)-flavor QCD mostly on $32^3 \times 96$ lattices using the highly improved staggered quark action with $M_{\pi} \approx $ 220 MeV at zero temperature.  We find that the masses of neutral pseudoscalar mesons monotonously decrease as the magnetic field strength grows and then saturate at a nonzero value, while there exists a nonmonotonous behavior of charged pion and kaon masses in the magnetic field. We observe a $qB$ scaling of the up and down quark flavor components of neutral pion mass, neutral pion decay constant as well as the quark chiral condensates at 0.05 $\lesssim eB\lesssim$ 3.35 GeV$^2$. We show that the correction to the Gell-Mann-Oakes-Renner relation involving the neutral pion is less than 6\% and the correction for the relation involving neutral kaon is less than 30\% at $eB\lesssim$ 3.35 GeV$^2$. We also derive the Ward-Takahashi identities for QCD in the magnetic field in the continuum formulation including the relation between integrated neutral pseudoscalar meson correlators and chiral condensates.
\end{abstract}


\maketitle

\section{Introduction} 
\label{sec:intro}
The properties of strongly interacting matter in the external magnetic field have attracted many studies in recent years as strong magnetic fields appear in heavy-ion collisions~\cite{Kharzeev:2007jp,Skokov:2009qp,Deng:2012pc}, the early Universe~\cite{Vachaspati:1991nm}, and magnetars~\cite{duncan1992formation}.  The QCD thermodynamics in the presence of a background magnetic field is of particular interest. At zero temperature, it is found from lattice QCD studies using standard staggered fermions that the order parameter of the transition, the chiral condensate, increases with the magnetic field strength $eB$, which is so-called magnetic catalysis (MC)~\cite{DElia:2010abb,Shovkovy:2012zn}. The MC suggests that the magnitude of the chiral symmetry breaking becomes larger in the vacuum, which leads to an expectation that the chiral crossover transition temperature $T_{pc}$ increases with the magnetic field strength $eB$. The increasing behavior of $T_{pc}$ with $eB$ is also found in the lattice QCD studies in two-flavor and three-flavor QCD using standard staggered fermions at finite lattice cutoffs~\cite{DElia:2010abb, Ding:2020inp}. However, the surprise came later that $T_{pc}$ actually decreases with $eB$ as found from continuum extrapolated results in $N_f=2+1$ lattice QCD studies using improved staggered  (stout) fermions~\cite{Bali:2011qj}. The discrepancy of results in Ref.~\cite{DElia:2010abb} from those in Ref.~\cite{Bali:2011qj} is most likely due to the large discretization errors in the standard staggered fermions~\cite{Ding:2020inp}. Accompanied with the reduction of $T_{pc}$, a decreasing behavior of chiral condensate in $eB$ in the proximity of transition temperature, i.e., the so-called inverse magnetic catalysis (IMC), is found in Refs.~\cite{Bali:2011qj,Bali:2012zg}. The IMC has also been observed in further lattice QCD studies using improved discretization 
schemes~\cite{Ilgenfritz:2013ara,Bornyakov:2013eya,Bali:2014kia,Tomiya:2019nym}.  Many model and theoretical studies have been performed to understand the (inverse) magnetic catalysis, reduction of $T_{pc}$ as well as their relations~\cite{DElia:2011koc,Shovkovy:2012zn,Andersen:2014xxa,Kojo:2012js,Bruckmann:2013oba,Fukushima:2012kc,Ferreira:2014kpa,Yu:2014sla,Feng:2015qpi,Li:2019nzj,Mao:2016lsr,Gursoy:2016ofp,Xu:2020yag}. Recently it has been suggested from lattice QCD studies with heavier-than-physical pions that the IMC is not necessarily associated with the reduction of $T_{pc}$ as a function of $eB$~\cite{DElia:2018xwo,Endrodi:2019zrl}, and it is more like a deconfinement catalysis~\cite{Bonati:2016kxj}. 

Although the increase (reduction) of $T_{pc}$ is often connected with the increase (reduction) of chiral condensates, which suggests the increase (reduction) of the magnitude of the chiral symmetry breaking, the breaking of chiral symmetry in QCD is also related to the Goldstone pions. At vanishing magnetic field, the square of the Goldstone pion mass is proportional to the product of a sum of quark masses and quark condensates. The quark mass explicitly breaks the chiral symmetry of the QCD Lagrangian, while the chiral condensate measures the strength of spontaneous symmetry breaking.  This is the well-known 
Gell-Mann-Oakes-Renner (GMOR) relation~\cite{GellMann:1968rz}, whose validity has been confirmed on lattice QCD simulations at vanishing magnetic field and in the vacuum~\cite{Boucaud:2007uk,Engel:2014cka}. The GMOR relation has also been extended to the three-flavor case, including a strange quark~\cite{Gasser:1984gg}.  The next-to-leading order chiral corrections to the GMOR relation have also been studied at zero temperature and vanishing magnetic field~\cite{Jamin:2002ev,Bordes:2010wy,Bordes:2012ud}.  Extending to the case of low temperature and vanishing magnetic field~\cite{Gasser:1986vb}, weak magnetic field at zero temperature~\cite{Shushpanov:1997sf}, and in both low temperature and weak magnetic field~\cite{Agasian:2001ym}, it is found that the GMOR relation for neutral pions holds true in the leading order chiral perturbation theory ($\chi$PT) in the chiral limit of quark masses as well. It is known that, at vanishing magnetic field, the transition temperature decreases with lighter pions~\cite{Ding:2019prx,Ding:2020rtq,Bazavov:2017xul, Kotov:2021rah, Aarts:2020vyb,Bhattacharya:2014ara,Umeda:2016qdo}. One may expect that the reduction of the transition temperature in the magnetic field could be associated with a lighter Goldstone boson, which can be a neutral pion. 

Meson spectrum of QCD in the external magnetic field is of important interest by itself~\cite{Andersen:2014xxa,Hattori:2015aki,Avancini:2016fgq,Wang:2017vtn,Mao:2018dqe,Avancini:2018svs,Coppola:2019uyr,Cao:2019res,Xu:2020sui,Xu:2020yag,Chao:2020jjy}. Moreover, the mass of a neutral pion in the external magnetic field may be helpful to understand the reduction of $T_{pc}$ given that neutral pion is a Goldstone boson in the nonzero magnetic field. Besides that, as implied from the QCD inequality~\cite{Hidaka:2012mz}, the sum of masses $M_{\pi^0_u}$ and $M_{\pi^0_d}$ as obtained from the up and down quark flavor components of contributions from connected diagrams to neutral pion correlators is a lower bound of the mass of a charged $\rho$ meson. The condensation of $\rho$ meson is also of particular interest since it could signal the transition of the QCD vacuum into a superconducting state in a sufficiently strong magnetic field~\cite{Chernodub:2010qx,Chernodub:2011mc}.
Most of the lattice studies on the meson spectrum in external magnetic fields have been performed in the quenched approximation, e.g., in the quenched two-color QCD with overlap fermions as valence quarks \cite{Luschevskaya:2012xd}, quenched QCD with Wilson fermions~\cite{Hidaka:2012mz,Bali:2017ian}, and overlap fermions~\cite{Luschevskaya:2014lga,Luschevskaya:2015cko} as valence quarks in the computation of meson correlation functions.  In the earlier studies, there exists a discrepancy in the behavior of $eB$ dependence of the neural pion mass from lattice QCD studies in the quenched approximation. It is found that the neutral meson mass in lattice study using quenched unimproved Wilson fermions firstly decreases and then increases as $eB$ grows~\cite{Hidaka:2012mz}. However, in the lattice study using quenched overlap fermions, the neutral meson mass monotonously decreases~\cite{Luschevskaya:2014lga}. Later the discrepancy is resolved as it is pointed out in Ref.~\cite{Bali:2017ian} that the $eB$ dependence of hopping parameter $\kappa$ has to be taken into account in the discretization scheme using Wilson fermions, and a monotonous reduction of neutral pion mass with increasing $eB$ is finally established in the quenched QCD~\cite{Bali:2017ian}. For a charged pion, a monotonous increase of its mass $M_{\pi^-}$ with $eB$ is found in all the quenched QCD studies~\cite{Hidaka:2012mz,Luschevskaya:2014lga,Luschevskaya:2015cko,Bali:2017ian}. On the other hand, lattice studies in full QCD on the $eB$ dependence of mass spectrum focus only on the light pseudoscalar mesons  ($\pi^{0,\pm}$)~\cite{Bali:2011qj,Bali:2017ian}.  It is shown in Refs.~\cite{Bali:2011qj,Bali:2017ian} that the behavior of masses of $\pi^0$ and $\pi^{\pm}$ in external magnetic fields obtained from (2+1)-flavor QCD using stout fermions have similar trends as those from quenched QCD~\cite{Luschevskaya:2014lga,Luschevskaya:2015cko,Bali:2017ian}.

The similar decreasing trend of the neutral pion mass and $T_{pc}$ in the nonzero magnetic field shown in~\cite{Bali:2017ian} might indicate the connection between these two quantities given that the GMOR relation holds in the nonzero magnetic fields. As the chiral condensate measures the strength of spontaneous chiral symmetry breaking,  the validity of the GMOR relation could imply that the mechanism for explicit breaking of chiral symmetry by the light quark mass is not changed by the magnetic field~\cite{Agasian:2001ym}. Furthermore, the Goldstone pion mass gives the strength of both spontaneous and explicit breaking of chiral symmetry as seen in the GMOR relation. 
Thus, it is also interesting to see whether the magnetic catalysis at zero temperature and the reduction of $T_{pc}$, in other words, the increase of chiral condensate and restoration of the chiral symmetry at a lower temperature, could be reconciled as implied from the GMOR relation.
However, it is not known yet whether the GMOR relation holds in a strong magnetic field, although $\chi$PT suggests its validity in the weak magnetic field. On the other hand, model studies suggest that the GMOR relation holds for a neutral chiral pion at nonzero magnetic field while it is violated for the charged ones at $eB\gtrsim$ 0.2 GeV$^2$~\cite{Orlovsky:2013wjd}.   Studies on the GMOR relation in the nonzero magnetic field are intricate due to the explicit breaking of rational invariance caused by the magnetic field~\cite{Fayazbakhsh:2013cha}. 
Investigations on the charged pion decay constant have been performed on the lattice~\cite{Bali:2018sey}. It was found that a further pion decay constant exists due to the possibility of a nonzero pion-to-vacuum transition via the vector piece of the electroweak current. Both charged and neutral pion decay constants have been parameterized for the one-pion-to-vacuum matrix elements of the vector and axial vector hadronic currents in the background magnetic fields~\cite{Coppola:2018ygv} and studied in the Nambu-Jona-Lasinio model~\cite{Coppola:2019uyr,Coppola:2019idh}. It is worth noting that early studies on pion decay constants in, e.g., Refs.~\cite{Shushpanov:1997sf,Agasian:2001ym,Orlovsky:2013wjd,Andreichikov:2018wrc} involve only the one for neutral pion fields related to the axial vector current parallel to the magnetic field.

In this paper, we focus on the chiral properties of QCD vacuum by studying the light and strange meson masses in the pseudoscalar channel, chiral condensates as well as the neutral pion and kaon decay constants related to the axial vector current in a wide range of magnetic field strength from 0 to $\sim 3.35$ GeV$^2$ in $N_f=2+1$ QCD. We will present a first lattice QCD study on the GMOR relation in the external magnetic field, show a novel decreasing behavior of charged pseudoscalar meson mass, and discuss the magnetic polarizabilities of light and strange pseudoscalar mesons. We will also present the first observation of $qB$ scaling of up and down quark flavor components of neutral pion mass, neutral pion decay constant, and chiral condensates.  The results are obtained based on lattice QCD simulations performed on $32^3\times96$ and $40^3\times96$ lattices at a single lattice cutoff $a=0.117$ fm using highly improved staggered fermions with a heavier-than-physical pion mass of $\sim220$ MeV at zero temperature. 

The paper is organized as follows. In Sec.~\ref{sec:basics}, we introduce the basic quantities we are going to study. In Sec.~\ref{sec:setup}, we will explain our simulation parameters in lattice QCD and our methodology to extract the meson masses and the amplitudes to compute the decay constants, show the volume dependence of correlation functions, and discuss the $\pi-\rho$ mixing in the magnetic field via the generalized Ward-Takahashi identities as well as the extraction of infrared contributions from the chiral condensates. In Sec.~\ref{sec:results}, we will demonstrate the $qB$ scaling, and then present our results on masses of pseudoscalar mesons, magnetic polarizabilities, chiral condensates, neutral pion and kaon decay constants as well as corrections to the GMOR relation as a function of $eB$. Finally, we will summarize our results in Sec.~\ref{sec:conclusions}. In Appendix~\ref{sec:app_WI} we will show the extended Ward-Takahashi identities in the nonzero magnetic field, in Appendix~\ref{A:GMOR} we will show the GMOR relation for up and down quark components of neutral pion at $eB=0$, and in Appendix~\ref{sec:app_HISQ}, we will show the details of the implementation of the magnetic field in the lattice QCD simulations using the highly improved staggered fermions. Some of our previous results on the masses of pseudoscalar mesons have been reported in conference proceedings~\cite{Ding:2020jui,Ding:2020pao}.

\section{Temporal correlators, masses of pseudoscalar mesons, chiral condensates and neutral pion and kaon decay constants}
\label{sec:basics}
Hadron spectrum in the vacuum can be extracted from two-point temporal correlation functions in the Euclidean space
\begin{equation}
G(\tau) = \int d^3\vec{x} \left\langle 
\mathcal{M}(\vec{x},\tau)
\Big( \mathcal{M}(\vec{0},0)\Big)^\dagger \right\rangle
\;,
\end{equation}
where $\mathcal{M}=\bar{\psi}(\tau,\vec{x}) \,\Gamma \,\psi(\tau,\vec{x}) $ is a meson operator that projects to a certain quantum channel $\Gamma=\Gamma_D\bigotimes t^a$ with Dirac matrices $\Gamma_D$ and a flavor matrix $t^a$. For instance, $\Gamma=\gamma_5$ and $\gamma_\mu$ correspond to the pseudoscalar and vector channel, respectively.
The angular brackets $\langle\cdots\rangle$ stand for the expectation value over the gauge field ensembles. The temporal correlator decays exponentially at a large distance $\tau$
\begin{equation}
\lim_{\tau \rightarrow\infty} G(\tau) \sim e^{-m_{\Gamma}\tau},
\end{equation}
which defines the mass $m_{\Gamma}$ of the corresponding ground state.
In the case of staggered fermions, the corresponding meson operators are of the form  $\bar{\psi}(x)(\Gamma_D\bigotimes\Gamma_T^*)\psi(x)$ with $\psi(x)$ a 16-component hypercubic spinor and $\Gamma_D$ and $\Gamma_T$  Dirac matrices in spin and taste space, respectively~\cite{Bazavov:2019www,Durr:2005ax,Kilcup:1986dg}. In our work, we consider local operators only, and the meson operator thus reduces to $\mathcal{M}=\zeta(\vec{x})\bar{\chi}(\vec{x})\chi(\vec{x})$, where $\zeta({x})$ is the phase factor depending on the choice of $\Gamma=\Gamma_D=\Gamma_T$ and $\chi(\vec{x})$ is the staggered fermion field.

The connected part of the correlation function of the staggered bilinear can thus be written as
\begin{equation}
\mathrm{G}(\tau)=-\sum_{x,y,z} \,\zeta(\vec{n}) \mathrm{Tr} \left[\left(M^{-1}(\vec{x}, \tau;\vec{0},0)\right)^\dagger M^{-1}(\vec{x}, \tau;\vec{0},0)\right],
\end{equation}
where $M^{-1}(\vec{x}, \tau;\vec{0},0)$ is the staggered propagator from $(\vec{0},0)$ to $(\vec{x}, \tau)$.   
In this work, we focus on the mesons in the pseudoscalar channel built from $\bar{q}_iq_j$ flavor combinations, where $i,j=u,d,s$, and the phase factor for the pseudoscalar channel is $\zeta(\vec{n})=1$. Because of the presence of the magnetic field, the isospin symmetry of up and down quarks is broken by their different electric charges. The neutral pion is not an isovector state anymore. 
The computation of neutral pion correlation function at nonzero magnetic fields, extending from the case at zero magnetic fields, thus could include both connected and disconnected parts of the correlation function and a mixing factor between $u \bar{u}\left(\pi_{u}^{0}\right)$ and $d \bar{d}\left(\pi_{d}^{0}\right)$ components of the connected part.\footnote{In the presence of the magnetic field, the operator for $\pi^0$ could be $\alpha\,\bar{u}  \gamma_{5} u -\beta\,\bar{d} \gamma_{5} d$ with $\alpha^2+\beta^2=1$~\cite{Bali:2017ian}.  To extract the mass of $\pi^0$ and neutral pion decay constant $f_{\pi^0}$ from the neutral pion correlation function, we assume $\alpha=\beta=1/\sqrt{2}$ as the case for the vanishing magnetic field.}
It has been shown in Ref.~\cite{Luschevskaya:2015cko} that the quark-line disconnected part in quenched QCD is negligible in the nonzero magnetic fields. In our current study of neutral pions, we thus neglect the disconnected contributions which could be small as well (cf. discussions in Sec.~\ref{sec:setup_mixing}).

A typical staggered meson correlator, for a fixed separation (in lattice unit) between the source and sink,  is an oscillating correlator that simultaneously couples to two sets of mesons with the same spin and opposite parities. It thus can be parameterized as~\cite{Bazavov:2019www}
\begin{eqnarray}
G\left(n_{\tau}\right)=\sum_{i=1}^{N_{nosc}} A_{nosc, i} \exp \left(-M_{nosc, i} \,n_{\tau}\right)-(-1)^{n_{\tau}} \sum_{i=0}^{N_{osc}} A_{osc, i} \exp \left(-M_{ osc, i} \,n_{\tau}\right),
\label{eq:fit_ansatz}
\end{eqnarray}
where $N_{nosc}$ ($N_{osc}$) is the number of nonoscillating (oscillating) meson states whose masses are denoted by $M_{nosc,i}$ ($M_{osc,i}$), and $n_\tau=\tau/a\in
\mathbb{Z}$ with lattice spacing $a$. Note that both amplitudes, $A_{nosc,i}$ and $A_{osc,i}$, are positive. In the current study, the mass $M_{nosc}$ is the mass of the pseudoscalar meson we are interested in.

It is well known that the energy of a pointlike charged particle in the nonzero magnetic field and at zero temperature are the Landau levels,
\begin{equation}
E_n^2 = M^2 + (2n+1)|eB| -g\,s_z\,qB +  p_z^2, ~~n\in \mathbb{Z}^+_0,
\end{equation}
where $M$ is the mass of a charged particle at zero magnetic field, $q$ is the electric charge of the particle, $s_z$ is the spin polarization in the $z$ direction, and the magnetic field is assumed to go along with the $z$ direction. For pseudoscalar mesons, the gyromagnetic ratio $g=0$ while for vector mesons $g=2$. In the case of the lowest Landau level and zero momentum along the $z$ direction, i.e., $n$=0 and $p_z$=0, the mass of a charged pointlike pseudoscalar meson in the external magnetic fields can be expressed as
\begin{equation}
M^{\pm}_{\mathrm {ps}}(B) = \sqrt{\left(M^{\pm}_{\mathrm {ps}}(B=0)\right)^2+ |eB|}.
\label{eq:LLL}
\end{equation}

While the charged particle becomes heavier with increasing $eB$, a neutral particle is supposed to remain independent of the magnetic field if it remains as a pointlike particle.  The lightest pseudoscalar mesons like pions are of particular interest as they are Goldstone bosons at the vanishing magnetic field.  And their masses are connected to the quark chiral condensate in the vacuum known as the Gell-Mann-Oakes-Renner relation~\cite{GellMann:1968rz}. The corrected GMOR relation including a correction term $\delta_\pi$ is expressed as follows:
\begin{equation}
(m_u + m_d) ~\left(\pbp_u + \pbp_d\right)=2 f_\pi^2 M_\pi^2\, (1-\delta_\pi) .
\label{eq:GMOR}
\end{equation}
The above corrected GMOR relation in the two-flavor theory has been extended to the three-flavor case including a strange quark as follows~\cite{Gasser:1984gg}:
\begin{equation}
(m_s + m_d) ~\left(\pbp_s + \pbp_d\right) =  2 f_K^2 M_K^2 \,(1-\delta_K),
\label{eq:GMOR_K}
\end{equation}
where $m_\pi$ and $m_K$ are the masses of pion and kaon, respectively, $f_\pi  (f_K)$ is the pion (kaon) decay constant, $m_{f=u,d,s}$ is the mass of  up, down, and strange quarks, and the corresponding quark chiral condensate is denoted by $\pbp_{f=u,d,s}$. The single flavor quark chiral condensate $\pbp_f$ is obtained as follows:
\begin{equation}
\pbp_f = \frac{1}{4} \frac{1}{V} \frac{\partial \ln Z}{\partial m_f}= \frac{1}{4} \frac{1}{V} \mathrm{Tr} D^{-1}_f, 
\label{eq:pbp_q}
\end{equation}
where $Z$ is the partition function of QCD, $V$ is the full volume of space-time, and the factor $1/4$ accounts for the fourth root of the Dirac matrix $D_f$ in the staggered theory.  $\delta_\pi$ and $\delta_K$ are the next-to-leading order chiral corrections, and both of them are related to certain low-energy constants and have a relation of $\delta_K=M_K^2/M_\pi^2\delta_\pi$~\cite{Gasser:1984gg}. The estimates on these two quantities using $\chi$PT combining with QCD sum rules are $\delta_\pi=(6.2\pm1.6)\%$ and $\delta_K=(55\pm5)\%$ at the physical-mass point and the vanishing magnetic field~\cite{Bordes:2010wy,Bordes:2012ud}. As $\delta_\pi$ and $\delta_{K}$ go to zero, Eq.~\ref{eq:GMOR} and Eq.~\ref{eq:GMOR_K} recover the two-flavor and three-flavor GMOR relations obtained from the leading order chiral perturbation theory.

As mentioned in the Introduction, some additional pion decay constants related to the vector and axial vector currents can be defined in the external magnetic field~\cite{Fayazbakhsh:2013cha,Bali:2018sey,Coppola:2018ygv}. In the current paper, we focus on the original neural pion and kaon decay constants related only to the axial vector current parallel to the magnetic field at zero momentum. This decay constant has the same definition as that at the vanishing magnetic field~\cite{Fayazbakhsh:2013cha,Coppola:2018ygv,Kilcup:1986dg}, which is written as follows:
 \begin{eqnarray}
 \label{eq:fpi}
 \sqrt{2} f_{\pi^0} \,M_{\pi^0}^{2}&=&\left(m_{u}+m_{d}\right)\left\langle 0\left|\frac{1}{\sqrt{2}}\left(\bar{u}  \gamma_{5} u -\bar{d} \gamma_{5} d\right) \right| \pi^{0}(\boldsymbol{p}=0)\right\rangle,\\
 \label{eq:fk}
 	\sqrt{2} f_{K^0} \,M_{K^0}^{2}&=&\left(m_{d}+m_{s}\right)\left\langle 0\left|\bar{d} \gamma_{5} s\right| K^{0}(\boldsymbol{p}=0)\right\rangle.
 \end{eqnarray}
We also look into the up and down quark flavor components of the connected part of the neutral pion correlation functions which lead to decay constants ($f_{\pi^0_u}$ and $f_{\pi^0_d}$) and masses ($M_{\pi^0_u}$ and $M_{\pi^0_d}$), whose relations are expressed as follows:
  \begin{eqnarray}
  \label{eq:fpiu}
 \sqrt{2} f_{\pi^0_u} M_{\pi^0_u}^{2}&=&2m_{u}\left\langle 0\left|\bar{u} \gamma_{5} u\right| \pi^{0}_u(\boldsymbol{p}=0)\right\rangle, \\
   \label{eq:fpid}
 \sqrt{2} f_{\pi^0_d} M_{\pi^0_d}^{2}&=&2m_{d}\left\langle 0\left|\bar{d} \gamma_{5} d\right| \pi^{0}_d(\boldsymbol{p}=0)\right\rangle.
 \end{eqnarray}
 The corresponding GMOR relations 
 \begin{eqnarray}
  \label{eq:GMOR_u}
  4 m_u ~\pbp_u &=&2 f_{\pi^0_u}^2 M_{\pi^0_u}^2\, (1-\delta_{\pi^0_u}) \, ,\\
    4 m_d ~\pbp_d &=&2 f_{\pi^0_d}^2 M_{\pi^0_d}^2\, (1-\delta_{\pi^0_d})\, , 
 \label{eq:GMOR_d}
 \end{eqnarray} 
 are derived from the Ward-Takahashi identity at zero magnetic fields in Appendix \ref{A:GMOR}. Here, the corrections $\delta_{\pi^0_{u,d}}$ include contributions from the anomalous part as seen from Eqs.~\ref{eq:app_GMORpi0u} and~\ref{eq:app_GMORpi0d}.
 The matrix elements in Eqs.~\ref{eq:fpi}, \ref{eq:fk}, \ref{eq:fpiu}, and~\ref{eq:fpid} can be extracted from the correlation function of the pseudoscalar meson at zero spatial momentum $\left\langle O_S(\tau) \mathrm{P}_W(0)\right\rangle$~\cite{Aoki:1999av}. The amplitude of the correlation function can be written as
 \begin{equation}
 	C_{O_{S} {\rm P}_{W}}=\frac{\left\langle 0\left|O_{S}\right| \mathrm{P}(\vec{p}=0)\right\rangle\left\langle \mathrm{P}(\vec{p}=0)\left|\mathrm{P}_{W}\right| 0\right\rangle}{2 M_{\mathrm{P}} V_{s}},
 	\label{eq:cpp}
 \end{equation}
 where $V_s$ is the spatial lattice volume and $\rm P$ denotes the operator for $\pi^0$, $K^0$, and $\pi^0_{u,d}$ with $M_P$ the corresponding meson masses.  By inserting the corresponding operator ($O_S=\mathrm{P}_W$) into Eq.~\ref{eq:cpp} and combining Eqs.~\ref{eq:fpi}, \ref{eq:fk}, \ref{eq:fpiu} and~\ref{eq:fpid}, the pion and kaon decay constants can be obtained as follows~\cite{Aoki:1999av}:
 \begin{eqnarray}
 \label{eq:fpi_det}
 	f_{\mathrm{P}^\pi}&=&2 m_l \, \sqrt{\frac{V_{s}}{4}} \sqrt{\frac{C_{\mathrm{P}^\pi_W \mathrm{P}^\pi_W}}{M_{\mathrm{P}^\pi}^{3}}},\\
 		f_{K^0}&=& (m_d+m_s) \, \sqrt{\frac{V_{s}}{4}} \sqrt{\frac{C_{K^0_W K^0_W}}{M_{K^0}^{3}}},
 	\label{eq:fK_det}
 \end{eqnarray}
where the factor of $1/\sqrt{4}$ on the right-hand sides of the above equations accounts for the case in staggered fermions, $m_l\equiv m_u=m_d$, and $\mathrm{P}^\pi$ denotes the cases for $\pi^0$, $\pi^0_u$, and $\pi^0_d$ . The amplitudes $C_{\mathrm{P}^\pi_W \mathrm{P}^\pi_W}$ and $C_{K^0_W K^0_W}$ as well as the masses $M_{\mathrm{P}^\pi}$ and $M_{K^0}$ will be obtained from the mass fit (cf. Eq.~\ref{eq:fit_ansatz}) to the correlators in the pseudoscalar channel.  We emphasize that $\pi_u^0$ and $\pi^0_d$ are not physical states in the sense that they do not directly correspond to real particles like the neutral pion. However, $M_{\pi_u^0}$ and $M_{\pi_d^0}$, as determined from the large-time behavior of up and down quark components of the connected part of the neutral pion correlator (i.e., $\bar{u} \gamma_5 u$ and $\bar{d}\gamma_5 d$ correlators, respectively),  are useful quantities since their average provides a lower bound of charged $\rho$ mass due to the QCD inequality as mentioned in Sec.~\ref{sec:intro}. On the other hand, $f_{\pi^0_u}$ and $f_{\pi^0_d}$ are determined from $\bar{u}\gamma_5u$ and $\bar{d}\gamma_5d$ correlators according to Eq.~\ref{eq:fpi_det}, respectively.

\section{Numerical setups}
\label{sec:setup}

\subsection{Lattice setup}
\label{sec:setup_lqcd}
Most of the previous lattice QCD studies for $(2+1)$-flavor QCD in the external magnetic field were performed using the stout staggered fermions. In our current simulations, we use $N_f=2+1$ highly improved staggered quarks (HISQ)~\cite{Follana:2006rc}. At a given value of the lattice spacing, the HISQ action achieves better taste symmetry than two-stout actions~\cite{Ding:2015ona}. The HISQ action is constructed by the Kogut-Susskind one-link action and Naik improvement term with smeared links.
The smeared links are obtained in the following way.
First, level-one smeared links $V_\mu$ are constructed by fat-7 from thin \suthree  ~links $U_\mu$.
Next, reunitarized links $W_\mu$ are constructed by projecting $V_\mu$ on U(3).
Finally, level-two smeared links $X_\mu$ are constructed by fat-7 from thin \suthree ~links $W_\mu$ with
the Lepage term. The HISQ Dirac operator is built by the Kogut-Suskind term with $X_\mu$ and
the Naik term with $W_\mu$. The magnetic field only couples directly to quarks; thus, the implementation of the magnetic field is done just by replacing $X_\mu \to u_\mu X_\mu$ in the Kogut-Susskind term
and $W_\mu \to u_\mu W_\mu$ in the Naik term~\cite{Tomiya:2019nym}.  Details about the implementation of magnetic fields in the lattice QCD simulations using the HISQ action are summarized in Appendix~\ref{sec:app_HISQ}.

The external magnetic field pointing along the $z$ direction $\vec{B}=(0,0,B)$ is described by a fixed factor  $u_\mu(n)$ of the U(1) field, and 
$u_\mu(n)$ is expressed as follows in the Landau gauge \cite{AlHashimi:2008hr,Bali:2011qj},
\begin{eqnarray} 
\label{eq:BC_extB}
u_x(n_x,n_y,n_z,n_\tau)&=&
\begin{cases}
\exp[-iq a^2 B N_x n_y] \;\;&(n_x= N_x-1)\\
1 \;\;&(\text{otherwise})\\
\end{cases}\notag\\
u_y(n_x,n_y,n_z,n_\tau)&=&\exp[iq a^2 B  n_x],\label{eq:def_mag_u} \notag\\
u_z(n_x,n_y,n_z,n_\tau)&=&u_t(n_x,n_y,n_z,n_\tau)=1.
\end{eqnarray}
Here the lattice size is denoted as $(N_x,\; N_y,\; N_z,\; N_\tau)$ and coordinates as $n_\mu=0,\cdots, N_\mu-1$ ($\mu = x,\; y,\; z,\; \tau$).
The external magnetic field applied along the $z$ direction $\mathbf{B}=(0,0,B)$ is quantized in the following way:
\begin{equation} 
q B= \frac{2 \pi N_{b}}{N_{x} N_{y}}a^{-2},
\end{equation} 
where $q$ is the electric charge of a quark, and $N_x$ $(N_y)$ is the number of points in the $x(y)$ direction on the lattice. Since the quantization has to be satisfied for all the quarks in the system, a greatest common divisor of the electric charge of all the quarks, i.e., $|q_d|=|q_s|=e/3$ with $e$ the elementary electric charge, is chosen in our simulation. In practice, the strength of the magnetic field $eB$ is expressed as follows
\begin{equation}
eB= \frac{6 \pi N_{b}}{N_{x} N_{y}} a^{-2},
\end{equation}
where $N_b \in \mathbf{Z}$ is the number of magnetic fluxes through a unit area in the $x$-$y$ plane. The periodic boundary
 condition for U(1) links is applied for all directions except for
 the $x$ direction, as shown in Eq.~\ref{eq:BC_extB}. As limited by the boundary condition, $N_b$ is constrained in the range of $0\leq N_b<\frac{N_xN_y}{4}$. In our study $N_\sigma\equiv N_x=N_y=N_z$.

For the gauge part, we use a tree-level improved Symanzik gauge action. The simulations have been performed on lattices with temporal size $N_\tau$=96 at zero temperature. The inverse lattice spacing is $a^{-1}\simeq1.685$ GeV, and strange quark mass $m_s$ is tuned to its physical value by tuning the mass of the $\eta_{s}^0$ meson, $M_{\eta_{s}^0}\simeq 684$ MeV, which is based on the leading order chiral perturbation theory relation $M_{\eta_{s}^0}=\sqrt{2M_K^2-M_\pi^2}$ between masses of $\eta^0_s,\ \pi$ and $K$. The light quark mass is then set to be $m_l=m_s/10$ corresponding to a pion mass $M_\pi\simeq$ 220 MeV in the vacuum. Details on the scale setting, which is extensively used by the HotQCD Collaboration, can be found in Refs.~\cite{Bazavov:2011nk,Bazavov:2019www}. In the HISQ discretization scheme, so-called taste symmetry violations give rise to a distortion of the light pseudoscalar (pion) meson masses. These discretization effects are commonly expressed in terms of a root-mean-square (RMS) pion mass, which approaches the Goldstone pion mass in the continuum limit. The computational setup with the three different lattice cutoff values has been discussed in~\cite{Bazavov:2011nk}. For the lattice spacing used in our simulation, one finds $M_{RMS}\approx$ 240 MeV for physical quark masses ~\cite{Ding:2015ona,Bazavov:2011nk}.

In our simulations, 16 different values of $N_b$ on $32^3\times96$ lattices have been chosen, i.e., 0, 1, 2, 3, 4, 6, 8, 10, 12, 16, 20, 24, 32, 40, 48, and 64 whose corresponding magnetic field strength $eB$ ranges from 0 to around 3.35 GeV$^2$. To check the volume effects, simulations with $eB\simeq1.67$ GeV$^2$ on $40^3\times96$ have also been performed. If not mentioned explicitly, most of the results shown in this paper are obtained from $32^3\times96$ lattices. To have small discretization errors for $B$, the magnetic field implemented in the lattice simulations should be small in lattice units, i.e., $a^2q_dB\ll 1$ or $N_b/N_\sigma^2\ll1$~\cite{Endrodi:2019zrl}. In our work, the largest number of magnetic fluxes $N_b^{max}=64$ resulting in $N_b^{max}/N_\sigma^2\approx6\%$. Thus, the discretization errors for $B$ should be small. All configurations have been produced using the rational hybrid Monte Carlo algorithm and saved by every 5 time units. The statistics for each $N_b$ are about $O(10^3)$ which are listed in Table~\ref{tab:statistics} in detail.

\begin{table}[htp!]
		\begin{minipage}{0.7\hsize}
			\centering
	\begin{tabular}{c|c|c|c|c|c|c|c|c}
		\hline
		$eB$ [GeV$^2$] & 0 & 0.052 & 0.104 & 0.157 & 0.209 & 0.314 & 0.418 & 0.523   \\
		\hline
		$N_b$ & 0 & 1 & 2 & 3 & 4 & 6 & 8 & 10 \\
		\hline
		\# of conf. & 2548 & 2651 & 3135 & 2444 & 2224 & 3142 & 2935 & 3302 \\
		\hline
		
		\hline
		$eB$ [GeV$^2$] & 0.627 & 0.836 &  1.045  & 1.255 & 1.673 & 2.09& 2.510 & 3.345    \\
		\hline
		$N_b$ & 12 & 16 & 20 & 24 & 32 & 40 & 48 & 64 \\
		\hline
		\# of conf. & 3432 & 1993 &  3160 & 1994 & 2174 & 3385& 2234 & 3263  \\
		\hline
	\end{tabular}\\
\vspace{0.1cm}
{Lattice size: $32^3\times96$}
\end{minipage}
	\begin{minipage}{0.25\hsize}
	\begin{tabular}{c|c}
		\hline
		$eB$ [GeV$^2$] & 1.673    \\
		\hline
		$N_b$ & 50  \\
		\hline
		\# of conf. & 1260  \\
		\hline
	\end{tabular}\\
\vspace{0.1cm}
	{Lattice size: $40^3\times96$}
\end{minipage}
	\caption{The statistics of analyzed $32^3\times96$ and $40^3\times96$ lattices with $\beta=6.68$ and $a\simeq0.117$ fm ($a^{-1}\simeq1.685$ GeV) produced using the HISQ fermion action and a tree-level improved Symanzik gauge action in $N_f=2+1$ QCD. The light quark mass in lattice spacing is tuned to be $am_l=0.00506=am_s/10$ with $m_s$ the physical strange quark mass. The resulting masses of the pion, $\eta^0_s$, $K^0$, and $\rho$ at $eB=0$ are $M_\pi=220.61(6)$ MeV, $M_{\eta^0_s}$=684.44(6) MeV, $M_K=507.0(7)$ MeV, and $M_\rho$=779(35) MeV, respectively.
	}
	\label{tab:statistics}
\end{table}

\subsection{Meson correlation functions}
\label{sec:setup_fits}
\begin{figure}[htbp]
	\centering
	\includegraphics[width=0.47\textwidth]{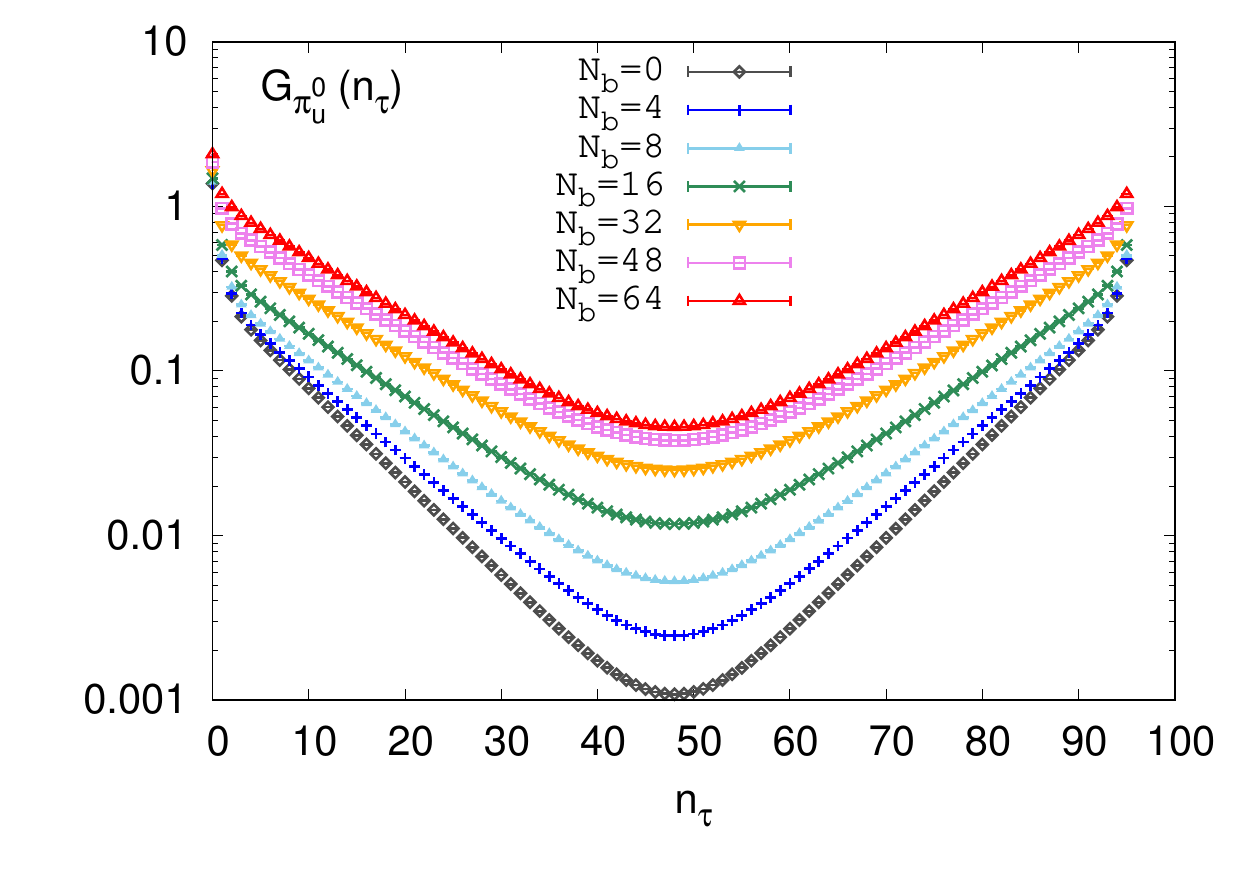}
	\includegraphics[width=0.47\textwidth]{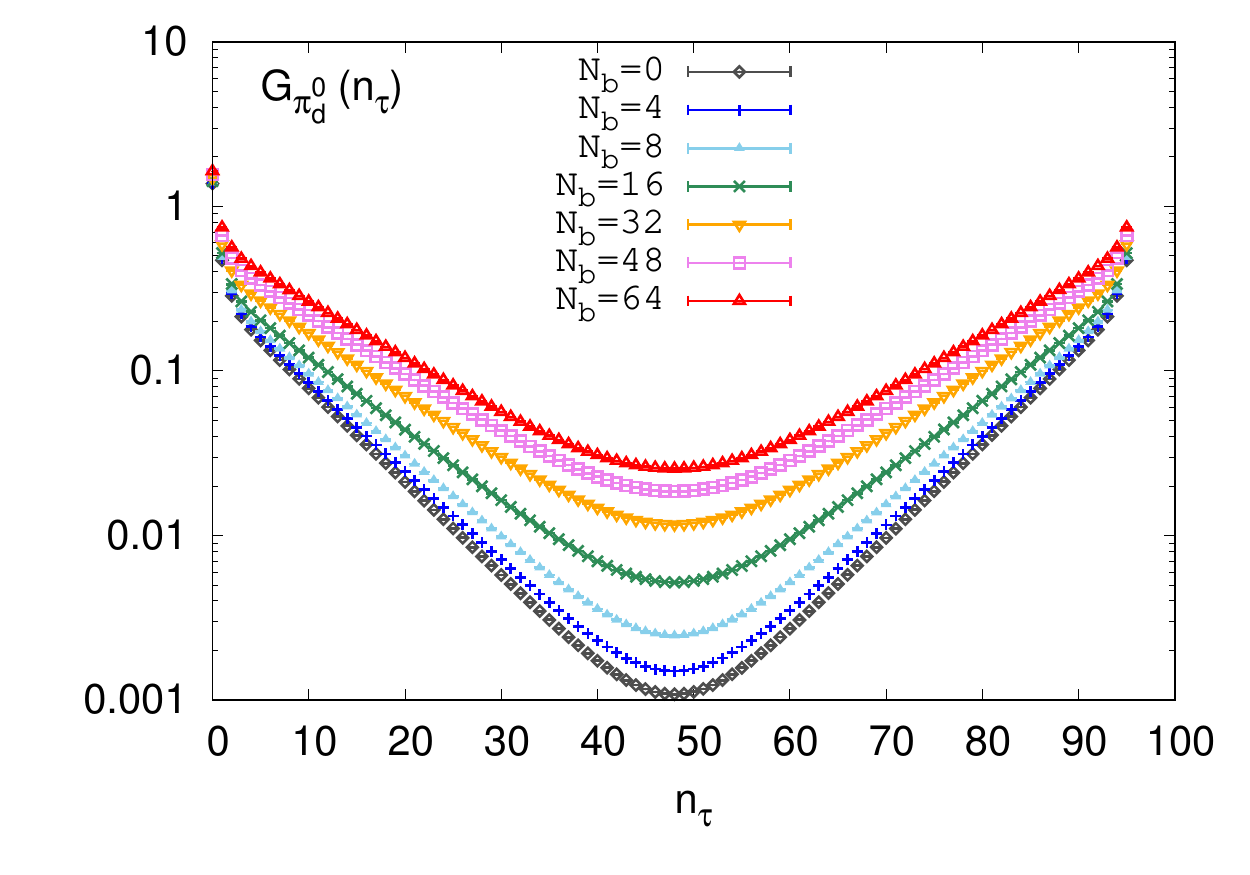}
	\caption{Left: The $u\bar{u}$ component of the neutral pion correlation function $G_{\pi^0_u}(n_\tau)$ at different values of $N_b$ calculated using 12 corner wall sources. Right: Same as the left one but for the $d\bar{d}$ component $G_{\pi^0_d}(n_\tau)$. $G$ shown here are rescaled with lattice spacing $a$ such that they are dimensionless quantities.}
	\label{fig:corr}
\end{figure}

We show our results of correlation functions for $\pi_u^0~(u\bar{u})$ and $\pi_d^0 ~(d\bar{d})$ as an example in the left and right plots of Fig.~\ref{fig:corr}, respectively. It can be seen clearly from the plots that both correlation functions become larger with increasing magnetic field strength $eB$, and the $u\bar{u}$ component of the two-point correlation function increases faster. It may be understood that the internal structure of pion is probed, and the $u\bar{u}$ component is more affected due to the larger absolute value of the electric charge of the $u$ quark.

We then extract the mass of neutral and charged pseudoscalar mesons by fitting to the corresponding temporal correlation functions with the ansatz (cf. Eq.~\ref{eq:fit_ansatz}). The nonoscillating states are the physical states we need, and oscillating states are also necessary to be included in the fit, particularly in the case of correlation functions for charged particles. In the fits, we used various numbers of nonoscillating as well as oscillating states, i.e., ($N_{nosc}$\,, $N_{ osc}$) has been set to (1,0), (1,1), (2,1). All the fits have been performed in a given interval $[\tau_{min},N_\tau/2]$, where $\tau_{min}$ ranges from 0 to $N_\tau/2-1$. The final results are chosen as the best fit from all different fit modes through the corrected Akaike Information criterion (AICc) \cite{1100705, cavanaugh1997unifying}
\begin{equation}
\mathrm{AICc}=2 k-\ln (\hat{L})+\frac{2 k^{2}+2 k}{n-k-1},
\end{equation}
where $k$ is the number of parameters, $\hat{L}$ is the likelihood function, and the last term is needed to correct overfitting if the number of data points $n$ is not much larger than $k$. We then choose a plateau in the AICc selected results and obtain the final mass and uncertainty from the plateau using a Gaussian bootstrapping method~\cite{Bazavov:2019www}.

\begin{figure}[htbp]
	\centering
	\includegraphics[width=0.47\textwidth]{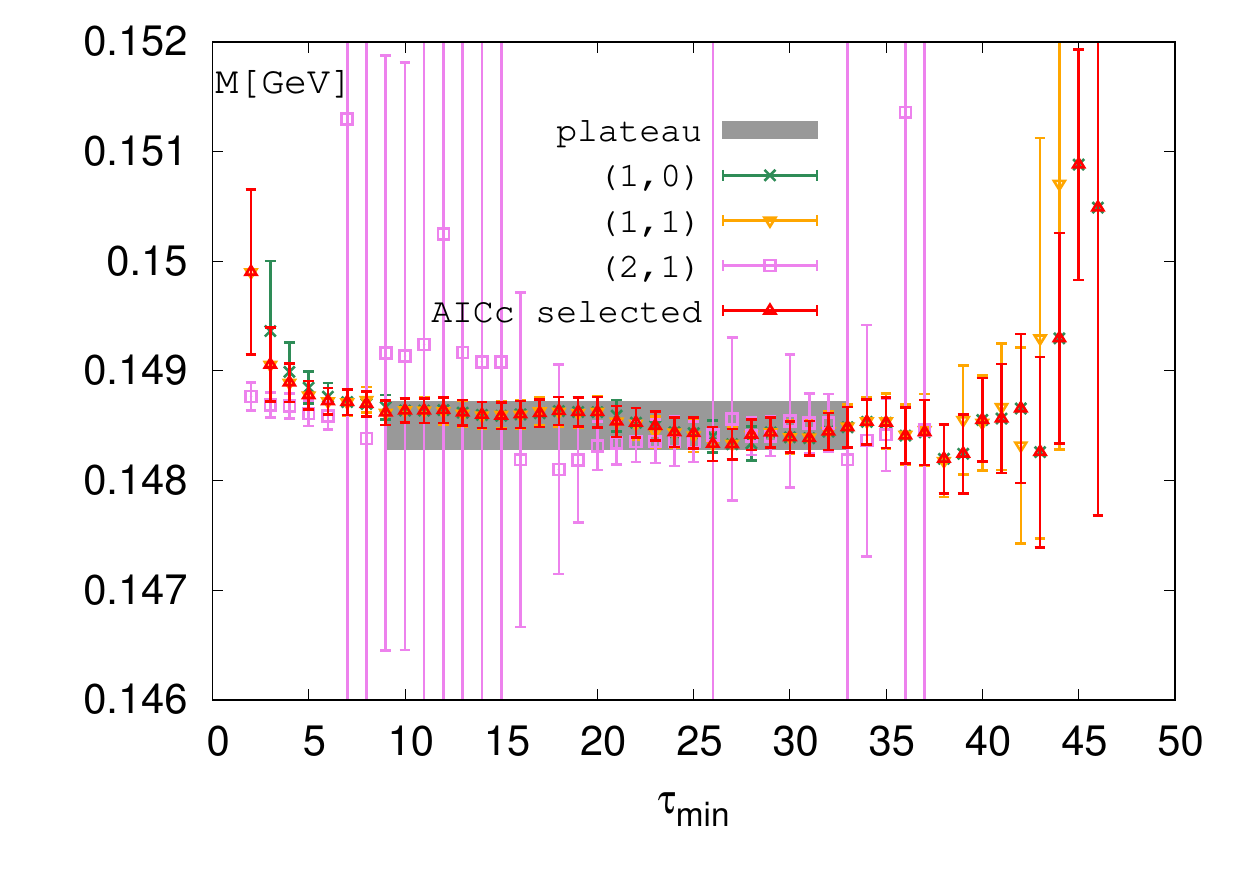} 
        \includegraphics[width=0.47\textwidth]{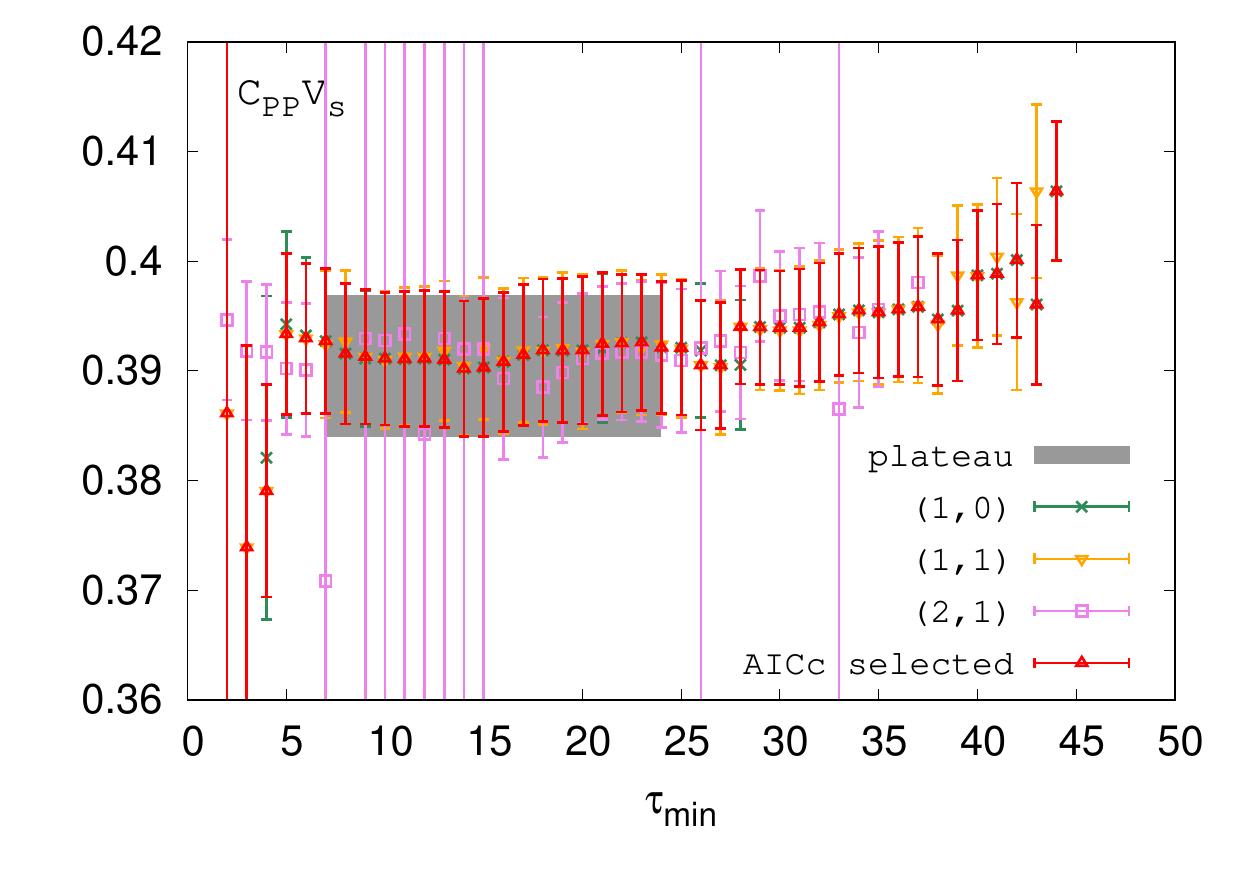} 
	\includegraphics[width=0.47\textwidth]{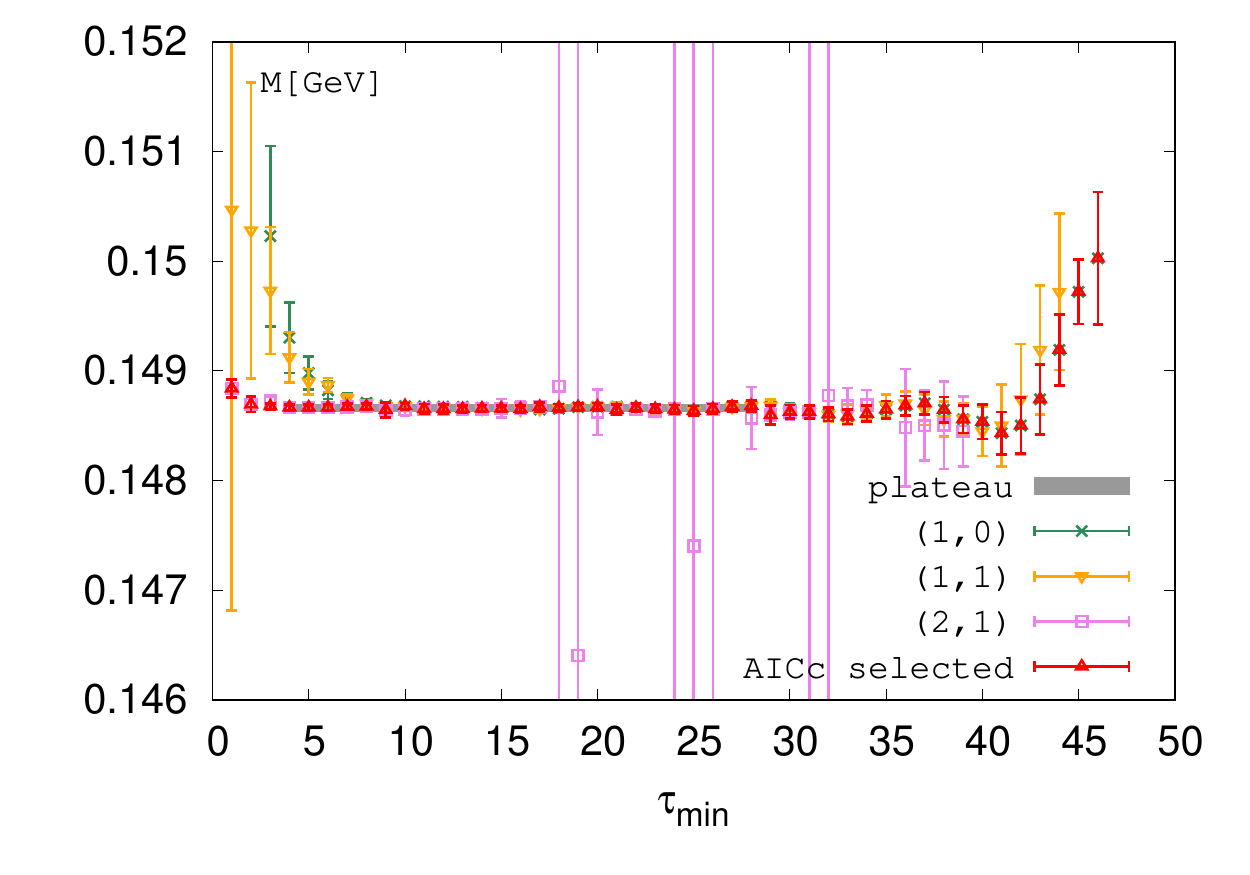} 
	\includegraphics[width=0.47\textwidth]{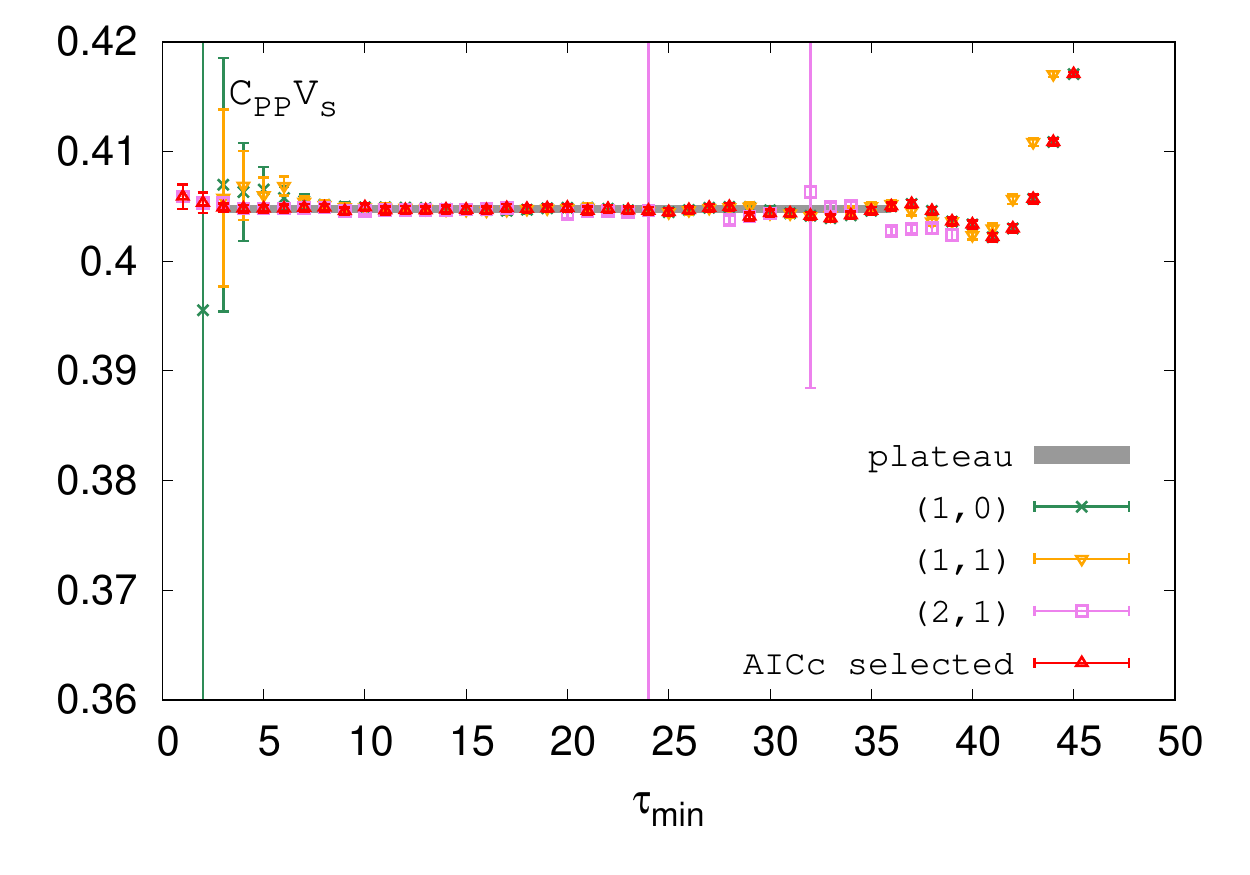} 
	\caption{Top: Mass (left) and its associated matrix element $C_{PP}$ multiplied by the spatial volume $V_s$  (right) of $\pi^0_u$ at $N_b=16$ ($eB$ = 0.84 GeV$^2$) with there different fit modes: (no. of nonoscillating states, no. of oscillating states) = (1,0), (1,1), (2,1) obtained from the fits to correlation functions measured using a single point source. Bottom: Same as the top two plots but for fits to correlation functions measured using 12 corner wall sources.}
	\label{fig:aicc}
\end{figure}
 The usage of a single point source in the computation of correlation functions does not suppress the excited states and could make the isolation of the ground state from excited states difficult unless the states are well separated or the lattice extent is large. The usage of smeared (extended) sources can help to suppress the excited states and make the extraction of ground state mass and amplitude more reliable. We thus also compute correlation functions using corner wall sources. When applying a corner wall source in temporal correlators, a unit source is needed at the origin of each $2^3$ cube  on a chosen $z$ slice~\cite{Bernard:1991ah,Bernard:1993an,Bernard:2001av,Bazavov:2019www}. We 
 further improve the signal by putting 12 corner wall sources at (0,0,0,0), (0,0,0,8),...,(0,0,0,88). The signal-to-noise ratio $\delta G(\tau) / \langle G(\tau)\rangle$ obtained using a single corner wall source is reduced by a factor of 6 compared to the results obtained using a single point source. The signal-to-noise ratio obtained using multiple corner wall sources is $\sqrt{\#\ \mathrm{of\ sources}}$ times better than using a single corner wall source. We provide a typical example of the fit results, whose procedure was illustrated earlier, for the $u\bar{u}$ component of a neutral pion correlation function at $eB$=0.84 GeV$^2$ measured using a single point source (top two plots) and 12 corner wall sources (bottom two plots) in Fig.~\ref{fig:aicc}. We find that the usage of the corner wall sources yields a much better signal in the ground states than   that of the point source and has a longer plateau with much smaller uncertainty. The value of ground state mass extracted using the corner wall source is consistent with that extracted using the point source. As seen from Fig.~\ref{fig:aicc}, the corner wall source also works better for the extraction of the amplitude, i.e., the amplitude obtained with the corner wall source is about 3\% larger than the amplitude obtained with the point source while the relative error is reduced by $\sim$20 times. The amplitude obtained from the corner wall sources is then used in the calculation of decay constants in Sec.~\ref{sec:results}.

\begin{figure}[tbph]
	\centering
	\includegraphics[width=0.47\textwidth]{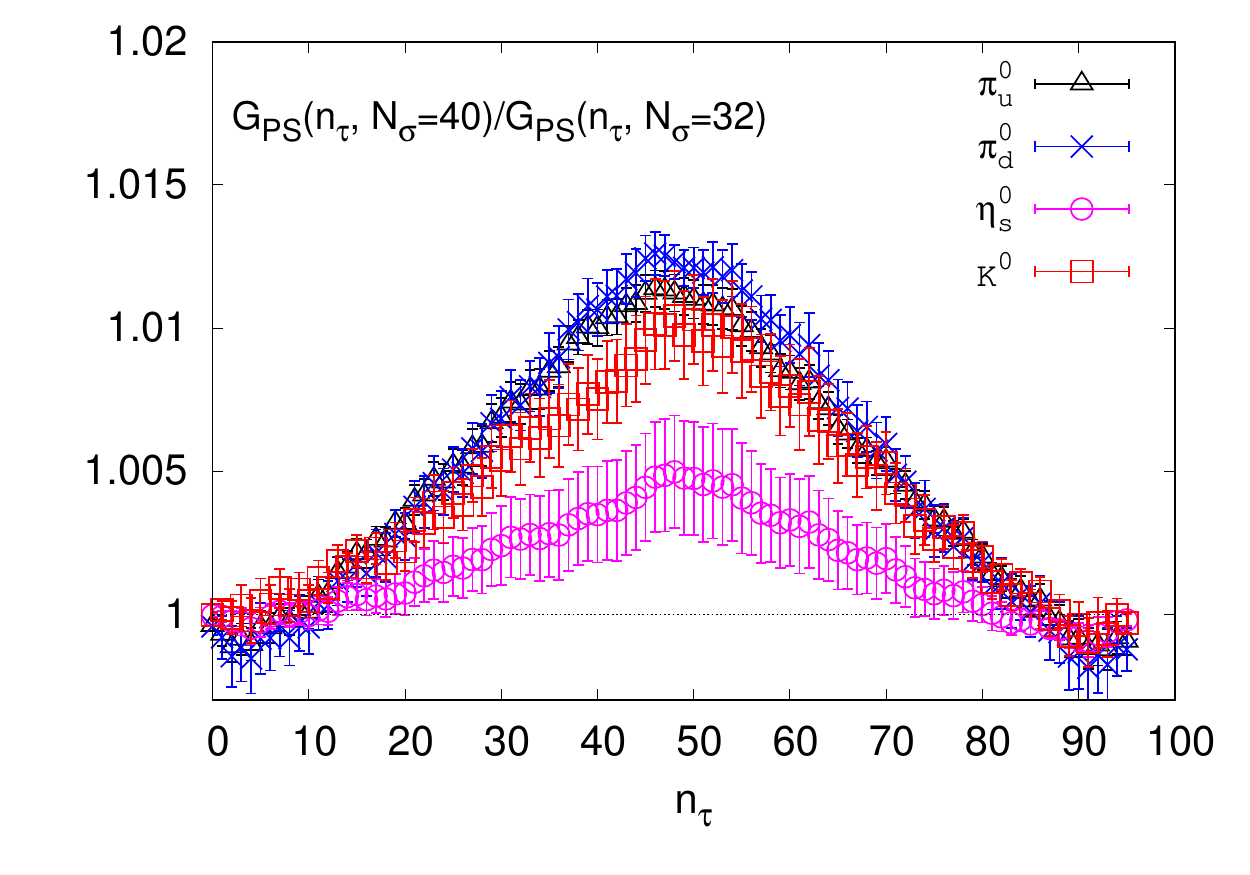}
	\includegraphics[width=0.47\textwidth]{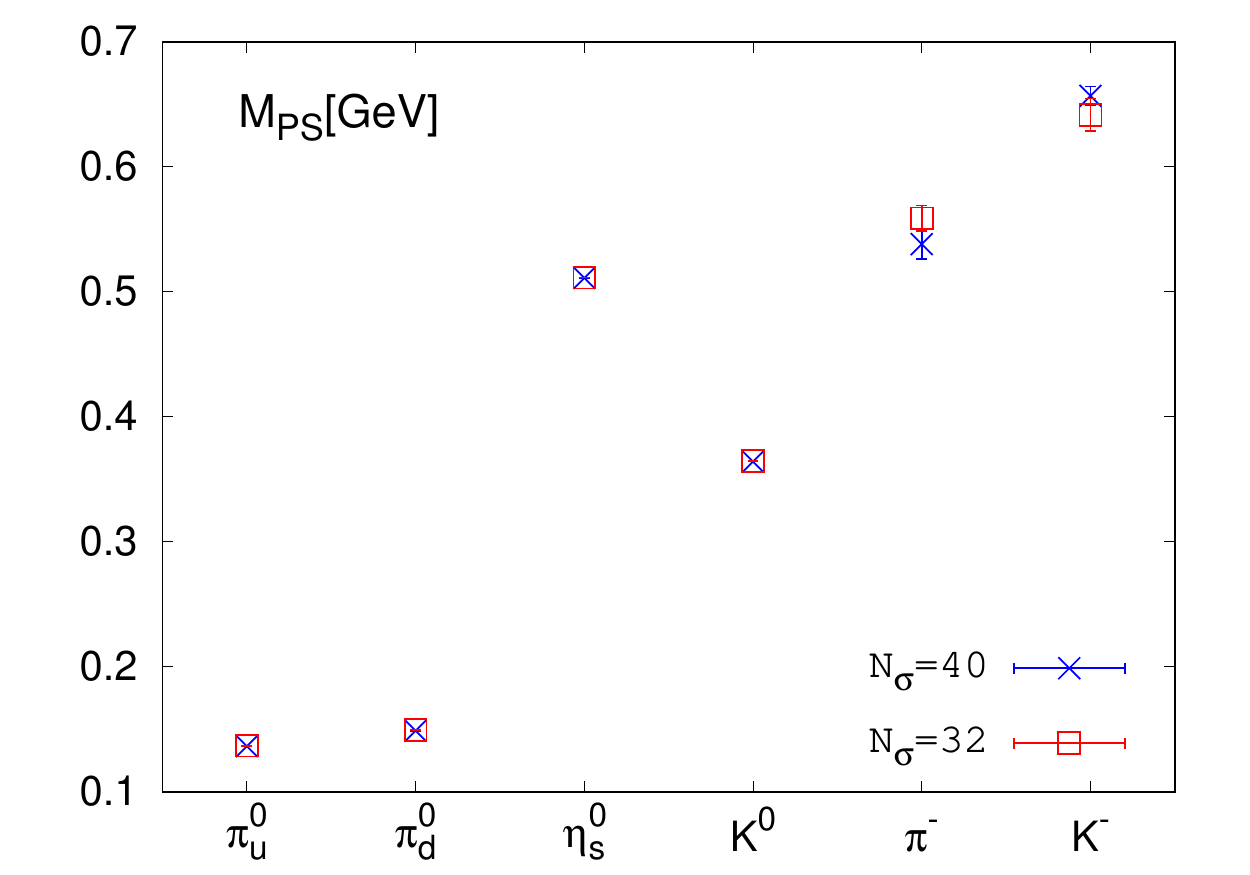}
	\caption{Left: Ratio of correlation functions of neutral pseudoscalar mesons obtained from lattices with spatial extent $N_\sigma=40$ to those with $N_\sigma=32$. Right: Pseudoscalar meson masses extracted from correlation functions obtained from lattices with two different volumes. Both plots are obtained at $eB\simeq$1.67 GeV$^2$.
	} 
	\label{fig:corrM_V}
\end{figure}

To check the volume dependence of masses and decay constants of neutral pseudoscalar mesons we show the ratio of correlation functions obtained on $40^3\times96$ lattices to those obtained on $32^3\times96$ lattices at $eB\simeq1.67$ GeV$^2$ in the left plot of Fig.~\ref{fig:corrM_V}. The corresponding $M_\pi V_s^{1/3}$ increases from $\simeq2.6$ on $32^3\times96$ lattices to $\simeq3.3$ on $40^3\times96$ lattices. One can see that the correlation function obtained in a larger volume becomes larger at most by 1.5$\%$. This naturally leads to the negligible volume dependences of $M_{\pi_{u,d}^0,K^0,\eta^0_s}$ as shown in the right plot of Fig.~\ref{fig:corrM_V}, and of related decay constants as well. Because of the relation between chiral condensates and corresponding correlation functions as will be shown in the next subsection, chiral condensates consequently also have mild volume dependences.

\begin{figure}[tbph]
	\centering
	\includegraphics[width=0.45\textwidth]{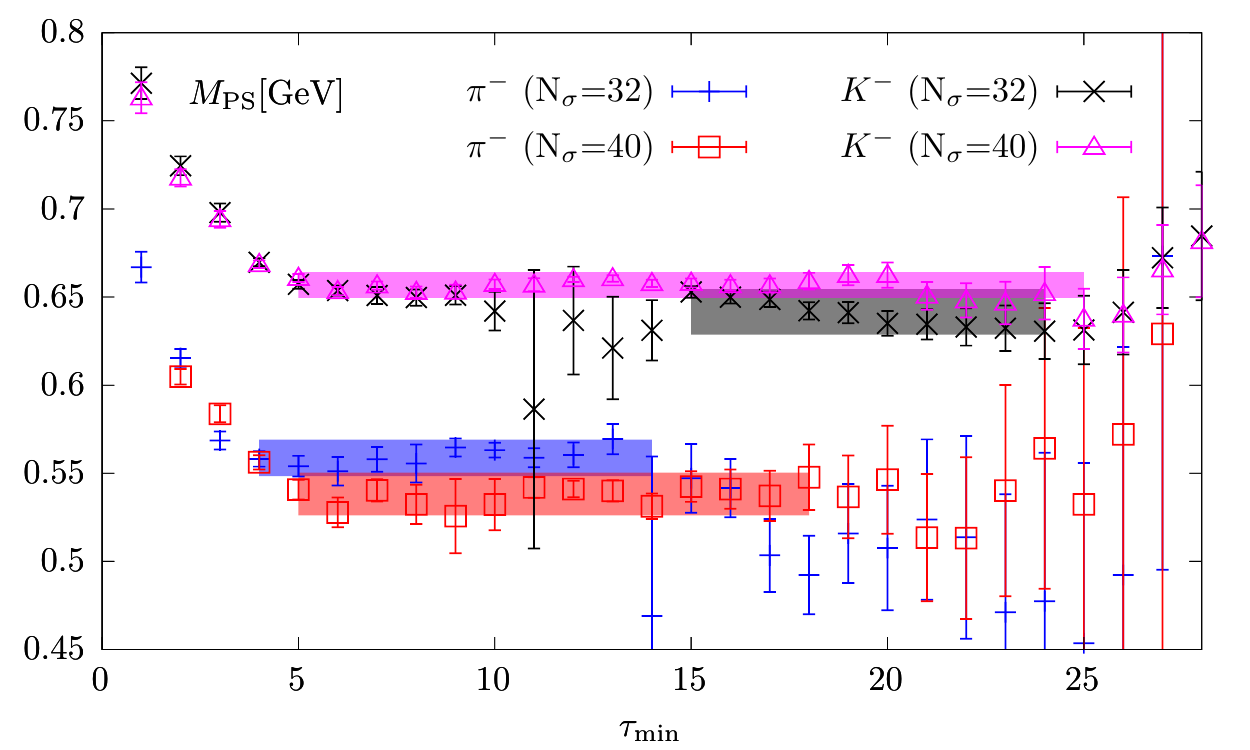}
	\includegraphics[width=0.45\textwidth]{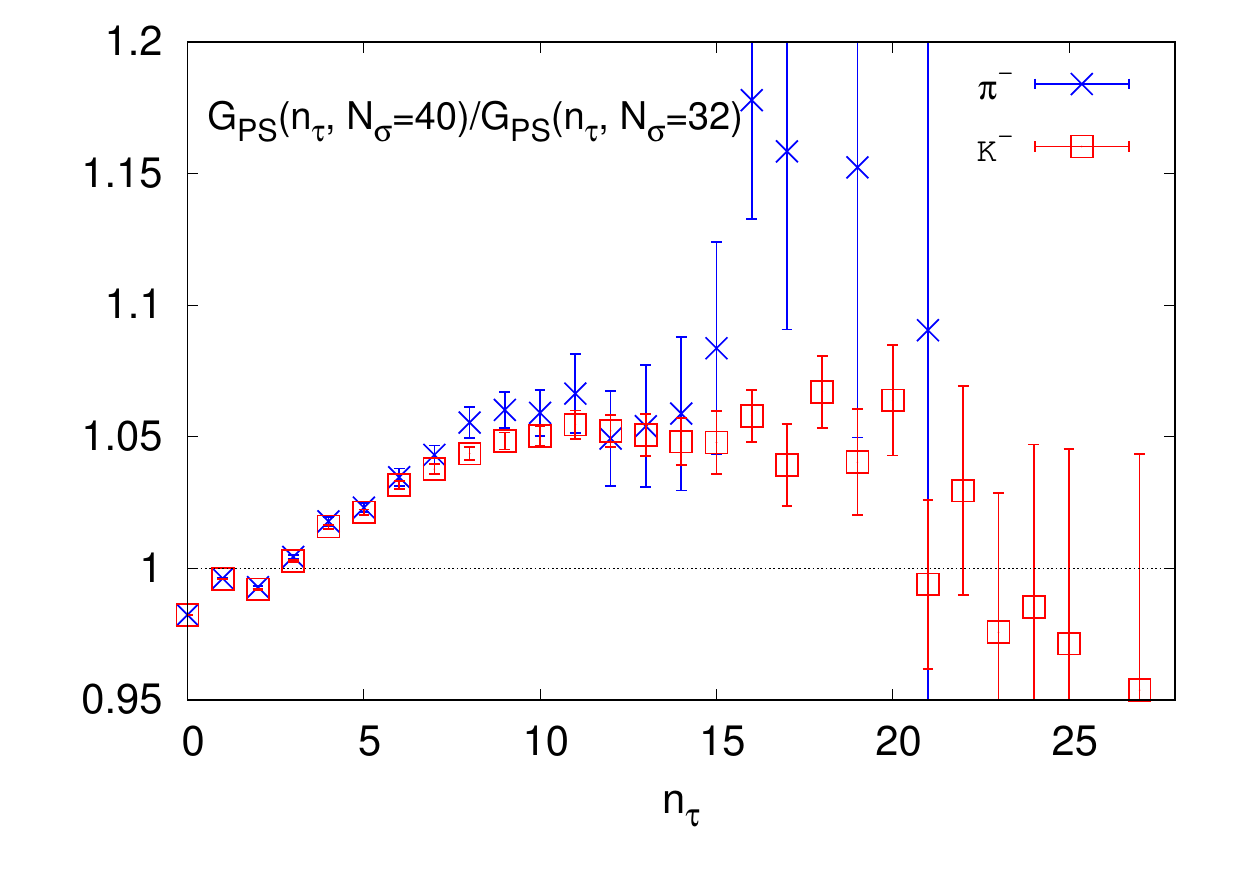}
	\caption{Left: Mass plateaus of $\pi^-$ and $K^-$ extracted from correlation functions obtained from lattices with spatial extent $N_\sigma=40$ and 32. Right: Ratio of correlation functions of $\pi^-$ and $K^-$ obtained from lattices with two different volumes. Both plots are obtained at $eB\simeq$1.67 GeV$^2$.
	} 
	\label{fig:chargedM_V}
\end{figure}

The charged pseudoscalar meson correlation function receives contributions from both oscillating and nonoscillating states, 
and generally has smaller signal-to-noise ratio compared to the neutral one. As an example we show in the left plot of Fig.~\ref{fig:chargedM_V} the extracted mass plateau of $\pi^-$ and $K^-$ obtained from lattices with both $N_\sigma=32$ and 40 at $eB\simeq$1.67 GeV$^2$. The 12 corner wall sources are also used in the computation of correlation functions of $\pi^-$ and $K^-$, and the mass plateau is selected by the AICc using the same procedures as mentioned before. One can observe that on lattices with $N_\sigma=40$ a longer and more stable plateau can be obtained, and $M_{\pi^-}$ ($M_{K^-}$) extracted from lattices with two different volumes are consistent within errors (cf. Fig.~\ref{fig:corrM_V} right). The ratios of corresponding correlation functions obtained using two volumes are shown in the right plot of Fig.~\ref{fig:chargedM_V}, and differ by less than 10\% in the region where mass plateaus are extracted.

Since the volume dependence of observables we are interested in is mild, most of the results shown in the following sections are obtained from $32^3\times96$ lattices if not mentioned explicitly. More discussions on the volume effects are presented in Sec.~\ref{sec:res_mass}.

\subsection{Mixing of pion states and Ward-Takahashi identities at $eB>0$}
\label{sec:setup_mixing}
Since the states of vector meson $\rho$ with spin polarization $s_z=0$, i.e., $\rho_{s_z=0}$ has the same quantum number of the states of $\pi$ in the presence of a magnetic field, it is supposed that the mixing between the pion and $\rho_{s_z=0}$ states is enabled~\cite{Bali:2017ian}, i.e., neutral $\pi^0$ mixes with neutral $\rho^0_{s_z=0}$ while charged $\pi^\pm$ mixes with charged $\rho^\pm_{s_z=0}$.\footnote{Note that the mixing with $s_z=\pm 1$ is not permitted due to the conservation of angular momentum.} This is to say that once the mixing is enabled, correlation functions in the vector channel and pseudoscalar channels receive mutual contributions and their corresponding ground state masses should be the same. The mixing has been investigated in detail in quenched QCD in Ref.~\cite{Bali:2017ian}, and it was found that the influence to the ground state of $\pi$ is negligible. Here we show evidence that the influence of mixing to neutral pseudoscalar meson states as well as the contribution from disconnected diagrams to the neutral pion correlation function are mild. This can be seen as follows. At nonzero magnetic fields, as derived in Appendix~\ref{sec:app_WI}, the following Ward identity (cf.~\ref{eq:app_WI_PS1}) also holds as that at $eB=0$:
\begin{equation}
2m_l\,\tilde{\chi}_{\pi^0} = \pbp_u + \pbp_d,
\label{eq:WI_np}
\end{equation}
where $\tilde{\chi}_{\pi^0}$ is the space-time sum of the neutral pion correlation function which includes contributions from both connected and disconnected diagrams. As the computation of disconnected diagrams in the correlation function is beyond the scope of current paper, we rather look into the up and down quark components of the connected part of the neutral pion correlation function. In practice we check whether two following relations hold true at nonzero magnetic fields:
\begin{align}
  \begin{split}
\pbp_u &= m_u\chi_{\pi_u^0}, \\
\pbp_d &= m_d\chi_{\pi_d^0}.
\label{eq:WI_pbp-ps}
\end{split}
\end{align}
Here $\chi_{\pi_u^0}$ ($\chi_{\pi_d^0}$) is the space-time sum of the up (down) quark component of the connected part of neutral pion correlation function, and $m_u=m_d$ in our setup. If the above two relations held true at nonzero magnetic field, this leads to $\chi_{\pi_u^0}+\chi_{\pi_d^0}$=$2\tilde{\chi}_{\pi^0}$, which is the same as the case at $eB=0$.\footnote{Note that the disconnected diagram does not contribute to $\tilde{\chi}_{\pi^0}$ at $eB=0$.}

\begin{figure}[htbp]
	\includegraphics[width=0.47\textwidth]{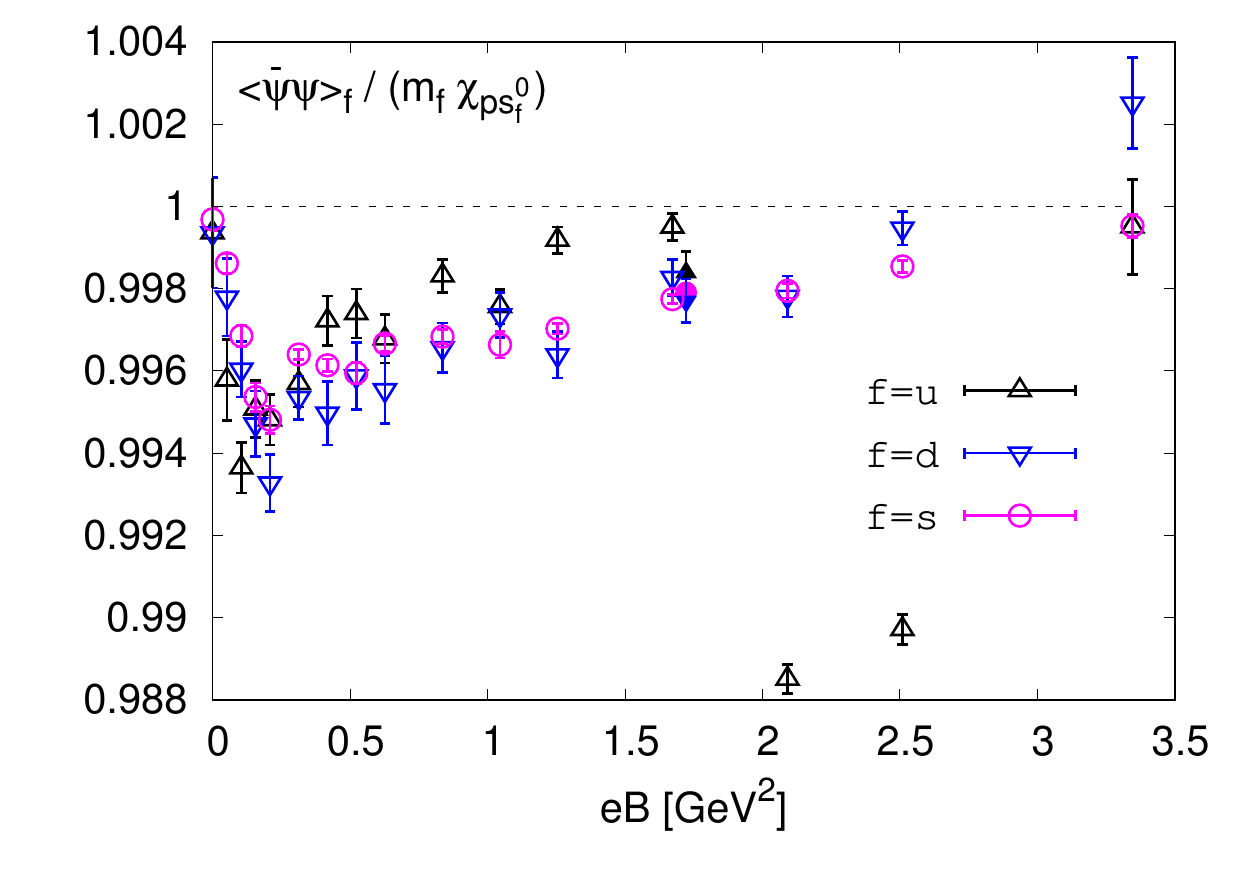}
	\includegraphics[width=0.47\textwidth]{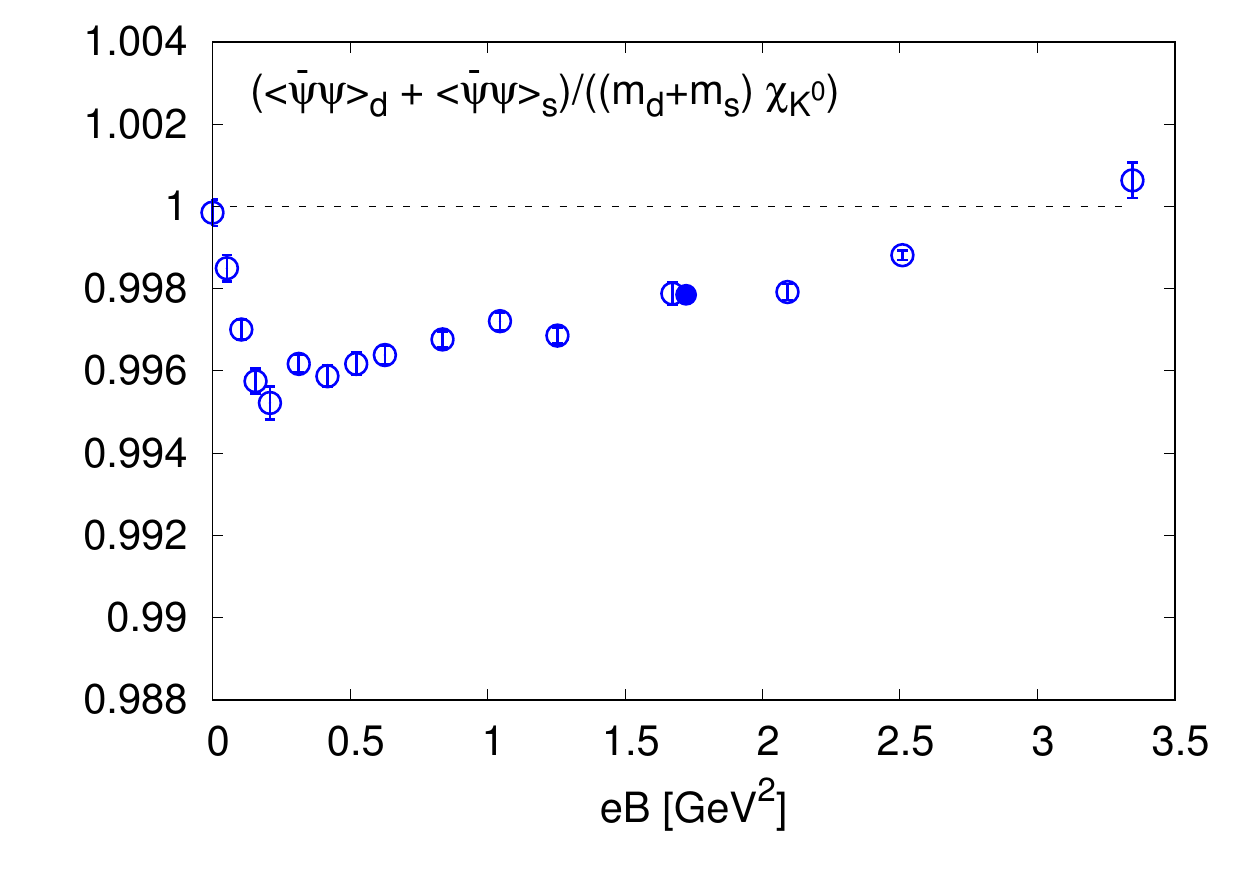}
	\caption{Left: Ratio $\pbp_f/(m_f\chi_{\mathrm{ps}^0_f})$ as a function of $eB$ for $f$ as up ($u$), down ($d$) and strange ($s$) quark flavors. Here $\chi_{\rm ps_f^{0}}$ is the space-time sum of the correlation function of neutral mesons consisting only the connected $f\bar{f}$ component in the pseudoscalar channel, $\chi_{\rm ps_f^{0}}=\sum_{\tau=0}^{N\tau-1}G_{\mathrm{ps}^0_f}(\tau)$ with $f=u,~d$, and $s$ quark flavors. For $f=u,~d$ and $s$, $\chi_{\rm ps_f^{0}}$ =  $\chi_{\rm \pi_u^{0}}$, $\chi_{\rm \pi_d^{0}}$, and $\chi_{\rm \eta_s^{0}}$, respectively. Right: Ratio $(\pbp_d+\pbp_s)/((m_d+m_s)\chi_{K^0})$ as a function of $eB$. In both plots open points denote results obtained from $32^3\times96$ lattices, while filled points shifted horizontally denote results at $eB\simeq$1.67 GeV$^2$ obtained from $40^3\times96$ lattices.}
	\label{fig:WIandMixing}
\end{figure}

We show the ratios, $\pbp_u/(m_u\chi_{\pi_u^0})$ and $\pbp_d/(m_d\chi_{\pi_d^0})$, as functions of $eB$ in the left plot of Fig.~\ref{fig:WIandMixing}. We found that both ratios agree with unity with deviations less than 1.2\% at all the values of the magnetic field strength we studied. This suggests that the summed contribution from the anomalous part $\Delta_{\mathcal{J}}^{u,d}$ and the disconnected diagram in $\pi^0_{u,d}$ is negligible (cf. Eq.~\ref{eq:WIpiu} and~\ref{eq:WIpid}) at both zero and nonzero magnetic fields. Since chiral condensates do not contain any information of $\rho$, the influence of the mixing of $\rho$ to neutral pseudoscalar states should be negligible. Moreover, $\pbp_u\approx m_u\chi_{\pi^0_u}$ and $\pbp_d\approx m_d\chi_{\pi^0_d}$ naturally lead to $\pbp_u+\pbp_d\approx m_l(\chi_{\pi^0_u}+\chi_{\pi^0_d})$. Considering that the contribution from disconnected diagrams to $\pi^0$ is negligible, one then has $\pbp_u+\pbp_d\approx 2 m_l\tilde{\chi}_{\pi^0}$ at both vanishing and nonzero magnetic fields. This thus resembles Eq.~\ref{eq:WI_np}, indicating that the disconnected part can be ignored in the integrated neutral pion correlation function.
Although it is found in the quenched QCD that the large distance behavior of correlators arising from disconnected diagrams is negligible in the determination of neutral pion mass~\cite{Luschevskaya:2015cko}, the exact influence of the disconnected part to the large distance behavior of the neutral pion correlation function in full QCD requires further investigation, which is beyond the scope of current study. We also check the volume dependence at $eB\simeq1.67$ GeV$^2$, and the results obtained from $40^3\times96$ lattices shown as filled points (shifted horizontally) almost overlap with the results obtained from $32^3\times96$ lattices.

We also check the Ward identities involving strange quarks at nonzero magnetic fields (cf.~\ref{eq:app_WI_PS2} and~\ref{eq:app_WI_PS3} in Appendix~\ref{sec:app_WI}),
\begin{eqnarray}
\label{eq:WI_pbp-pss}
\pbp_s+ \Delta^s_{\mathcal{J}} &=& m_s \tilde{\chi}_{\mathrm{\eta_s^0}} , \\
\pbp_d + \pbp_s &=& (m_d+m_s) \chi_{K^0},
\label{eq:WI_pbp-psK0}
\end{eqnarray}
where $\tilde{\chi}_{\eta_s^0}$ and $\chi_{K^0}$ are the space-time sum of $\eta_s^0$ and the neutral kaon correlators. The former includes contributions from both connected and disconnected diagrams. In our study, we only investigate the connected contribution to $\tilde{\chi}_{\eta_s^0}$, which is denoted by ${\chi}_{\eta_s^0}$. The ratios, $\pbp_s/(m_s\chi_{\mathrm{\eta}_s^0})$ and $(\pbp_d + \pbp_s)/((m_d+m_s) \chi_{K^0})$ as functions of $eB$, are shown in the left and right plot of Fig.~\ref{fig:WIandMixing}, respectively. The deviation from unity is less than 0.6\% in $eB\in[0,3.35)$ GeV$^2$. One can clearly see that for Eq.~\ref{eq:WI_pbp-pss}, the overall contributions from $\Delta^s_{\mathcal{J}}$ and the disconnected diagrams are tiny at all the magnetic fields, while Eq.~\ref{eq:WI_pbp-psK0} holds well with marginal deviations in the current window of magnetic fields.

\subsection{UV divergence of quark chiral condensates}
\label{sec:setup_UV}
To investigate the GMOR relation, we need to take care of the UV divergence in the light and strange quark chiral condensates at zero and nonzero magnetic fields. Since it has been shown in Ref.~\cite{Bali:2011qj} that the UV-divergence part of the chiral condensate is independent of the magnetic field,  we can obtain the UV-free chiral condensate at nonzero magnetic fields by subtracting the UV divergence in the chiral condensate, i.e., $\pbp_{f=l,s}^{\rm UV}$ obtained at the zero magnetic field. To obtain $\pbp_{l,s}^{\rm UV}$, we thus look into the Dirac spectrum representation of the subtracted chiral condensate at the zero magnetic field,
\begin{eqnarray}
\pbp_{sub}\equiv\pbp_{l} - \frac{m_l}{m_s} \pbp_s = \int_0^\infty \frac{2m_l\,(m_s^2-m_l^2)\rho(\lambda)}{(\lambda^2+m_l^2)(\lambda^2+m_s^2)}\,\mathrm{d}\lambda,
\label{eq:pbp_sub_rho}
\end{eqnarray}
where $\rho(\lambda)$ is the eigenvalue spectral density of fermion matrix $D_f$ (cf. Eq.~\ref{eq:pbp_q}), and the light (up or down) quark and strange quark chiral condensate, $\pbp_l$ and $\pbp_s$, are connected to $\rho(\lambda)$ through the following relation:
\begin{equation}
\pbp_{l,s}= \int_0^\infty \,\frac{2\,m_{l,s}\,\rho(\lambda)}{\lambda^2 + m_{l,s}^2}\,\mathrm{d}\lambda.
\label{eq:pbp_rho}
\end{equation}

The UV-divergence part of the quark chiral condensate linear in quark mass is thus absent in $\pbp_{sub}$, while a logarithm UV divergence in the light quark chiral condensate should be negligible. 
Thanks to the Chebyshev filtering technique combined with the stochastic estimate method~\cite{Ding:2020eql,Tomiya:2019nym,Giusti:2008vb,Cossu:2016eqs,Fodor:2016hke}, we can compute the complete Dirac eigenvalue spectrum $\rho(\lambda)$ and then reproduce $\pbp_{sub}$ as well as $\pbp_l$ and $\pbp_s$ through Eqs.~\ref{eq:pbp_sub_rho} and~\ref{eq:pbp_rho}, respectively.\footnote{With the order of Chebyshev polynomials being 24000 and the bin width of the Dirac eigenvalue spectrum in lattice spacing being 0.002 $\pbp_{sub}$, $\pbp_l$ and $\pbp_s$ obtained from the stochastic noise method (cf. Eq.~\ref{eq:pbp_q}) are reproduced within an accuracy of 1\% via Eqs.~\ref{eq:pbp_sub_rho} and~\ref{eq:pbp_rho}.}  In the Dirac eigenvalue spectrum, the UV-divergence part should be represented by $\rho(\lambda)$ with $\lambda\geq \lambda_{cut}^{UV}$. Thus the UV-divergence part of the light quark condensate can be expressed as
\begin{equation}
\pbp_{l,s}^{\rm UV} = \int_{\lambda_{cut}^{\rm UV}}^\infty \,\frac{2\,m_{l,s}\,\rho(\lambda)}{\lambda^2 + m_{l,s}^2}\,\mathrm{d}\lambda.
\label{eq:pbp_UV}
\end{equation}
\begin{figure}[tbph]
	\centering
	    \includegraphics[width=0.47\textwidth]{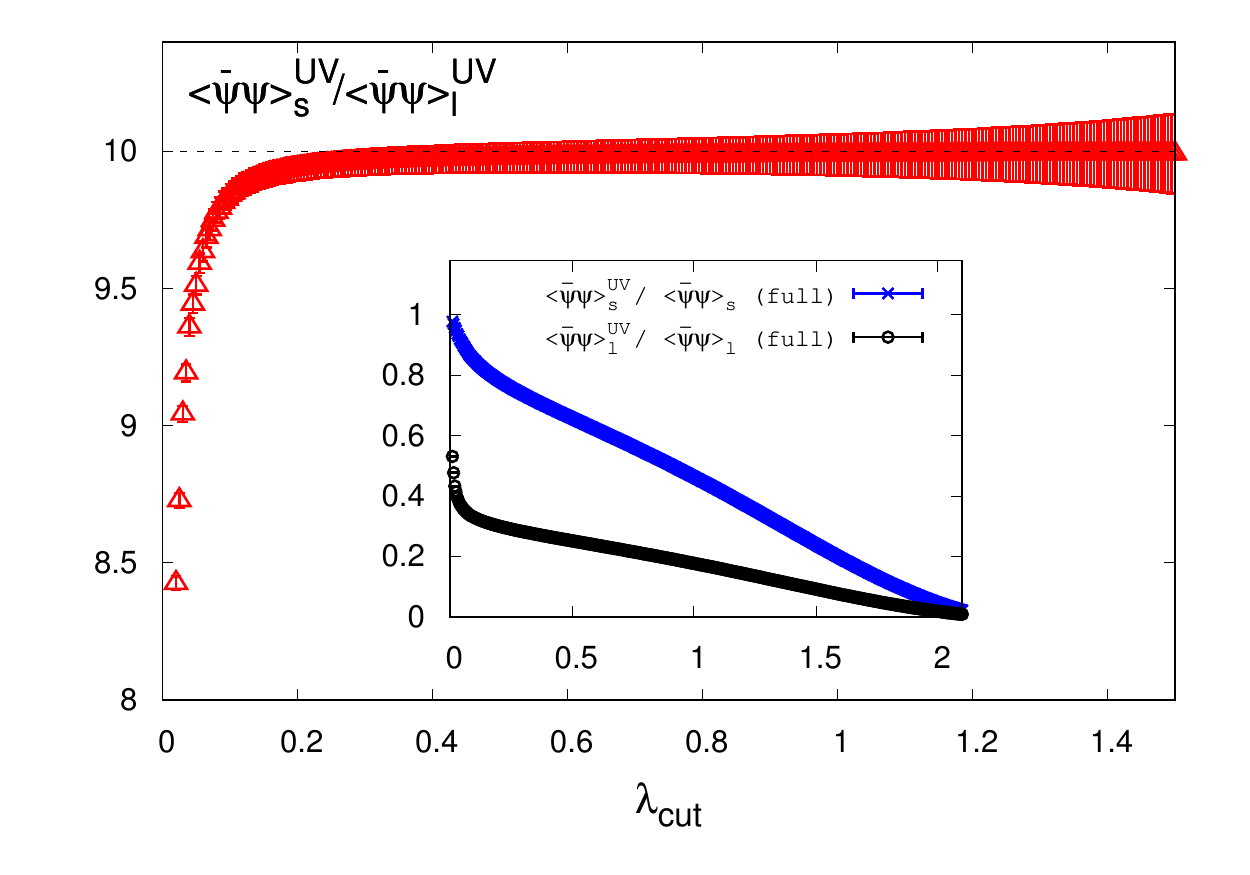}
	\includegraphics[width=0.47\textwidth]{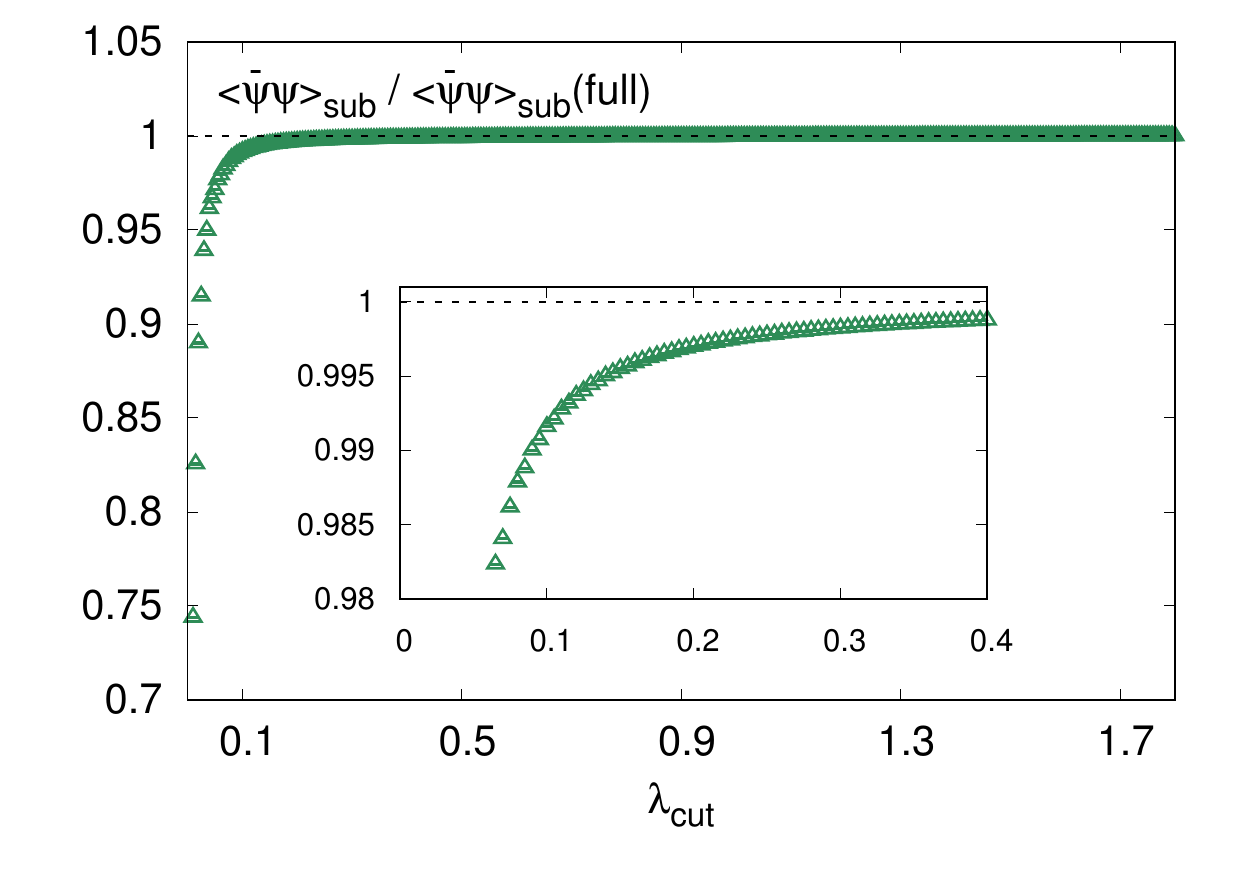}
	\caption{Left:  Ratio $\pbp_{s}^{\rm UV}/\pbp_{l}^{\rm UV}$ as a function of $\lambda_{cut}$. The inset shows the ratios of $\pbp_{l,s}^{\rm UV}$ to the corresponding full quark condensates as a function of $\lambda_{cut}$. Right: Ratio of the subtracted chiral condensates with an upper cutoff $\lambda_{cut}$ in the integration of $\lambda$ in Eq.~\ref{eq:pbp_sub_rho} to that with a full spectrum of $\rho(\lambda)$ as a function of $\lambda_{cut}$. The inset shows a blowup in the $y$-axis region close to 1.}
	\label{fig:lambda_cut}
\end{figure}
Given that the logarithm divergence in the quark mass is negligible, the UV-divergence part in the strange quark chiral condensate should be $m_s/m_l$ times as that in the light quark chiral condensate, i.e., $\pbp^{\rm UV}_s=10 ~\pbp^{\rm UV}_l$ in our case. Thus, $\pbp^{\rm UV}_{l,s}$ can be determined with the smallest value of $\lambda^{\rm UV}_{cut}$ which makes $\pbp^{\rm UV}_s/\pbp^{\rm UV}_l$=10.

To determine the value of $ \lambda_{cut}^{UV}$, we show the ratio $\pbp_s^{\rm UV}/\pbp_l^{\rm UV}$ as a function of $\lambda_{cut}$ in the left plot of Fig.~\ref{fig:lambda_cut}. Here $\lambda_{cut}$ is the lower limit in the integration of $\lambda$ in Eq.~\ref{eq:pbp_UV}. We see that the ratio approaches 10 rapidly at $\lambda_{cut}\lesssim0.2$ and then saturates at 10 at larger values of $\lambda_{cut}$. We thus pick up a value of 0.24 to be $\lambda^{\rm UV}_{cut}$, which makes the ratio start to approach 10 by less than 0.5\%.  As seen from the inset in the left plot of Fig.~\ref{fig:lambda_cut}, $\pbp_{f}^{UV}({\lambda_{cut}})$ itself is a rapidly decreasing function of $\lambda_{cut}$. We also check the uncertainty in the determination of $\lambda^{\rm UV}_{cut}$ by looking to the subtracted chiral condensate. We compute $\pbp_{sub}(\lambda_{cut})$ as a function of $\lambda_{cut}$, where $\lambda_{cut}$ is also the upper bound of the integration variable $\lambda$ in Eq.~\ref{eq:pbp_sub_rho} and show the ratio of $\pbp_{sub}(\lambda_{cut})$ to $\pbp_{sub}(\lambda_{cut}=\infty)$ as a function $\lambda_{cut}$ in the right plot of Fig.~\ref{fig:lambda_cut}. Thus, $\lambda_{cut}^{\rm UV}$ should be the smallest value of $\lambda_{cut}$ at which $\pbp_{sub}(\lambda_{cut})/\pbp_{sub}(\lambda_{cut}=\infty)\simeq 1$.  As seen from the plot, the ratio approaches unity at very small values of $\lambda_{cut}$, e.g., the ratio is 0.5\% deviation from unity at $\lambda_{cut}\simeq 0.12$.

The logarithm divergence in the quark mass is expected to be less than about 1.25\% of the UV part that is linear in quark mass in the free case, which should be negligible in our case. Here to study the uncertainty of the UV part in the chiral condensate, we adopt a rather wide window for the values of $\lambda^{\rm UV}_{cut}$ from 0.12 to 0.36 with the central value 0.24 which gives $\pbp^{\rm UV}_s/\pbp^{\rm UV}_l\simeq$ 10. Then the obtained $\pbp_l^{\rm UV}$ ranges from about 32$\%$ to 27$\%$ of $\pbp_l$ at $eB=0$, while $\pbp_s^{\rm UV}$ ranges from about 83$\%$ to 71$\%$ of $\pbp_s$ at $eB=0$.

\section{Results}
\label{sec:results}

\subsection{Masses and magnetic dipole polarizabilities of light and strange pseudoscalar mesons }
\label{sec:res_mass}
\begin{figure}[!htbp]
	\includegraphics[width=0.47\textwidth]{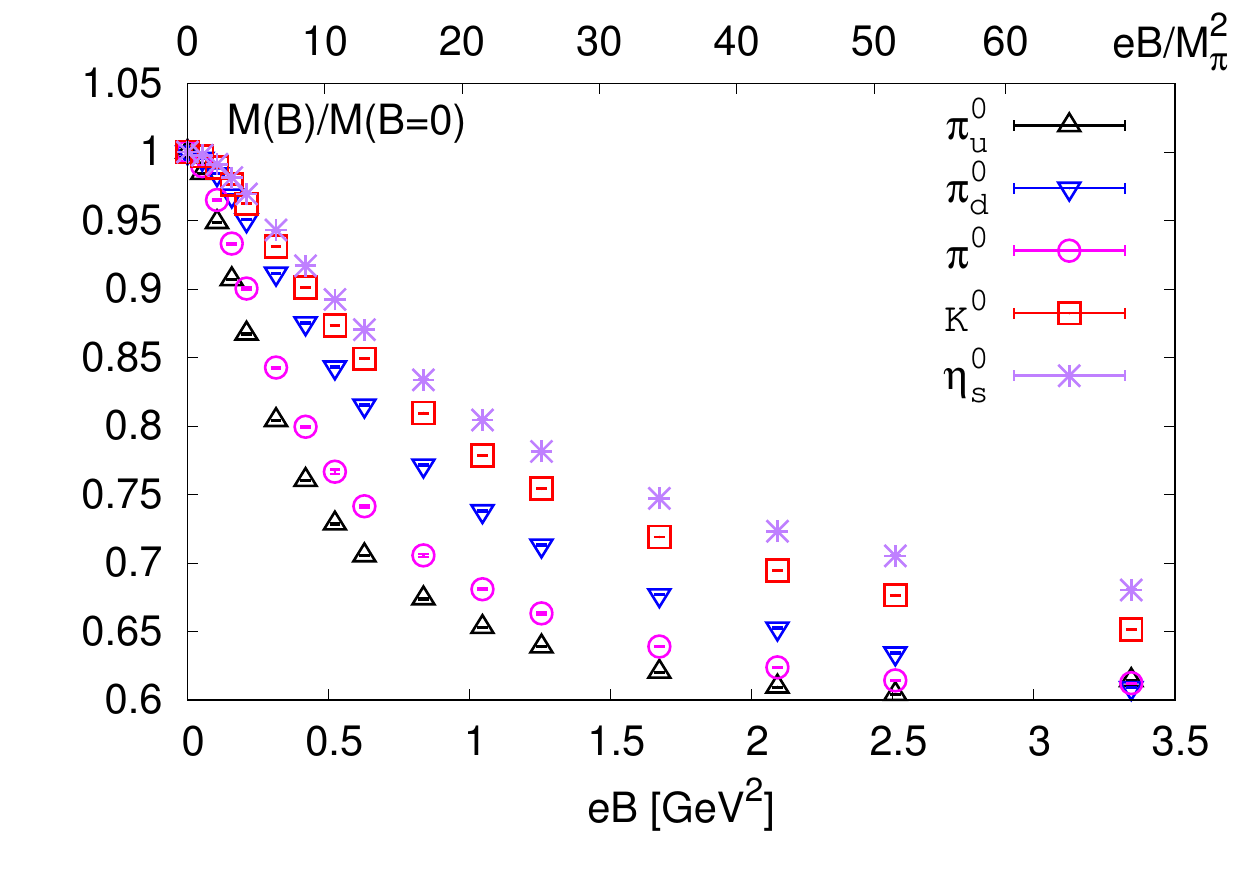}
	\includegraphics[width=0.47\textwidth]{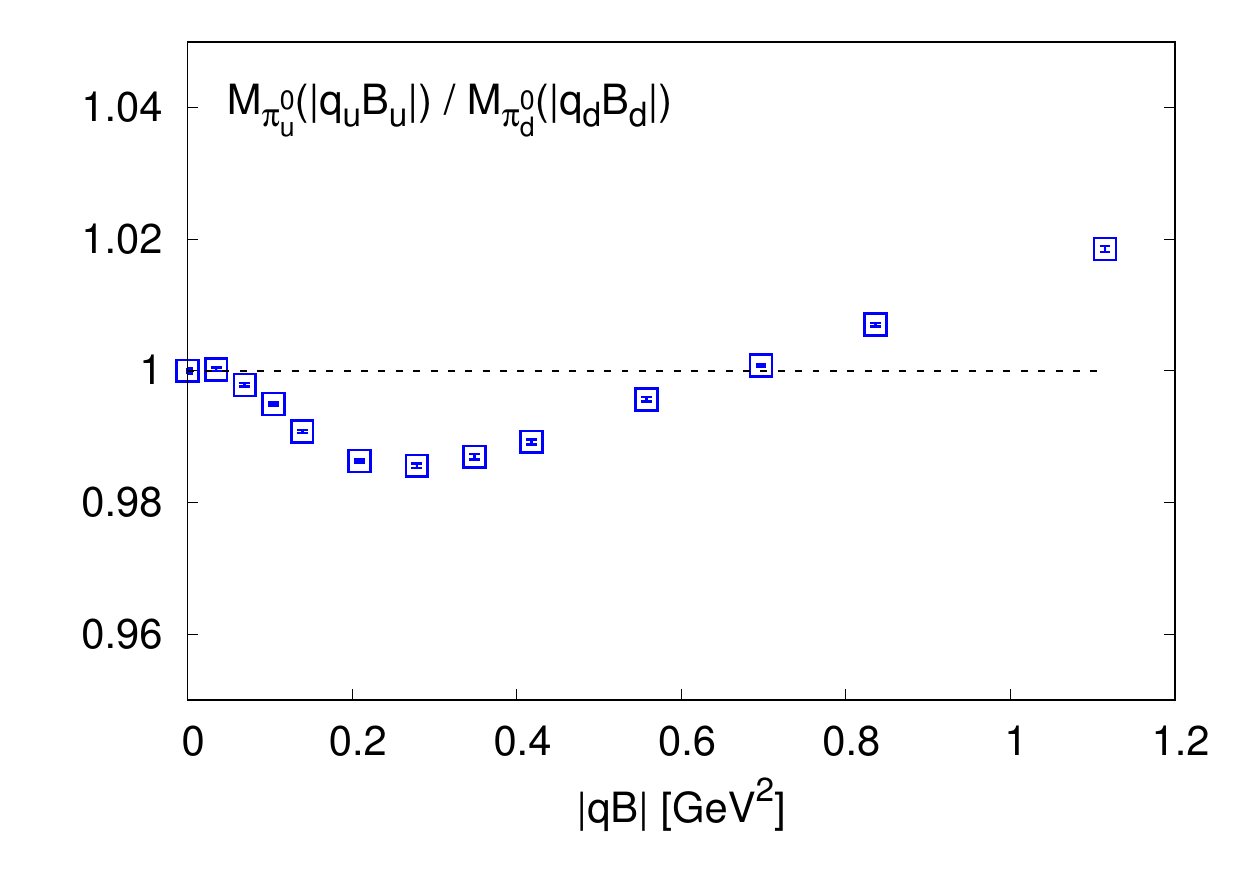}
	\caption{Left: Masses of $\pi^0_u$, $\pi^0_d$, $K^0$, $\eta^0_s$ normalized by their corresponding masses at $eB$=0 as a function of $eB$. Right: Ratio $M_{\pi^0_u}(|q_uB_u|)/M_{\pi^0_d}(|q_dB_d|)$ as a function of $|qB|$. Here $q_u$ and $q_d$ stand for the electric charges of $u$ and $d$ quarks, and $B_{u}$ and $B_{d}$ are different values of $B$, which make $|qB|\equiv|q_uB_u|=|q_dB_d|$.}
	\label{fig:neutral_mass}
\end{figure}

We now present our results for masses of pseudoscalar mesons calculated at 16 different values of $eB$ ranging from 0 to $\sim 3.35$ GeV$^2$. In the left plot of Fig.~\ref{fig:neutral_mass}, we show the ratio of masses of neutral pseudoscalar mesons at nonzero magnetic fields to those at a zero magnetic field as a function of $eB$. We found that the masses of all neutral mesons decrease with increasing $eB$ and tend to saturate at $eB\gtrsim 2.5$ GeV$^2$. By comparing the normalized masses of $\pi^0_u$, $\pi^0_d$, $K^0$, $\eta_s$, it is obvious that the lighter hadrons are more affected by the magnetic field. For instance in the strongest magnetic field ($eB\simeq$ 3.35 GeV$^2$) we have, it can be seen that $M_{\eta_s^0}$ and $M_{\pi^0_u}$ ($M_{\pi^0_d}$) are about 70\% and 60\% of their values at $B$=0, respectively.  The amount of reduction in $M_{\pi^0_u}$ and $M_{\pi^0_d}$ is roughly consistent with results presented in SU(2) gauge theory~\cite{Luschevskaya:2012xd} and SU(3) quenched QCD~\cite{Bali:2017ian} as well as in $N_f=2+1$ QCD with stout fermions and the physical pion mass in the vacuum~\cite{Bali:2011qj}.\footnote{The $eB$ dependence of $M_{\pi^0_u}$ obtained in $N_f=2+1$ QCD using stout fermions is shown in Fig. 20 in Ref.~\cite{Bali:2017ian}, where the determination of $M_{\pi^0_u}$ using stout fermions is based on the gauge ensembles produced in Ref.~\cite{Bali:2011qj}.} In the former case $M_{\pi^0_u}$ in quenched QCD~\cite{Bali:2017ian} decreases slower while the latter $M_{\pi^0_u}$ in $N_f=2+1$ QCD~\cite{Bali:2011qj,Bali:2017ian} decreases faster with $eB$ compared to our current study.  This could be partly due to the fact that hadrons with larger masses are less affected by the magnetic field, as the pion mass in the vacuum is about $415$ MeV in \cite{Bali:2017ian}, while it is 135 MeV in~\cite{Bali:2011qj,Bali:2017ian}. Because of the presence of a nonzero magnetic field, the $SU_V$(2) symmetry is broken, and the mixture of the $u\bar{u}$ and $d\bar{d}$ flavor contents in the neutral pion could depend on $eB$~\cite{Bali:2017ian}. To determine the mixture coefficient is beyond the scope of our current paper. However, as discussed in Sec.~\ref{sec:setup_mixing}, the mixture mostly likely is similar as that at $eB=0$. For a demonstration we nevertheless show in the left plot of Fig.~\ref{fig:neutral_mass} the ground state mass of $\pi^0$ extracted from the averaged correlation functions of $u\bar{u}$ and $d\bar{d}$ in the pseudoscalar channel, i.e., $G_{\pi^0}=(G_{\pi^0_u}+G_{\pi^0_d})/2$, assuming that the contribution of the disconnected diagram is negligible and the mixture coefficients are the same as the $B=0$ case~\cite{Luschevskaya:2015cko}. As seen from the plot, the ratio for $\pi^0$ is between those for $\pi^0_u$ and $\pi_d^0$ as expected.

As discussed in Sec.~\ref{sec:basics}, the mass of a neutral point particle should be independent of the magnetic field due to its zero electric charge. However, the mesons we studied are composite particles consisting of two constituent quarks. When the magnetic field is weaker than the inverse meson size squared, mesons remain pointlike, which is the case for charged pseudoscalar mesons at $eB\lesssim0.31$ GeV$^2$ as shown in Fig.~\ref{fig:charged_mass}. The $eB$ dependence of neutral pseudoscalar meson masses, on the other hand, suggests that the internal constituents of these neutral mesons, i.e., constituent quarks, are probed by the magnetic field we simulated. Thus, these neutral pseudoscalar mesons cannot be considered as point particles in all the simulated magnetic field strengths.
Also, the different magnitudes of the mass reduction between $\pi^0_u$ and $\pi^0_d$ may come from the different electric charges of up and down quarks, indicating that the meson's inner structure has been revealed.  Since the internal structure of the neutral pion is probed within our current window of magnetic field, we intend to investigate the influence of the electric charge of quarks on the mass of neutral pion. We thus show the ratio of  $M_{\pi^0_u}$ to $M_{\pi^0_d}$ as a function of $qB$ instead of $eB$ in the right plot of Fig.~\ref{fig:neutral_mass}. We find for the first time to our knowledge that, after rescaling the $x$ axis from $eB$ into $qB$, $M_{\pi^0_u}(|q_uB_u|)$ is almost the same as $M_{\pi^0_d}(|q_dB_u|)$ at $|qB|=|q_uB_u|=|q_dB_d|$ and differs at most by 2\%. Here $q_u$ and $q_d$ are the electric charges of $u$ and $d$ quarks, respectively, and $B_{u,d}$ stands for different magnetic field strengths that the quark feels to make $|qB|$ the same for up and down quarks. We call this behavior the $qB$ scaling. 

The $qB$ scaling should be exact in the quenched limit where the $eB$ ($qB$)  only enters into the Dirac operator. However, dynamical quarks could spoil the $qB$ scaling as they carry different electric charges and enter into the quark action and affect the probability of different background gauge fields in the path integral. The $eB$ dependence of $M_{\pi^0_u}$ and $M_{\pi^0_d}$ has been obtained in quenched QCD, and a qualitative consistency of the data with the $qB$ scaling can be read off from the top plot of Fig. 13 in Ref.~\cite{Bali:2017ian} showing the $eB$ dependence of $M_{\pi^0_{u,d}}$. The mild deviation of $M_{\pi^0_u}(|qB|)/M_{\pi^0_d}(|qB|)$ from unity observed in our study suggests that  influences from dynamical quarks are negligible at $M_\pi(eB=0)\simeq$ 220 MeV. On the other hand, the $qB$ scaling observed in our study also supports that the internal structure of the neutral pion is probed. This is due to the fact that the neutral pion cannot be considered as a point particle anymore and $M_{\pi_{u,d}^0}$ are functions of the electric charge of the quark ($q$) multiplied by the magnetic field strength ($B$). Note that the weakest magnetic field we simulated is $eB\approx 0.05$ GeV$^2$, which is about the value of $M^2_\pi(eB=0)$ in our simulation. It is expected that $eB$ needs to be larger to probe the internal structure of a heavier neutral pion to see the $qB$ scaling behavior. We will come back to this point in the discussion of the $qB$ scaling of chiral condensates in Sec.~\ref{sec:res_pbp}.

\begin{figure}[tbph]
	\centering
	\includegraphics[width=0.5\textwidth]{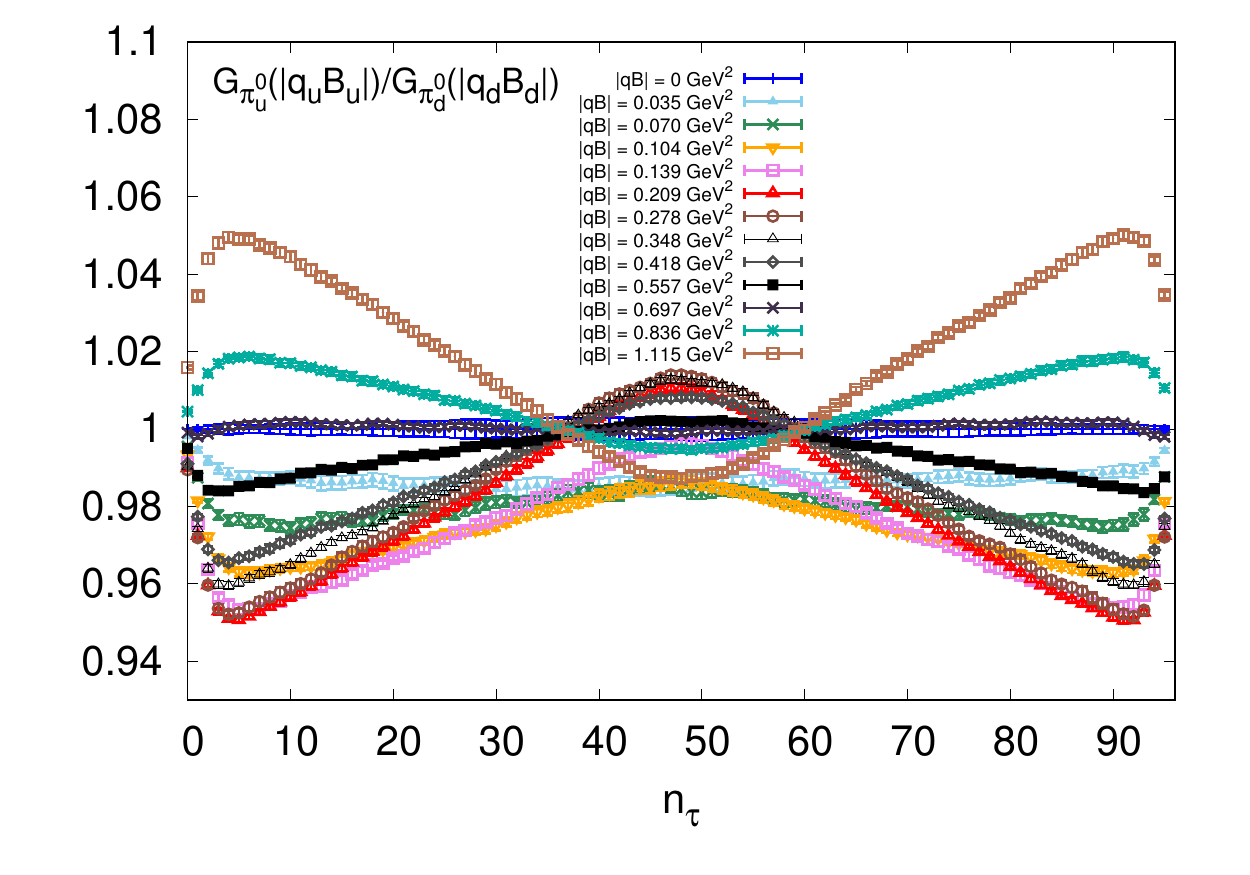}
	\caption{$G_{\pi_u^0}(n_\tau,|q_uB_u|)$/$G_{\pi_d^0}(n_\tau,|q_dB_d|)$ as a function of temporal distance $n_\tau$ at various  values of $|qB|=|q_uB_u|=|q_dB_d|$.  
	} 
	\label{fig:corr_qB}
\end{figure}

In Fig.~\ref{fig:corr_qB}, we show the ratio of $G_{\pi_u^0}(\tau,|q_uB_u|)$ to $G_{\pi_d^0}(\tau,|q_dB_d|)$ as a function of temporal distance $n_\tau$ at 13 different values of $|qB|$. We found that at large distances, i.e., $n_\tau$ close to 48 ($N_\tau/2$), the most relevant part for the extraction of $M_{\pi^0_u} (M_{\pi^0_d})$, the ratio deviates from unity at most by 2\%.  This naturally explains that the origin of $qB$ scaling behavior shown in $M_{\pi^0_u} (M_{\pi^0_d})$ is the $qB$ scaling behavior of correlation function $G_{\pi_u^0} (G_{\pi_d^0})$. We will see in Sec.~\ref{sec:res_pbp} and Sec.~\ref{sec:res_dc-GMOR} that according to Eq.~\ref{eq:fpi_det} and Eq.~\ref{eq:WI_pbp-ps} the $qB$ scaling of correlation functions also naturally leads to the $qB$ scaling of $\Sigma_u$($\Sigma_d$) and $f_{\pi^0_u}$ ($f_{\pi^0_d}$).

\begin{figure}[!htbp]
	\centering
	\includegraphics[width=0.47\textwidth]{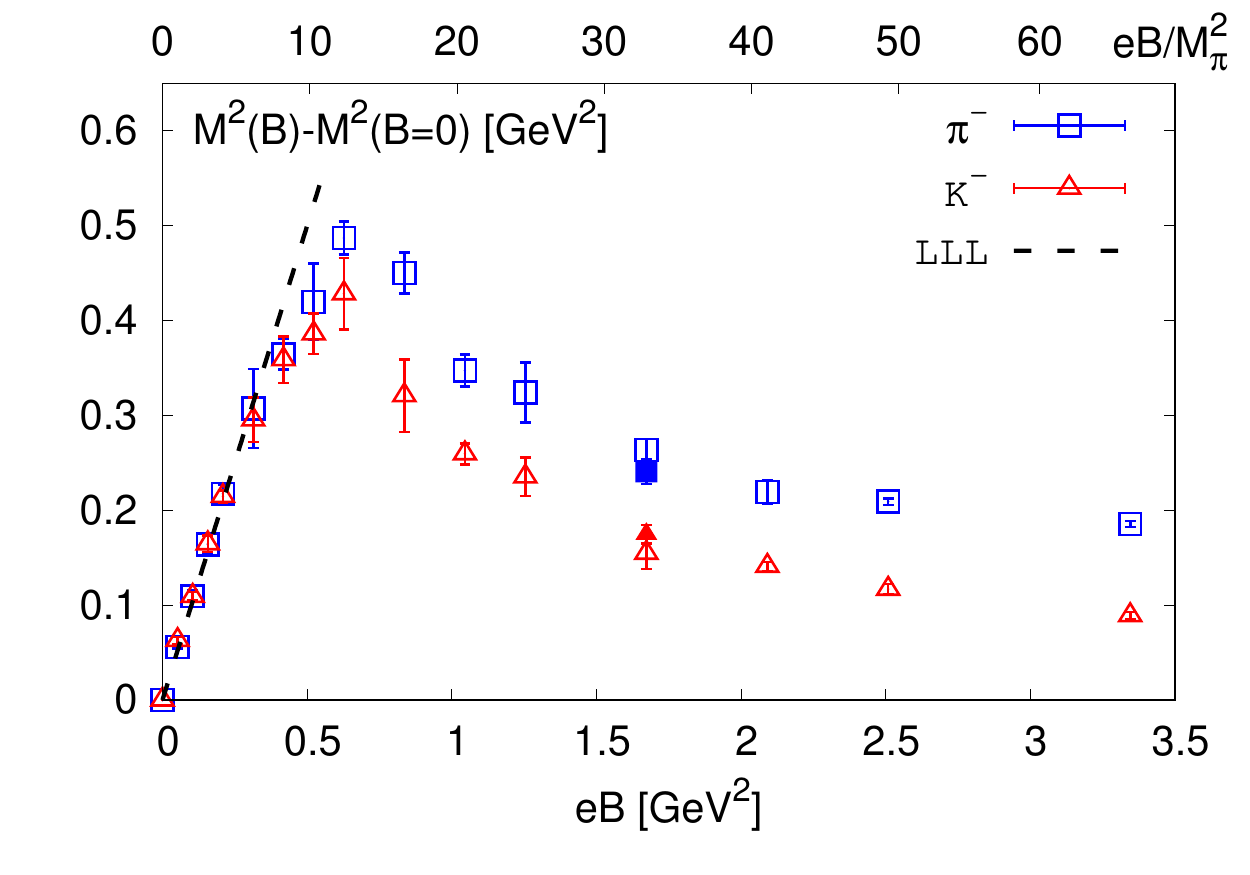}
      \includegraphics[width=0.47\textwidth]{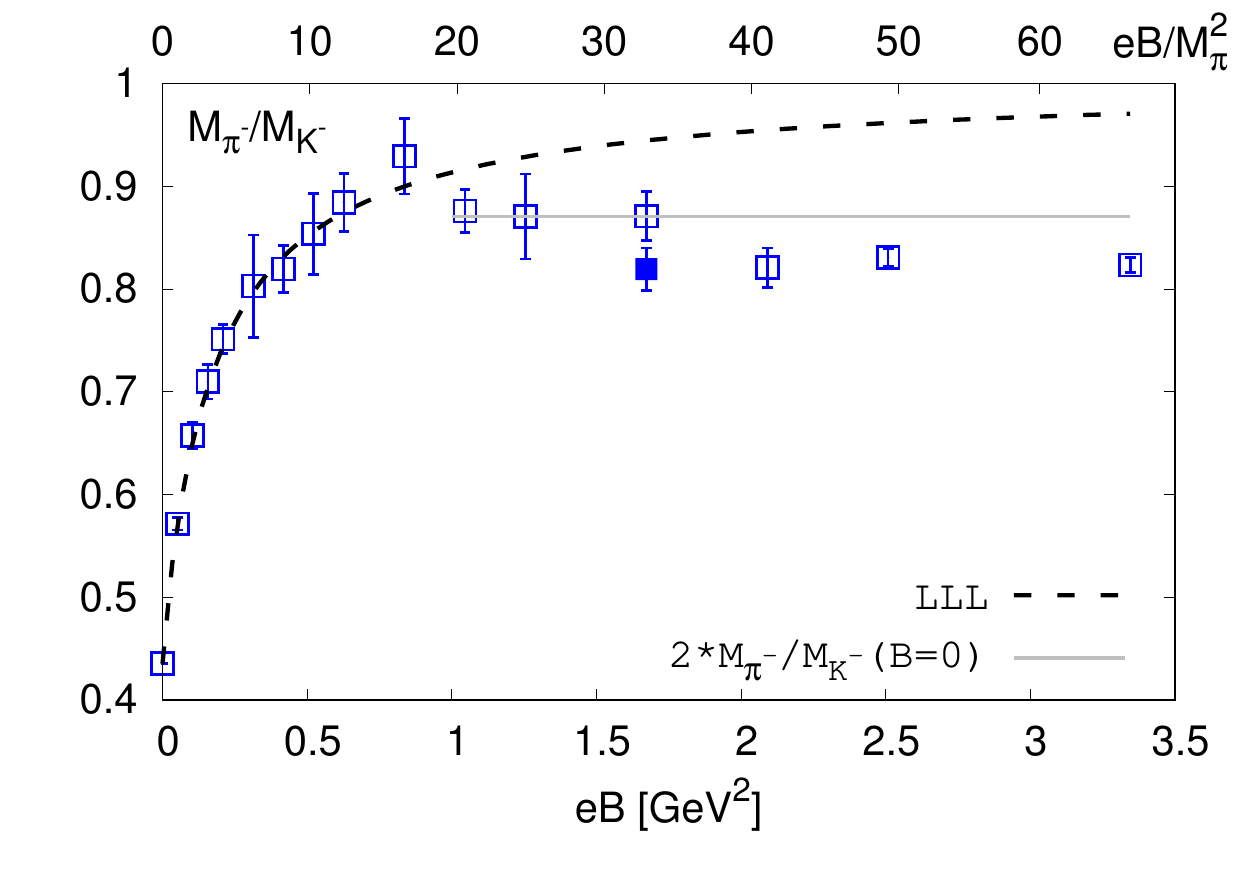}
	\caption{Left: Differences of squared masses between the case at $B\neq0$ and $B=0$, $M^2(eB) - M^2(eB=0)$ for $\pi^-$ and $K^-$ as a function of $eB$. Right: Ratio of the charged pion mass to charged kaon mass at nonzero magnetic fields. Dashed lines in both plots show the results of the lowest Landau level approximation, while the grey horizontal line in the right plot gives 2 times the value of $M_{\pi-}/M_{K^-}$ at $B=0$ to guide the eye. Data points shown in both plots are obtained from $32^3\times96$ lattices except that filled points are obtained from $40^3\times96$ lattices at $eB\simeq$1.67 GeV$^2$.}
	\label{fig:charged_mass}
\end{figure}

We now turn to the case of charged pseudoscalar mesons $\pi^{-}$ and $K^-$, and show the differences in their squared masses from the case of a zero magnetic field, i.e., $M^2(eB) - M^2(eB=0)$ 
in the left plot of Fig.~\ref{fig:charged_mass}.\footnote{Because of the parity in $eB$, the masses of their anti-particles should be the same.} We see that for both $\pi^{-}$ and $K^{-}$ the differences show a nonmonotonous behavior in the magnetic field, i.e., they first increase and then decrease with the magnetic field strength $eB$, and tend to saturate at $eB\gtrsim2.5$ GeV$^2$.  In the small magnetic field, i.e., $eB\lesssim0.31$ GeV$^2$ ($N_b\leq 6$), the differences, as labeled by blue circles and red triangles for $\pi^-$ and $K^-$ respectively, can be well described by the lowest Landau level (LLL) approximation (cf. Eq.~\ref{eq:LLL}) shown as the dashed line in the plot. At $eB>0.31$ GeV$^2$ ($N_b> 6$), the masses start to deviate from the results of the LLL approximation and then decrease with $eB$. The deviation of $M_{\pi^{-}}$ and $M_{K^{-}}$ from the LLL approximation suggests that $\pi^-$ and $K^{-}$ cannot be considered as point particles anymore at $eB\gtrsim$ 0.31 GeV$^2$. On the other hand, the decreasing behavior of $M^2(B)-M^2(B=0)$ in $eB$ at $eB\gtrsim 0.63$ GeV$^2$ is novel. In the quenched QCD, there exists no marked decreasing behavior of $\pi^+$ mass until $eB\sim$3.5 GeV$^2$~\cite{Bali:2017ian}, while the charged pion mass obtained in the previous $N_f=2+1$ QCD simulations do not possess a marked decreasing behavior as well until the largest available $eB\sim0.4$ GeV$^2$~\cite{Bali:2011qj}. 

The decreasing behavior of charged pseudoscalar meson masses at large $eB$, as observed in our study, might suffer from the finite volume effects as the lightest meson $\pi^0$ becomes lighter as $eB$ grows. To understand the finite volume effects, at a single point of $eB\simeq$1.67 GeV$^2$ lying in the region where masses decrease with $eB$, we also show the results of $M_{\pi^-}$ and $M_{K^-}$ obtained from lattices with a larger volume of $N_\sigma=40$ (denoted as filled points) in Fig.~\ref{fig:charged_mass}. As also shown in Fig.~\ref{fig:chargedM_V}, we thus consider that the finite size effects in our current study should be mild, because the mass of the lightest meson, i.e., the neutral pion, does not become much smaller at stronger magnetic fields. Thus the decreasing behavior of charged pseudoscalar meson masses as $eB$ grows at strong magnetic fields should be robust, and it may be due to the effects of dynamical quarks and strong magnetic fields in our current study.

One can also observe that at large magnetic fields, the mass of $K^-$ is less affected than $\pi^-$ by $eB$, which is probably due to the fact that the mass of $K^-$ is larger than $\pi^-$ in the vacuum as is the case for neutral mesons.  In the framework of the statistical hadron model~\cite{BraunMunzinger:2003zd}, the mass ratio of $K^-$ and $\pi^-$ could manifest itself in the difference of yields produced in the peripheral heavy-ion collisions given that the magnetic field lives sufficiently long. We further show the ratio $M_{K^-}(B)/M_{\pi^-}(B)$ in the right plot of Fig.~\ref{fig:charged_mass}. At vanishing magnetic field, the mass of $\pi^-$ is about 40\% of $K^{-}$. As the magnetic field strength $eB$ increases, $M_{\pi^-}$/$M_{K^-}$ first increases as it reaches up to $\sim$0.9 at $eB\sim$0.8 GeV$^2$, and then slightly decreases and becomes flat at a value of $\sim0.8$ at $eB\gtrsim$2 GeV$^2$.

\begin{figure}[!tbph]
	\centering
	\includegraphics[width=0.47\textwidth]{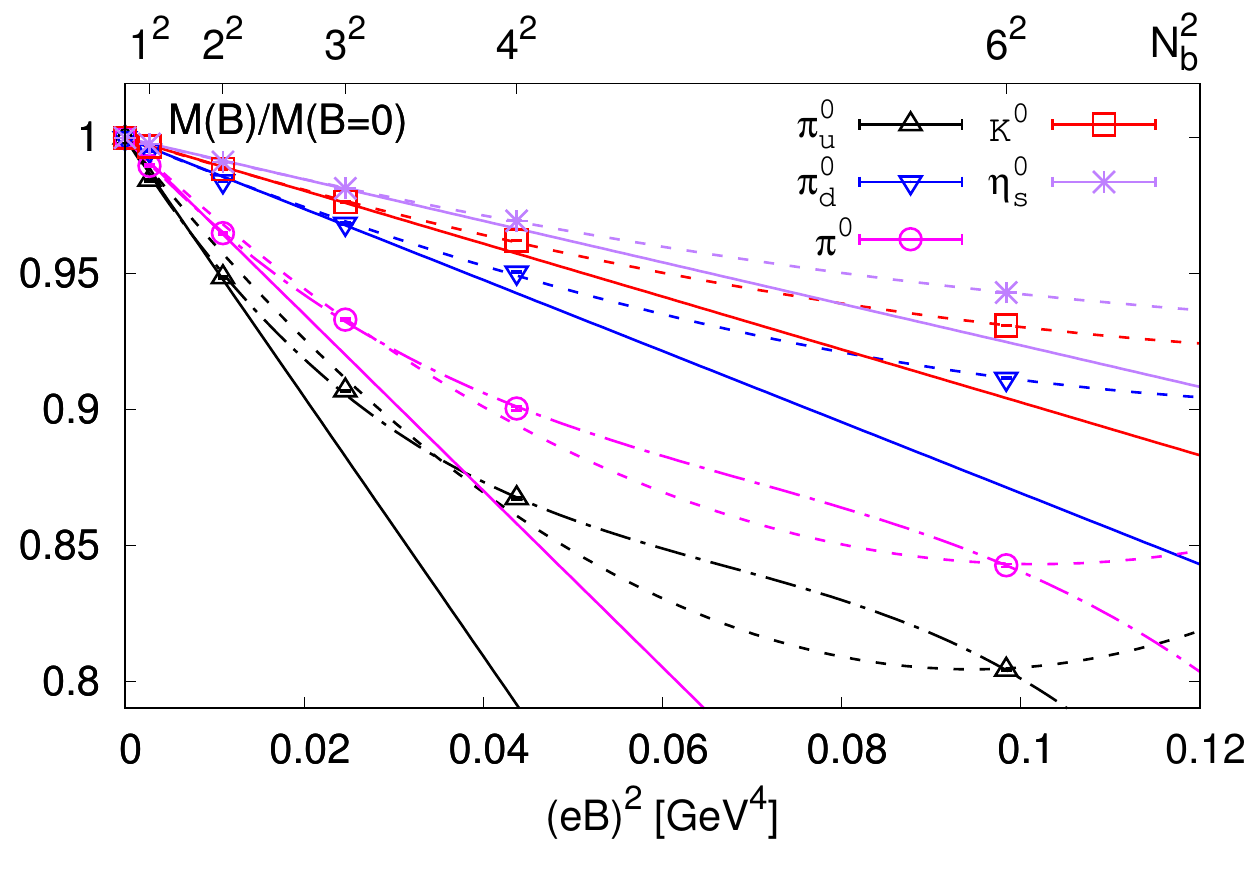}
	\includegraphics[width=0.47\textwidth]{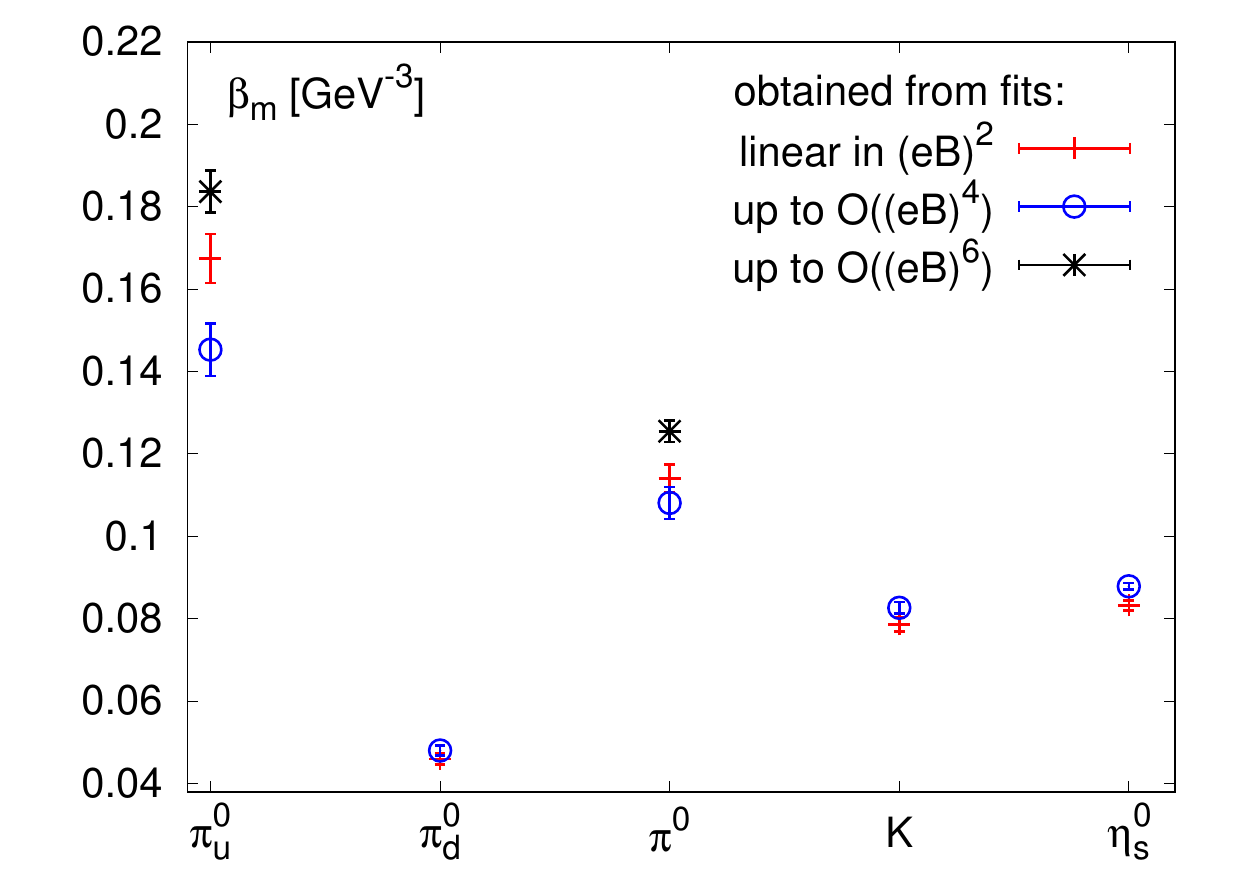}
	\caption{Left: Ratio of masses of various neutral pseudoscalar mesons to their values at the zero magnetic field as a function of $(eB)^2$ in a small magnetic field region. The solid lines represent the fits to data points with an ansatz of $M(B)/M(B=0)=1-2\pi\beta_{m}\, (eB)^2/M(B=0)$ with a fit range of $(eB)^2\in[0, 0.03]$ GeV$^4$ except that for $\pi_u^0$ and $\pi^0$ the fit range is [0, 0.02] GeV$^4$. The dashed lines and the dash-dotted lines denote fits with an ansatz of including higher-order terms up to $(eB)^4$ and $(eB)^6$, respectively, with a fit range of $(eB)^2\in[0, 0.1]$ GeV$^4$.  Right: The obtained magnetic polarizability for the $u\bar{u}$ and $d\bar{d}$ flavor components of neutral pion, $\pi^0$, neutral kaon and $\eta_s^0$ from the fits shown in the left plot. The fit results including fit ranges, fit ansatz and $\chi^2/d.o.f.$ are summarized in Table~\ref{tab:MP}.}
	\label{fig:neutral_polarizability}
\end{figure}

As seen from Fig.~\ref{fig:neutral_mass} and Fig.~\ref{fig:charged_mass}, the neutral and charged pseudoscalar mesons are not pointlike particles anymore in the strong magnetic field and their internal structures could be described by the magnetic dipole polarizability. In the relativistic case, the energy squared of a pseudoscalar meson has the following form~\cite{Luschevskaya:2015cko}
\begin{equation}
M^2(B)= M^2(B=0) + |qB| - 4\pi M(B=0)\, \beta_m \,(eB)^2 - 4\pi M(B=0)\,\beta_{m}^{1h}\, (eB)^4 + \mathcal{O}((eB)^6),
\label{eq:MP}
\end{equation}
where $q$ is the electric charge of the meson, $\beta_m$ is the  magnetic dipole polarizability, and $\beta_m^{1h}$ is the first-order magnetic hyperpolarizability.  In the weak magnetic field, we thus fit the ratio of the neutral pseudoscalar meson mass at nonzero magnetic fields to its value at a zero magnetic field using the following ansatz~\cite{Luschevskaya:2015cko}:
\begin{equation}
\frac{M(B)}{M(B=0)}=1-\frac{2 \pi\beta_{m}}{M(B=0)}\times (e B)^{2}-\frac{2 \pi}{M(B=0)} \Big(\beta_{m}^{1 h}+\pi(\beta_m)^2/M(B=0)\Big)\times(eB)^4+\mathcal{O}\left((e B)^{6}\right).
\label{eq:neutral_polarizability_ansatz}
\end{equation}

We show $M(B)/M(B=0)$ for the case of neutral pseudoscalar mesons in a small magnetic field range in the left plot of Fig.~\ref{fig:neutral_polarizability}. A clear linear behavior in $(eB)^2$ can be observed for $\pi_d^0$, $K^0$, and $\eta_s^0$ at $(eB)^2\lesssim$0.03 GeV$^2$ ($N_b\leq3$) while for $\pi_u^0$ in a smaller window, i.e., at $(eB)^2\lesssim$0.02 GeV$^2$ ($N_b\leq2$). We thus fit the data in the corresponding range with the ansatz, Eq.~\ref{eq:neutral_polarizability_ansatz}, including only the terms up to $(eB)^2$, and the fit results are denoted by solid lines in the plot. The obtained magnetic dipole polarizabilities $\beta_m$ are shown as red points in the right plot of Fig.~\ref{fig:neutral_polarizability}. To check the uncertainties of $\beta_m$, we also performed the fits to the data in a broader range of $(eB)^2<0.1$ GeV$^4$ ($N_b\leq6$) by adding higher-order terms in the fit ansatz. It can be seen from the left plot of Fig.~\ref{fig:neutral_polarizability} that the fit ansatz, including terms up to $(eB)^4$ (denoted as dashed lines), can describe the data for $\pi_d^0$, $K^0$, and $\eta_s^0$ fairly well, while an even higher-order term, i.e., $(eB)^6$, is needed to describe the data for $\pi_u^0$ (denoted as a dashed-dotted line). The corresponding results of $\beta_m$ from fits, including terms up to $(eB)^4$ and $(eB)^6$, are shown as blue and black points in the right plot of Fig.~\ref{fig:neutral_polarizability}, respectively. It can be seen that the uncertainties of $\beta_m$ for $\pi_d^0$, $K^0$, and $\eta_s^0$ are small while for $\pi_u^0$ they are relatively large. In the case of $\pi_u^0$, $\beta_m$ is about $0.167$ obtained from the linear fit in $(eB)^2$ and drops (increases) by about 15\% (10\%) obtained from fits including terms up to $(eB)^4$ ($(eB)^6$). The value of $\beta_m$ for $\pi_u^0$ is thus in the ballpark of 4 times the value of $\beta_m$ for $\pi_d^0$, which is consistent with the $qB$ scaling shown in the right plot of Fig.~\ref{fig:neutral_mass} due to $(q_u)^2=4(q_d)^2$. The $qB$ scaling can also be seen from the values of the first order hyperpolarizabilities for $\pi_u^0$ and $\pi_d^0$, i.e., $\beta_{m,\pi^0_u}^{1h}\simeq 16 ~\beta_{m,\pi^0_d}^{1h}$, as seen from Table~\ref{tab:MP} with the fit range $N_b\leq6$. We also show the results of $\pi^0$ obtained using the same fit policy as that for $\pi_d^0$ in Fig.~\ref{fig:neutral_polarizability} and Table~\ref{tab:MP}. The quality of the fit for $\pi^0$ is similar to that for $\pi^0_d$, and the obtained $\beta_m$ ($\beta_m^{1h}$) of $\pi^0$ from the best fit with smallest $\chi^2/d.o.f.$ is about 1.5 (5) times as those of both $K^0$ and $\eta_s^0$. Note that  $\chi^2/d.o.f.$ of the fits to neutral mesons are generally large which could be due to the fact that the statistics errors of neutral meson masses are tiny, i.e., at the order of $\sim 0.03\%$. This may suggest that even smaller values of $eB$ need to be included in the fits to extract reliable magnetic polarizabilities for neutral mesons.

\begin{figure}[!htbp]
	\centering
	\includegraphics[width=0.6\textwidth]{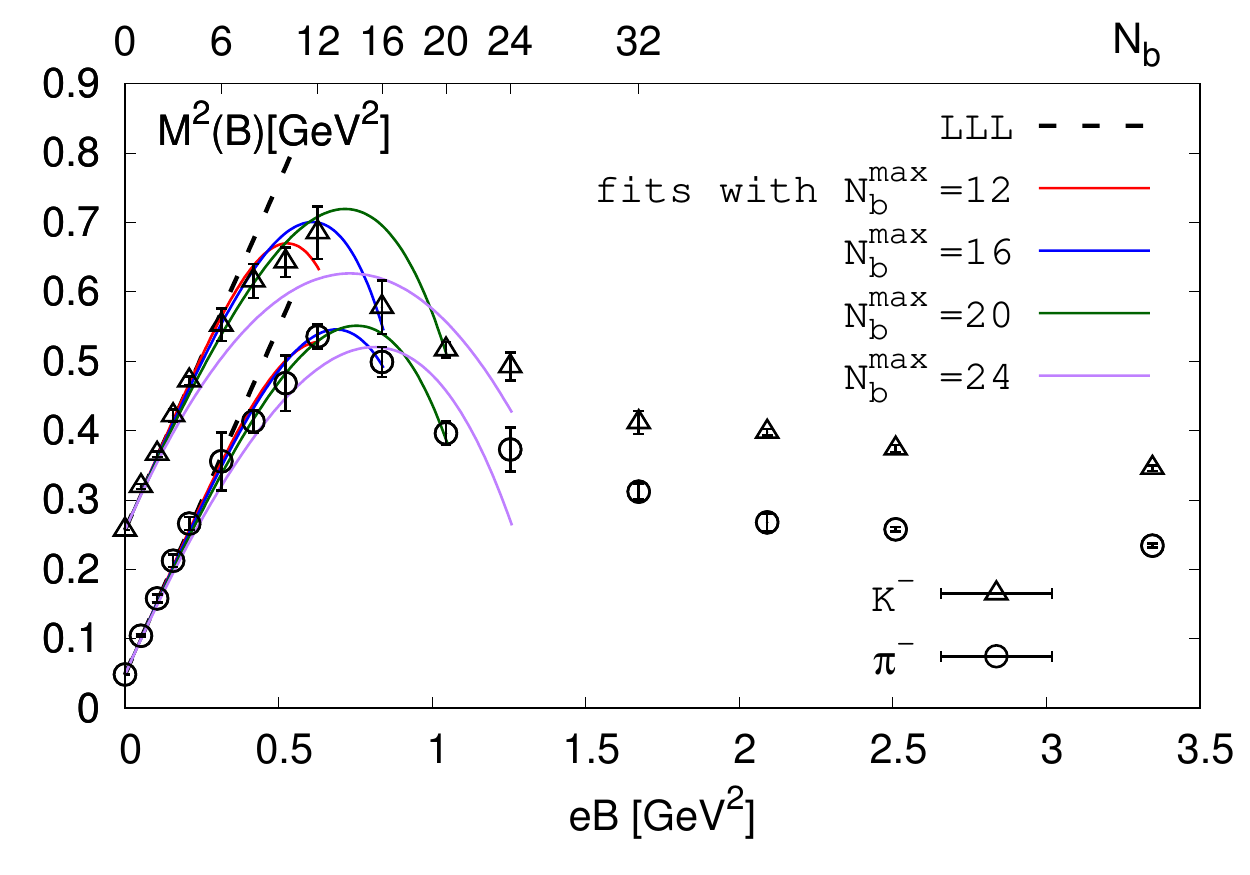}
	\caption{Fits to the squared masses of charged pion and kaon using the ansatz, Eq.~\ref{eq:MP}, to extract magnetic dipole polarizability. Solid lines represent fit results with different fit ranges of $N_b^{max}$=12, 16, 20, and 24, while black dashed lines represent the results from the LLL approximation. The fit results, including fit ranges, fit ansatz, and $\chi^2/d.o.f.$, are summarized in Table~\ref{tab:MP}. }
	\label{fig:charged_polarizability}
\end{figure}

We move forward to show the results of magnetic polarizabilities for charged pseudoscalar mesons, i.e., $\pi^{-}$ and $K^-$, based on a fit ansatz of Eq.~\ref{eq:MP} including terms up to $(eB)^4$. We performed fits to $M^2(B)$ using four different fit ranges, i.e., $N_b\leq12$, 16, 20, and 24, and the corresponding fit results are denoted by lines with $N_b^{max}$=12, 16, 20, and 24 in Fig.~\ref{fig:charged_polarizability}, respectively.\footnote{The reason we choose the smallest value of $N_b^{max}$ to be 12 is that the data can be well described by the LLL approximation at $N_b\leq6$ and at least three more data points are needed to accommodate a two-parameter fit.} The best fit is obtained with the narrowest fit range of $N_b^{max}$=12, whose $\chi^2/d.o.f.$ is closest to unity as listed in Table~\ref{tab:MP}. We find that the resulting $\beta_m$ is consistent with zero for both $\pi^-$ and $K^-$, while values of $\beta_m^{1h}$ for $\pi^-$ and $K^-$ are comparable to be around 0.3 GeV$^{-7}$.
As the fit range becomes broader, the quality of the fit becomes worse, and this indicates that higher-order hyperpolarizabilities are needed to describe the data, which is beyond the scope of the current paper. 
 \begin{table}[!htbp]
 	\begin{center}
 		\begin{tabular}{|c|c|c|c|c|c|c|}
 			\hline
 			$channel$  &$\beta_m$ [GeV$^{-3}$]  & $\beta_m^{1h}$ [GeV$^{-7}$]  &  $ \chi^2$/d.o.f.  & fit ansatz \& range \\
 			\hline
 			\multirow{ 3}{*}{$\pi_u^0$}              & 0.167(6)     & - & 31.3 &  up to $\mathcal{O}((eB)^2)$, $N_b\in[0,2]$ \\
 			&   0.145(6)       & -1.07(9)  & 410.3 & up to $\mathcal{O}((eB)^4)$, $N_b\in[0,6]$   \\
 			&  0.184(5)       &  -2.7(2) &  25.1 &up to $\mathcal{O}((eB)^6)$, $N_b\in[0,6]$   \\ \hline
 			\multirow{ 2}{*}{$\pi_d^0$}              & 0.046(1)  & -& 10.9 & up to $\mathcal{O}((eB)^2)$, $N_b\in[0,3]$ \\
 			&  0.048(1)    & -0.20(1)  &  11.0 &up to $\mathcal{O}((eB)^4)$, $N_b\in[0,6]$   \\ \hline
 			\multirow{ 3}{*}{$\pi^0$}              & 0.114(3)     & - & 9.3 &  up to $\mathcal{O}((eB)^2)$, $N_b\in[0,2]$ \\
 			&   0.108(4)       & -0.70(6)  & 49.7 & up to $\mathcal{O}((eB)^4)$, $N_b\in[0,6]$   \\
 			&  0.125(3)       &  -1.6(1) &  3.5 &up to $\mathcal{O}((eB)^6)$, $N_b\in[0,6]$   \\ \hline
 			\multirow{ 2}{*}{$K^0$}              &0.079(2)    & - & 7.8 & up to $\mathcal{O}((eB)^2)$, $N_b\in[0,3]$ \\
 			&  0.083(1)   & -0.31(2)    &  6.8 &up to $\mathcal{O}((eB)^4)$, $N_b\in[0,6]$   \\ \hline
 			\multirow{ 2}{*}{$\eta_s^0$}              & 0.083(1)  & - &  9.5& up to $\mathcal{O}((eB)^2)$, $N_b\in[0,3]$ \\
 			&  0.0878(8)    & -0.29(1)    &  5.8 &up to $\mathcal{O}((eB)^4)$, $N_b\in[0,6]$   \\ \hline
 			\multirow{ 4}{*}{$\pi^-$}              & -0.00(5)    & 0.4(1) & 1.3 & up to $\mathcal{O}((eB)^4)$, $N_b\in[0,12]$   \\
 			& 0.03(3)    & 0.25(5)     & 1.3  &$-$, $N_b\in[0,16]$ \\ 
 			& 0.08(2)    & 0.14(2)     & 1.9  &$-$, $N_b\in[0,20]$   \\ 
 			&  0.15(3)    & 0.06(3)    & 5.6  &$-$, $N_b\in[0,24]$  \\ \hline
 			\multirow{ 4}{*}{$K^-$}              & -0.02(2)     & 0.30(9) & 1.9 &  up to $\mathcal{O}((eB)^4)$, $N_b\in[0,12]$   \\
 			&  0.01(2)     & 0.16(4)  & 2.4  & $-$, $N_b\in[0,16]$   \\ 
 			&  0.05(2) & 0.06(2)  &  4.2 &$-$, $N_b\in[0,20]$   \\ 
 			&  0.11(2) & 0.00(2)  & 10.0  &$-$, $N_b\in[0,24]$   \\ \hline
 		\end{tabular}
 	\end{center}
 	\caption{The magnetic dipole polarizability $\beta_m$ and the first-order hyperpolarizability $\beta_{m}^{1h}$ of neutral mesons $\pi^0_u$, $\pi^0_d$, $\pi^0$, $K^0$ and $\eta_s^0$ as well as charged mesons $\pi^-$ and $K^-$,  fit ranges and fit ansatz as well as $ \chi^2$/d.o.f. obtained from fits shown in Figs.~\ref{fig:neutral_polarizability} and~\ref{fig:charged_polarizability}.
 	} \label{tab:MP}
 \end{table}

We close this subsection by comparing our results of magnetic polarizabilities (cf. Figs.~\ref{fig:neutral_polarizability} and ~\ref{fig:charged_polarizability} and Table~\ref{tab:MP}) to previous computations of $\beta_m$ for light pseudoscalar mesons in lattice QCD, which were done in quenched QCD~\cite{Lee:2005dq,Luschevskaya:2014lga, Luschevskaya:2015cko,Bali:2017ian} and dynamical QCD \cite{Bignell:2020dze}. 
In the quenched QCD studies, the corresponding pion mass tuned by the valence quark mass at a zero magnetic field is generally large, e.g., $M_\pi(eB=0)\gtrsim$ 512 MeV in~\cite{Lee:2005dq} , $M_\pi(eB=0)\gtrsim$ 320 MeV in~\cite{Luschevskaya:2014lga, Luschevskaya:2015cko}  and $M_\pi(eB=0)\gtrsim~400$ MeV in~\cite{Bali:2017ian}. On the other hand, electroquenched computations in (2+1)-flavor QCD, where no background magnetic field is present on the gauge field ensembles, are performed with $M_\pi(eB=0)$ ranging from 702 to 296 MeV \cite{Bignell:2020dze}.
One difference to be noted is that in Ref.~\cite{Bali:2017ian} $\beta_m$ for $\pi_u$ is about twice that for $\pi_d$, while in our case $\beta_m$ is about four times as that for $\pi_d$, which is consistent with the $qB$ scaling.  This could be due to the fact that a sufficiently small magnetic field needs to be used to extract $\beta_m$ and relatively larger weakest magnetic fields are applied in~\cite{Bali:2017ian}. As pointed in Ref.~\cite{Bali:2017ian}, the hopping parameter $\kappa$ used in the Wilson propagator needs to be along the line of constant physics in the magnetic field, i.e., $\kappa$ is $eB$ dependent,  $\beta_m$ obtained from~\cite{Lee:2005dq} with $\kappa$ as a constant in $eB$ might become smaller when the $eB$ dependence of $\kappa$ was taken into account as indicated from the study in~\cite{Bali:2017ian}. On the other hand, studies performed using overlap valence quarks in quenched QCD on $18^4$ and $20^4$ lattices give a negative value of $\beta_m$ for the charged pion, which is close to the experimental results obtained from the COMPASS Collaboration, i.e., $\beta_{m}=(-2.0\pm0.6_{stat}\pm0.7_{syst})\times10^{-4}$ fm$^3$~\cite{Adolph:2014kgj}. Note that the experiment value of $\beta_m$ is obtained under the assumption that electric polarizability $\alpha_\pi = -\beta_m$~\cite{Adolph:2014kgj}. However, the value of $\beta_m$ for the charged pion obtained in both Ref.~\cite{Bignell:2020dze} and our study is not negative. Moreover, the value of $\beta_m$ for the charged pion obtained from the electroquenched computation is in the range of 0.003$-$0.005 GeV$^{-3}$~\cite{Bignell:2020dze}, and is significantly different from our results. The discrepancies might be due to many issues, e.g., effects from the interaction of dynamical quarks with the magnetic field, a sufficiently weak magnetic field needed to compute the polarizability, etc. Thus, further studies in full QCD with continuum extrapolations at physical pion mass are crucially needed to have a better determination of magnetic polarizabilities.

\subsection{Light quark chiral condensates}
\label{sec:res_pbp}
As has been pointed out in Ref.~\cite{Bali:2011qj}, the presence of the external magnetic field does not introduce any new $eB$-dependent divergences. To take care of the additive divergences
as well as the multiplicative renormalization in the chiral condensate, we investigate the following dimensionless quantity~\cite{Bali:2012zg}:
\begin{equation}
\Sigma_{l} (B) = \frac{2 m_l}{M_\pi^2 f_\pi^2}\left (\pbp_l(B\ne0) - \pbp_l(B=0)\right ) +1 ,
\label{eq:SigmaB}
\end{equation}
where $m_l\equiv m_u\equiv m_d$ is the bare quark mass for up and down quarks, and $M_\pi$ and $f_\pi$ is the mass of pion and pion decay constant, respectively, at $eB=0$. In our study, $M_\pi$ is found to be 220.61(6) MeV while $f_\pi$ is 96.93(2) MeV, whose determination will be shown in Sec.~\ref{sec:res_dc-GMOR}. In Ref.~\cite{Bali:2012zg}, continuum extrapolated results of $(\Sigma_u+\Sigma_d)/2$ and $(\Sigma_u-\Sigma_d)$ have been obtained based on lattice simulations of $N_f=2+1$ QCD using stout fermions with lattice spacings of 0.29, 0.215, 0.15, 0.125, and 0.1 fm. It is found that the results obtained at the two finest lattice spacings are already quite close to the continuum limit~\cite{Bali:2012zg}. We are working at one single lattice cutoff of $a=0.117$ fm using the HISQ discretization scheme, which should also be close to the continuum limit. Moreover, the HISQ discretization scheme is expected to have a smaller taste symmetry breaking effect than the stout discretization scheme at the same lattice spacing~\cite{Ding:2015ona,Bazavov:2011nk}. Thus, the discretization error should be under control in the current computation of chiral condensates.

\begin{figure}[tbph]
	\centering
	\includegraphics[width=0.47\textwidth]{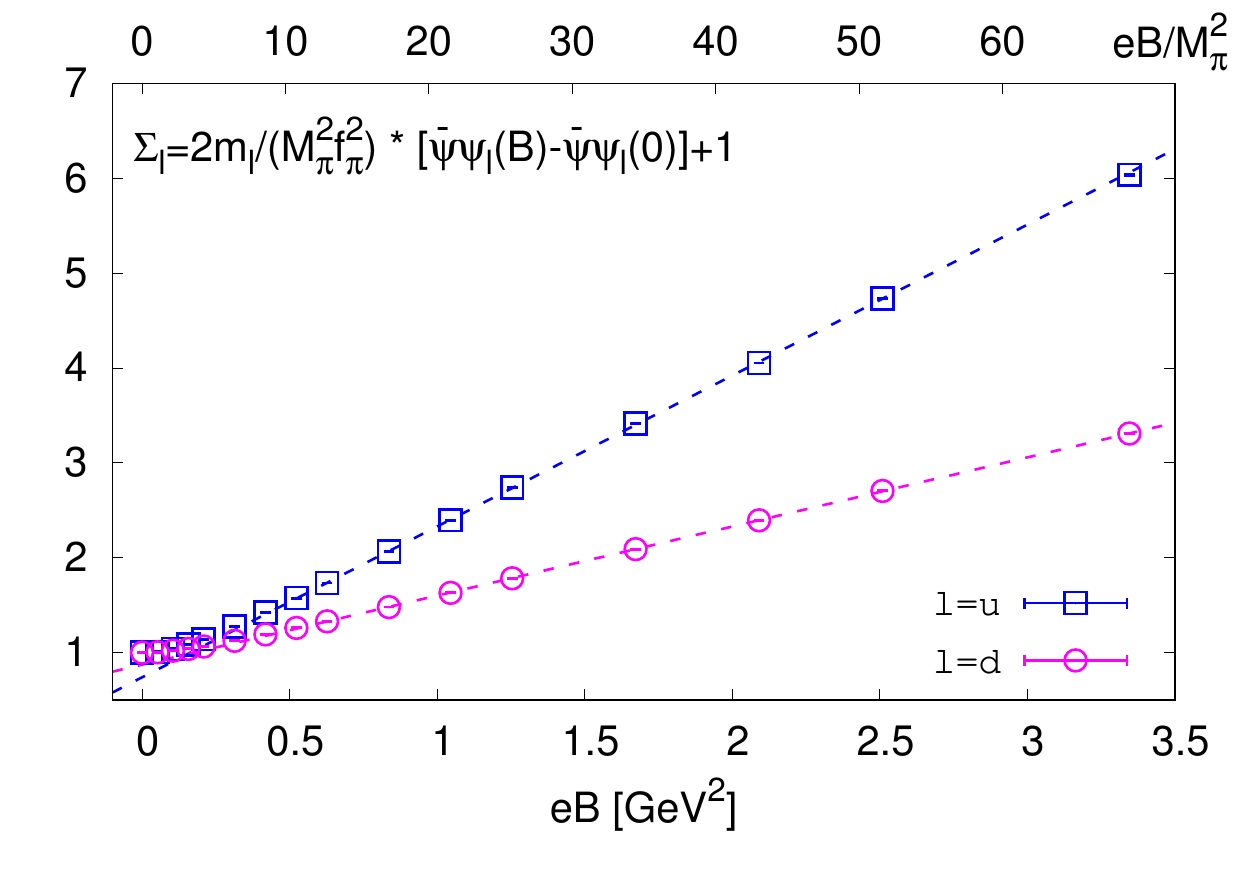}
		\includegraphics[width=0.47\textwidth]{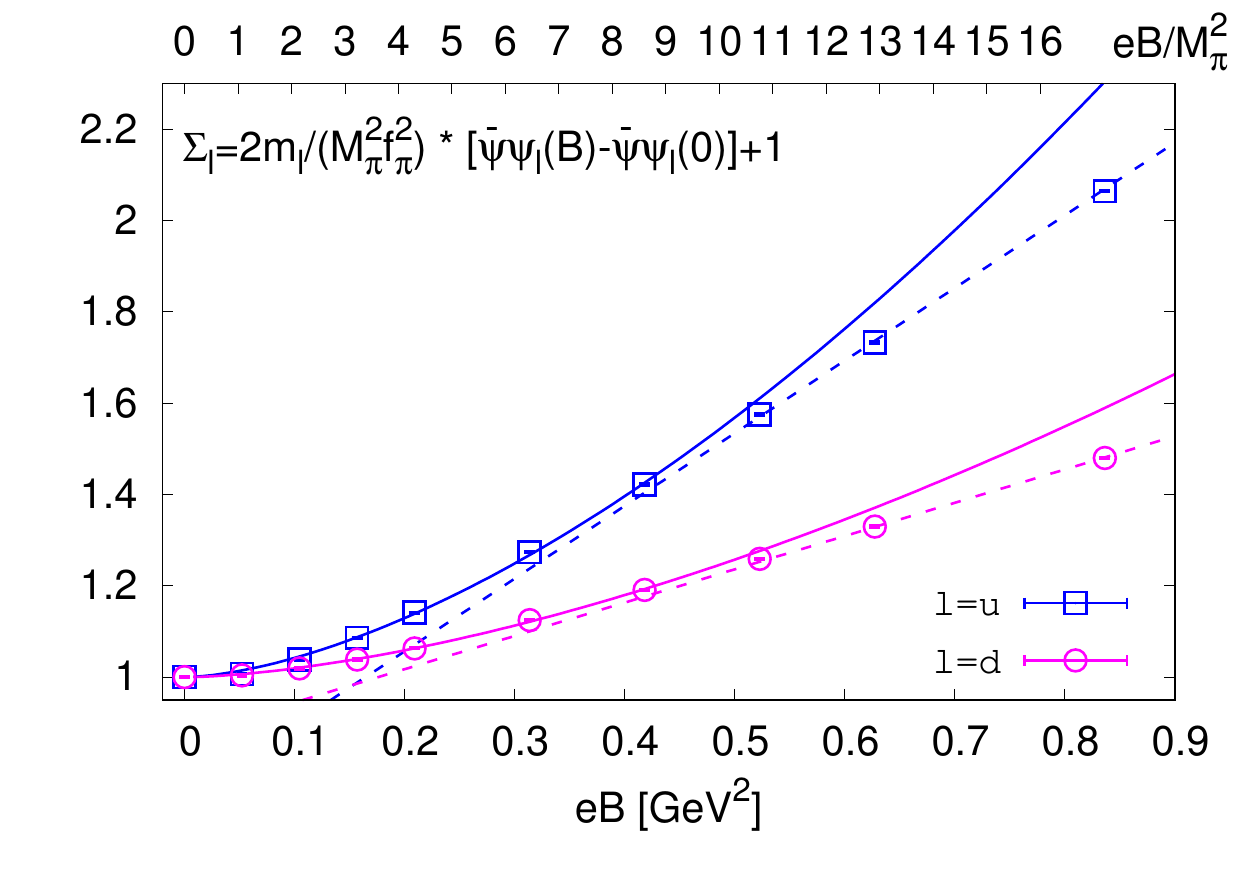}
	\caption{Left: Renormalized up and down quark chiral condensates as a function of magnetic field strength $eB$. The left plot shows chiral condensates in the whole $eB$ region while the right one shows those in a narrower $eB$ region.  The dashed lines in both plots denote two-parameter linear fits of chiral condensates at $eB\geq0.5 $ GeV$^2$ while the solid lines in the right plot represent power-law fits with ansatz of $h(eB)^\gamma$+1 at $eB\leq0.5 $ GeV$^2$.}
	\label{fig:pbp_ud}
\end{figure}
\begin{figure}
	\centering
	\includegraphics[width=0.47\textwidth]{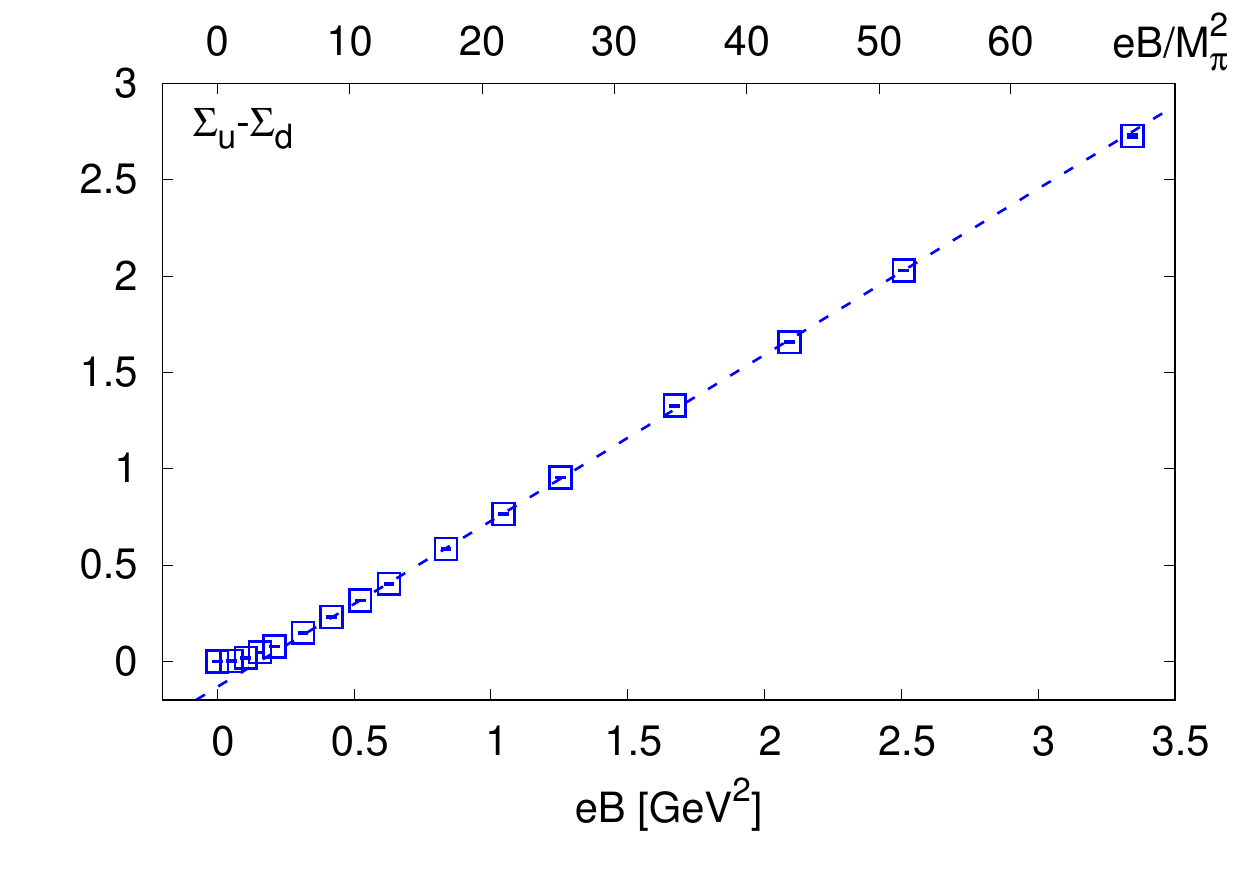}
	\includegraphics[width=0.47\textwidth]{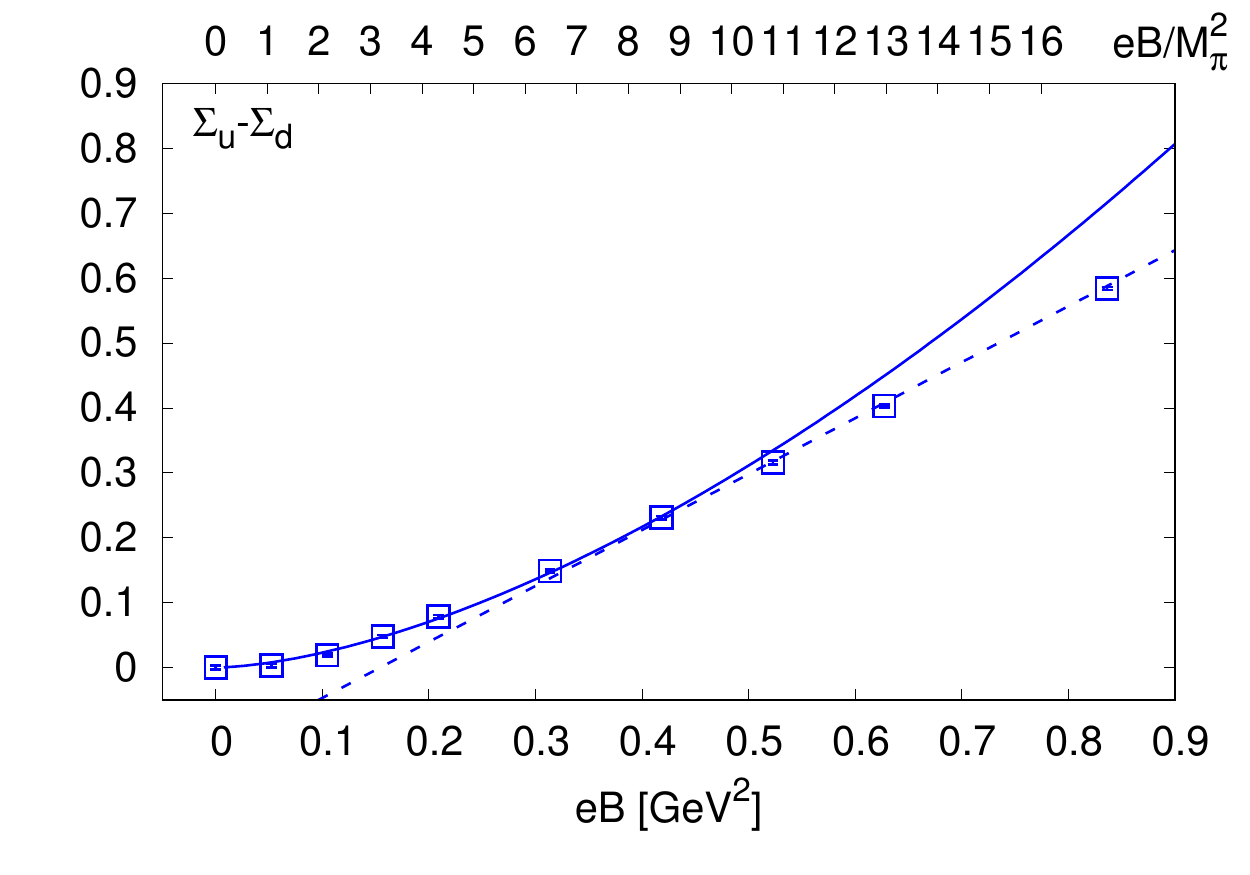}
	\caption{Similar as Fig.~\ref{fig:pbp_ud} but for the difference of renormalized quark chiral condensates $\Sigma_u - \Sigma_d$. The solid line represent a two-parameter power-law fit with an ansatz of $h|eB|^\gamma$, while the dashed lines denote two-parameter linear fits in $eB$. }
	\label{fig:pbp_ud-dif}
\end{figure}
We show $\Sigma_u$ and $\Sigma_d$ as a function of $eB$ in Fig.~\ref{fig:pbp_ud}. Because of different electric charges of up and down quarks, up and down quark chiral condensates become nondegenerate in the nonzero magnetic field. And the up quark chiral condensate is more affected by the magnetic field than that of the down quark chiral condensate, probably due to $|q_u|>|q_d|$. It is obvious to see that, at large magnetic fields, both up and down quark chiral condensates show linear behavior in $eB$. We performed linear fits for these two condensates at $(eB)\gtrsim 0.5$ GeV$^2$, and the corresponding fit results shown as dashed lines in Fig.~\ref{fig:pbp_ud} describe the data fairly well. We find that the slope for the up quark condensate obtained from the linear fit is 1.591(5), and it is about twice that for the down quark which is 0.729(1). While the linear fit works for strong magnetic fields, it is not the case anymore at $eB\lesssim$0.5 GeV$^2$. This can be seen from the right plot of Fig.~\ref{fig:pbp_ud} as a blowup plot of the left one. At $eB\lesssim$0.5 GeV$^2$, both chiral condensates increase faster than a linear behavior in $eB$. We thus adopt a two-parameter power-law fit ansatz of $h|eB|^{\gamma}+1$ to fit the chiral condensates. We found that this fit ansatz can describe the data at $eB\lesssim$0.5 GeV$^2$ well and the exponent $\gamma$ obtained from these two condensates are almost the same, i.e., $\gamma=1.62(4)$ for the up quark chiral condensate while $\gamma=1.61(4)$ for the down quark chiral condensate. We also show the difference of up and down quark chiral condensates $\Sigma_u-\Sigma_d$ in Fig.~\ref{fig:pbp_ud-dif}. Similar to the fits we showed in Fig.~\ref{fig:pbp_ud}, we performed two-parameter linear fits (shown as the dashed line) at $eB\gtrsim0.5$ GeV$^2$ and two-parameter power-law fits (shown as the solid line) at $eB\lesssim0.5$ GeV$^2$. It is expected that $\Sigma_u-\Sigma_d$ possesses a linear behavior in large $eB$ and a power-law behavior with the same exponent as that in Fig.~\ref{fig:pbp_ud} at $eB\lesssim$0.5 GeV$^2$.

\begin{figure}[tbph]
	\centering
	\includegraphics[width=0.47\textwidth]{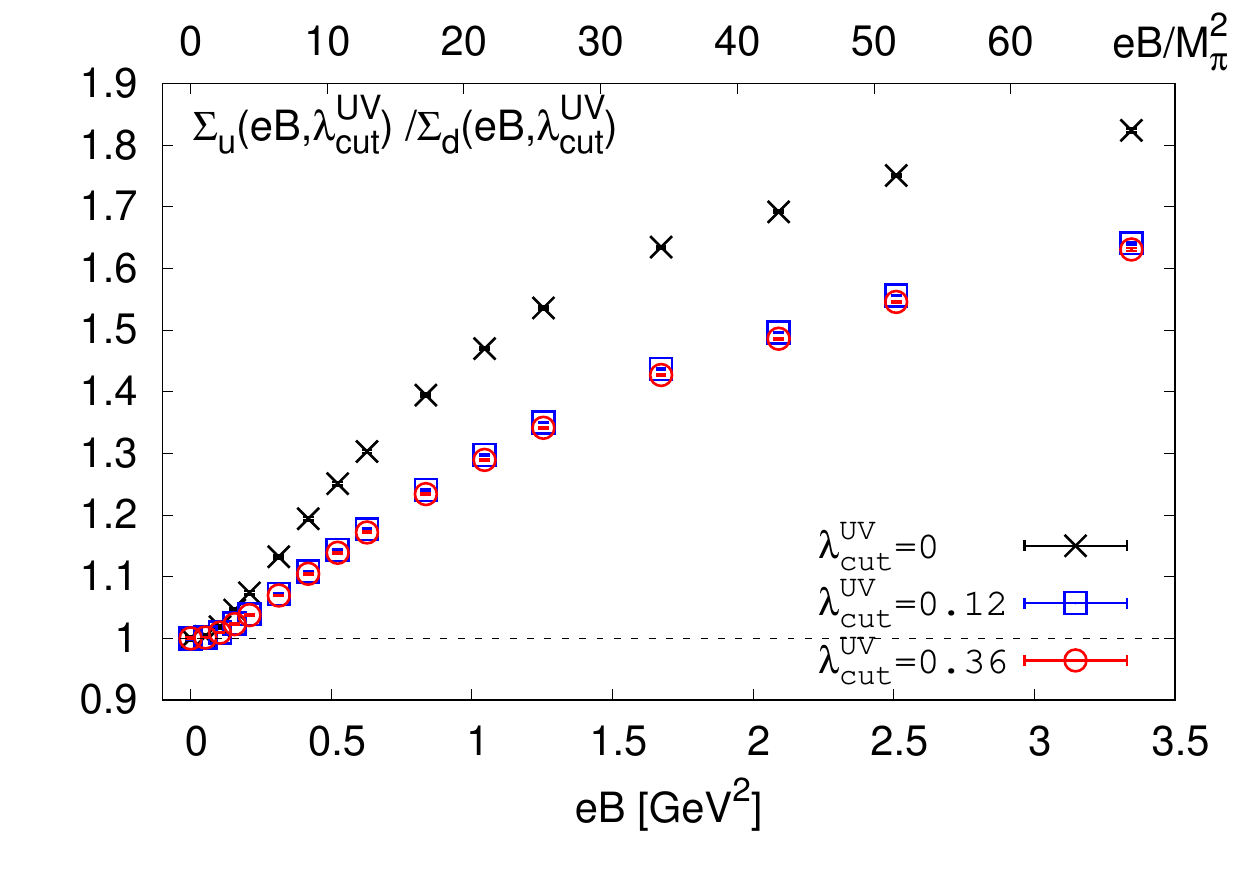}
	\includegraphics[width=0.47\textwidth]{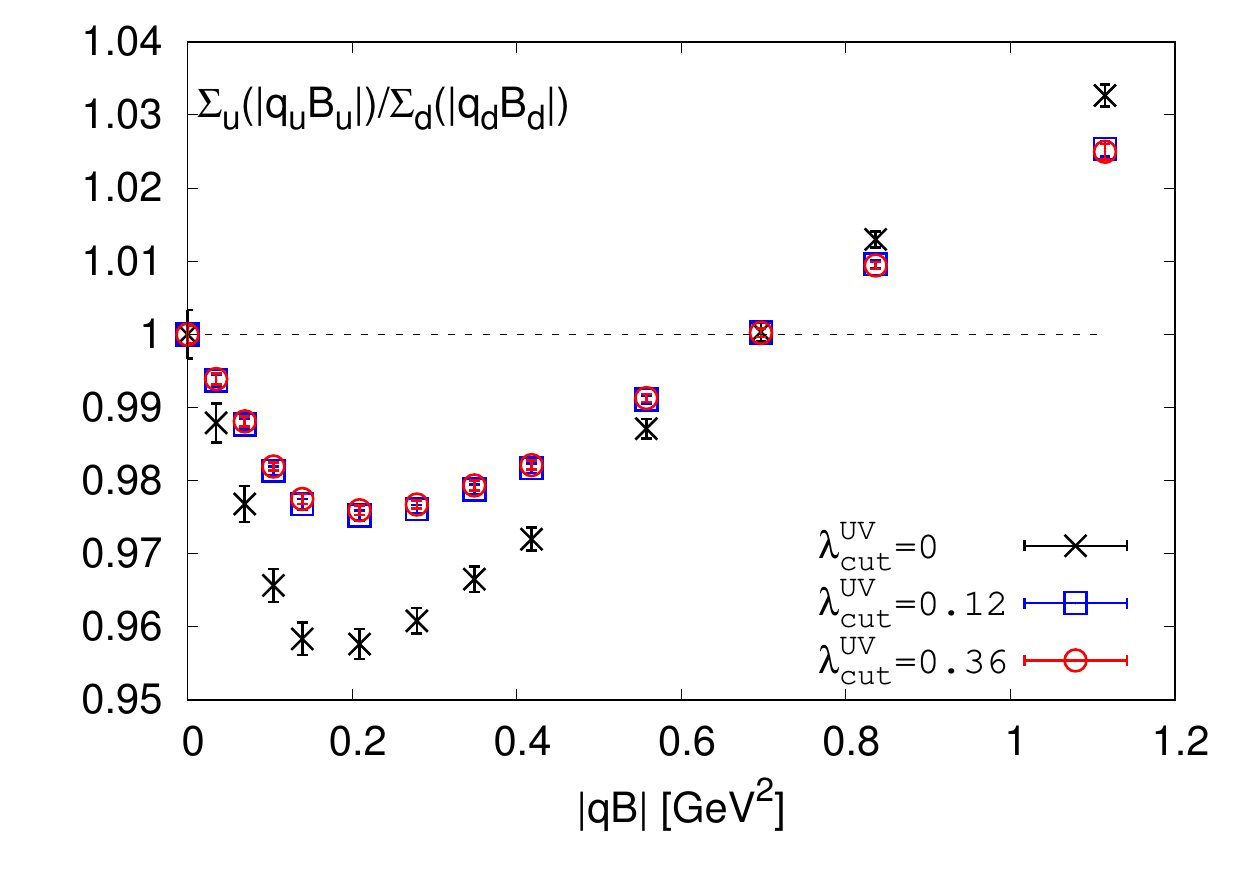}
	\caption{ Left: Ratio of renormalized up quark condensate to the down quark condensate $\Sigma_u(eB)/\Sigma_d(eB)$ as a function of $eB$. Right: Similar as the left plot but as a function of $|qB|=|q_uB_u|=|q_dB_d|$. In both plots, $\lambda^{\rm UV}_{cut}=0$ corresponds to the case that the subtrahend in the renormalized chiral condensate is $\pbp_l(B=0)$ (cf. Eq.~\ref{eq:SigmaB}), and $\lambda^{\rm UV}_{cut}=0.12$ and 0.36 gives the uncertainty on the estimate of UV-divergence part of the chiral condensates as the subtrahend in Eq.~\ref{eq:SigmaB_cut}.}
	\label{fig:pbp_ratio}
\end{figure}

We further show the ratio of $\Sigma_u/\Sigma_d$ in Fig.~\ref{fig:pbp_ratio}. To better understand the influence of the UV-divergence part of the chiral condensate in the ratio, we also investigate the following quantity:
\begin{equation}
\Sigma_{l} (B,\lambda^{\rm UV}_{cut}) = \frac{2m_l}{M_\pi^2 f_\pi^2}\left (\pbp_l(B) - \pbp_l^{\rm UV}(B=0,\lambda^{\rm UV}_{cut})\right ) +1,
\label{eq:SigmaB_cut}
\end{equation}
where $\lambda^{\rm UV}_{cut}$ is the lower limit of $\lambda$ in the integration in Eq.~\ref{eq:pbp_UV}  which gives the UV-divergence part of the chiral condensate, i.e., $\pbp_l^{\rm UV}(B=0,\lambda^{\rm UV}_{cut})$. Apparently, when $\lambda_{cut}^{\rm UV}=0$, Eq.~\ref{eq:SigmaB_cut} is the same as Eq.~\ref{eq:SigmaB}, i.e., $\Sigma_{l} (B,\lambda^{\rm UV}_{cut}=0) \equiv\Sigma_{l} (B) $. To estimate the UV-divergence contribution to $\pbp_l$, $\lambda^{\rm UV}_{cut}=0.12$ and 0.36 are adopted as discussed in Sec.~\ref{sec:setup_UV}.
In the left plot of Fig.~\ref{fig:pbp_ratio}, the ratio $\Sigma_u/\Sigma_d$ obtained using different values of $\lambda^{\rm UV}_{cut}$ increases with the increasing strength of the magnetic field $eB$, which is a consequence that $\Sigma_u-\Sigma_d$ increases faster in $eB$ than $\Sigma_d$. 

Keeping in mind the fact of $qB$ scaling for $M_{\pi^0_u}$ and $M_{\pi^0_d}$, we also show the ratio of $\Sigma_u$ and $\Sigma_d$ at the same values of $|qB|=|q_uB_u|$=$|q_dB_d|$ with three different $\lambda^{\rm UV}_{cut}$ in the right plot of Fig.~\ref{fig:pbp_ratio}. We find that the ratio $\Sigma_u(|q_uB_u|)/\Sigma_d(|q_dB_d|)$ with $\lambda^{\rm UV}_{cut}=0.12$ and 0.36 is very close to unity with deviation at most by 3\% with $qB$ up to about 1.1 GeV$^2$.  We thus conclude that the $qB$ scaling of light quark chiral condensates holds within an accuracy of 3\% in our current window of magnetic fields.

As we mentioned in the Sec.~\ref{sec:res_mass}, the $qB$ scaling of chiral condensates should also hold exactly in the quenched approximation and could be spoiled by the influence of dynamical quarks in the full QCD. However, one can find that the $qB$ scaling of chiral condensates holds in $N_f=2+1$ QCD with a physical pion mass by analyzing the continuum extrapolated values of $(\Sigma_u+\Sigma_d)/2$ and $\Sigma_u-\Sigma_d$ listed in Table I in Ref.~\cite{Bali:2012zg}. For the case of $N_f=2$ QCD with a much larger pion mass, one can find that the $qB$ scaling does not hold for up and down quark chiral condensates~\cite{DElia:2011koc} (cf. Table I in Ref.~\cite{DElia:2011koc}). It is also interesting to see in Ref.~\cite{DElia:2011koc} that the difference between up and down quark chiral condensates at the same values of $qB$ becomes smaller as $eB$ grows. 
In Ref.~\cite{DElia:2011koc}, the so-called valence quark contribution to the chiral condensate was also presented in order to understand the magnetic catalysis. 
It can be found that the $qB$ scaling holds in the valence contribution to the chiral condensate, which resembles the case in the quenched limit~\cite{DElia:2011koc}. Based on our current study with $M_\pi(eB=0)\simeq220$ MeV, and also the results mentioned above obtained with $M_\pi(eB=0)=140$ MeV in Ref.~\cite{Bali:2012zg} and $M_\pi^{\textrm{RMS}}(eB=0)>600$ MeV in Ref.~\cite{DElia:2011koc}, it is thus conceivable that the $qB$ scaling does depend on the mass of dynamical quarks, whose effects however are negligible in our current study.\footnote{In Ref.~\cite{DElia:2011koc}, simulations with the standard unimproved staggered fermions on lattices with $a\simeq0.3$ fm were adopted. The Goldstone pion mass is about 200 MeV in Ref.~\cite{DElia:2011koc} and the corresponding RMS pion mass in this setup should be larger than 600 MeV as inferred from, e.g., Fig. 1 in Ref.~\cite{Ding:2015ona}, where RMS pion masses as a function of lattice spacing are shown for improved staggered fermions and they are generally smaller compared to the case of standard unimproved staggered fermions.} We remark here that, based on the observation in Refs.~\cite{DElia:2011koc,Bali:2012zg} and our study, a sufficiently strong magnetic field, probably larger than the pion mass squared, is needed to observe the $qB$ scaling.

\begin{figure}[tbph]
	\centering
	\includegraphics[width=0.47\textwidth]{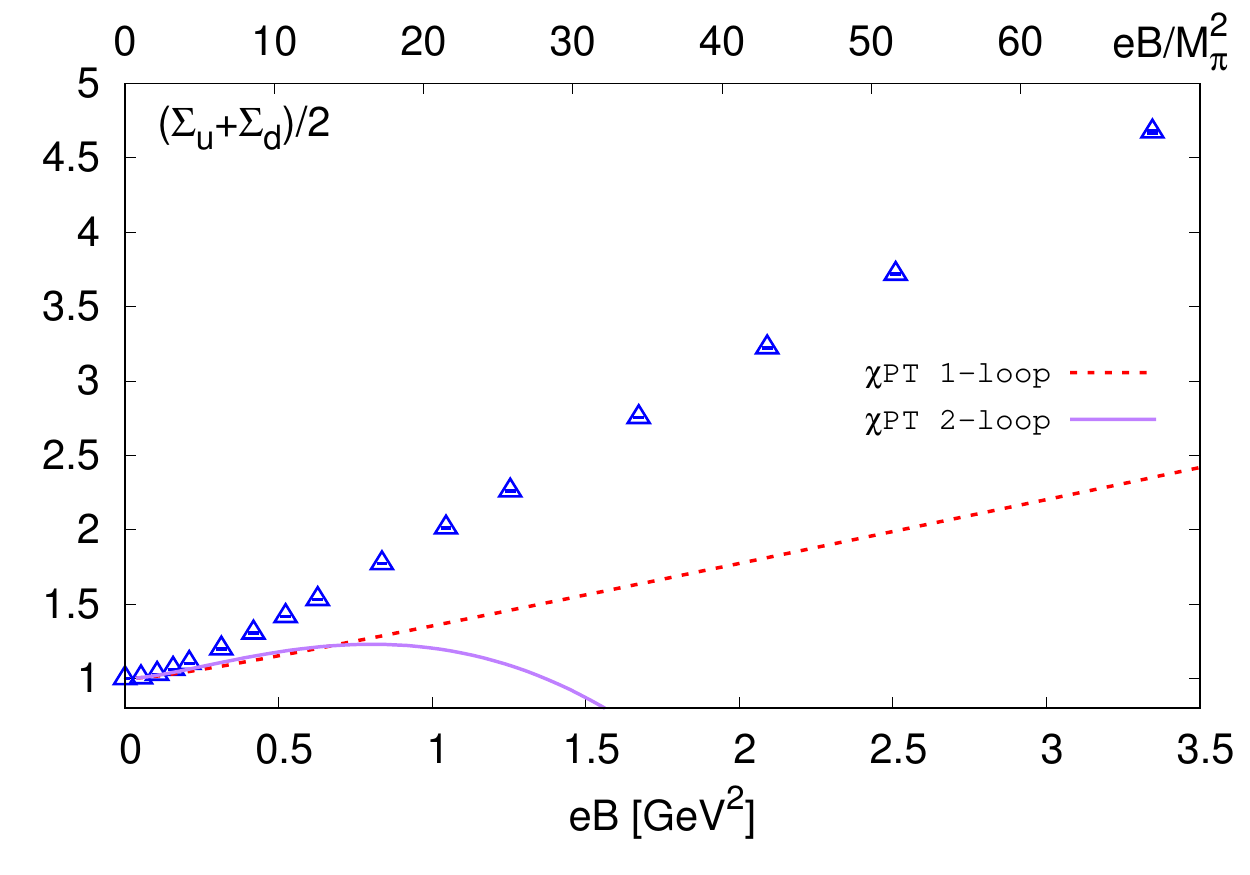}
	\includegraphics[width=0.47\textwidth]{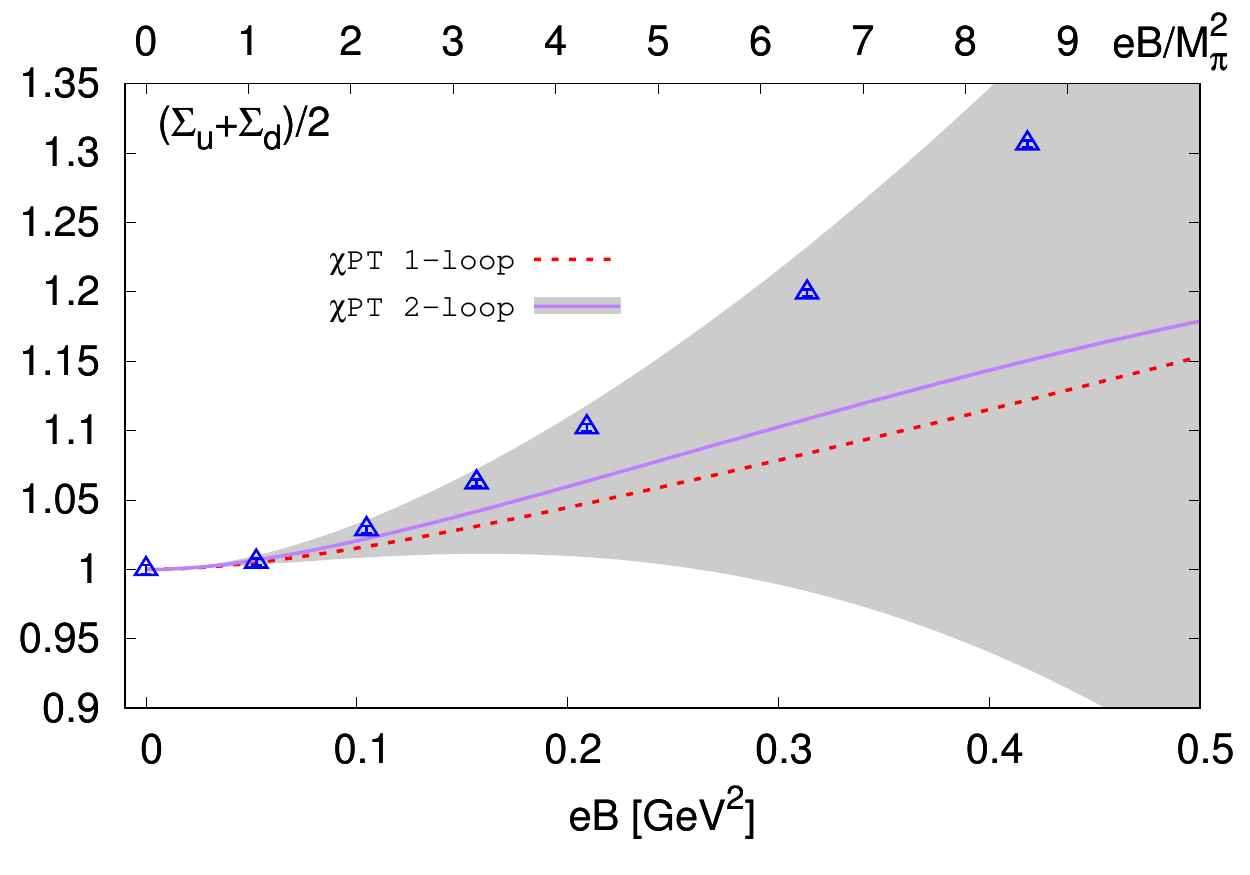}
	\caption{Left: Average of the up and down quark chiral condensates as a function of magnetic field strength $eB$. Right: Same as the left plot but a blow up in the small values of $eB$. The red dashed line stands for the one-loop $\chi$PT results~\cite{Cohen:2007bt,Andersen:2014xxa} while the purple solid line with a grey band represents the two-loop $\chi$PT results~\cite{Werbos:2007ym}.}
	\label{fig:pbp_avg}
\end{figure}

To compare with the results from $\chi$PT, we then show the average of chiral condensates, i.e., $\Sigma_{avg} = (\Sigma_u + \Sigma_d)/2$, in Fig.~\ref{fig:pbp_avg} together with the results from $\chi$PT.
By comparing to the one-loop $\chi$PT results extended to nonzero pion mass (dashed lines in the plot), we find that the results from $\chi$PT can only describe our lattice data at the weakest magnetic field of $eB=$ 0.052 GeV$^2$, which is already at the scale of $M_\pi^2(eB=0)$. 
While the two-loop $\chi$PT results are slightly larger than those from the one-loop, it has large uncertainties (denoted as the grey band) from the undetermined low-energy constants. It is worth noting that our pion mass at $eB=0$ is heavier than the physical one, and both one-loop and two-loop chiral perturbation theories give slightly smaller results for $(\Sigma_u+\Sigma_d)/2$ with a larger pion mass.

\subsection{Decay constants of neutral pion and kaon and the GMOR relation}
\label{sec:res_dc-GMOR}
\begin{figure}[tbph]
	\centering
	\includegraphics[width=0.47\textwidth]{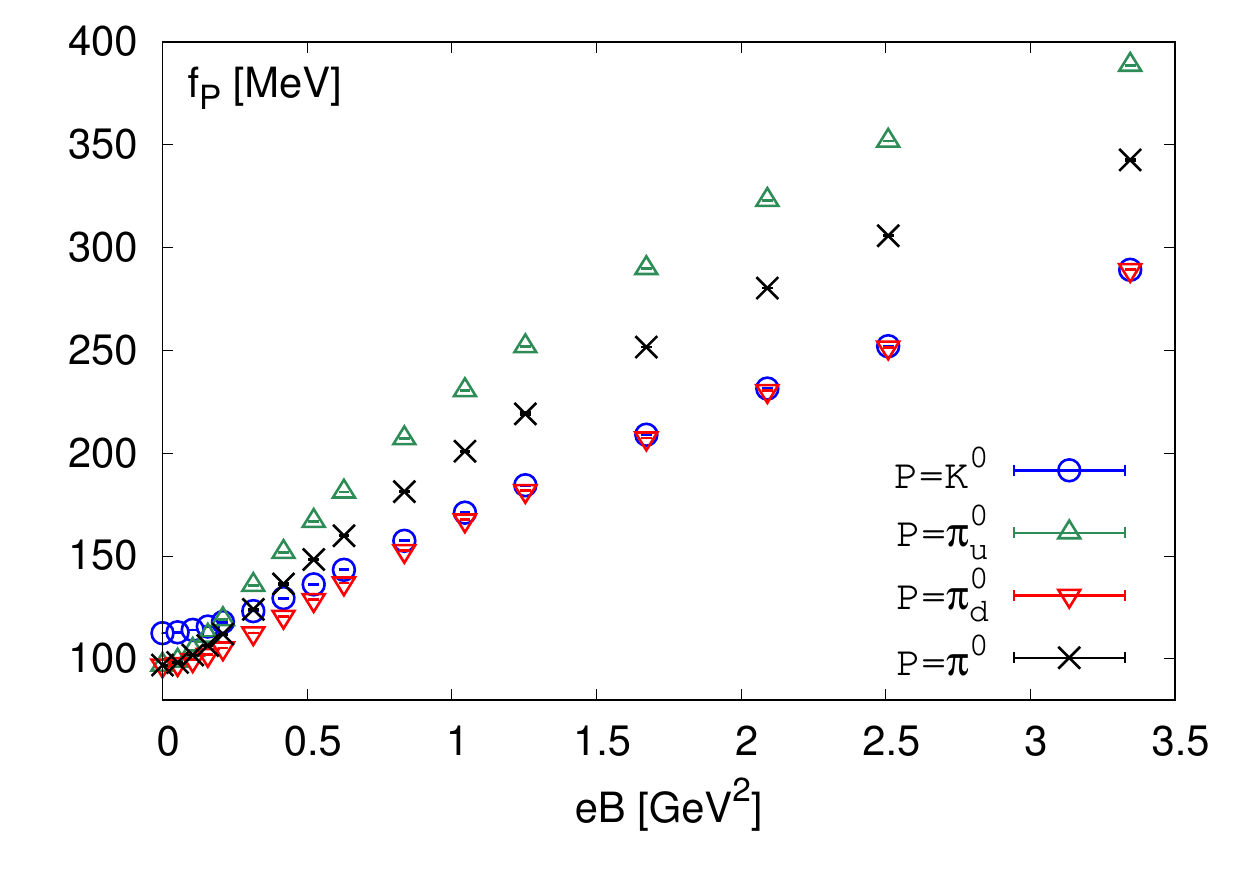}
	\includegraphics[width=0.47\textwidth]{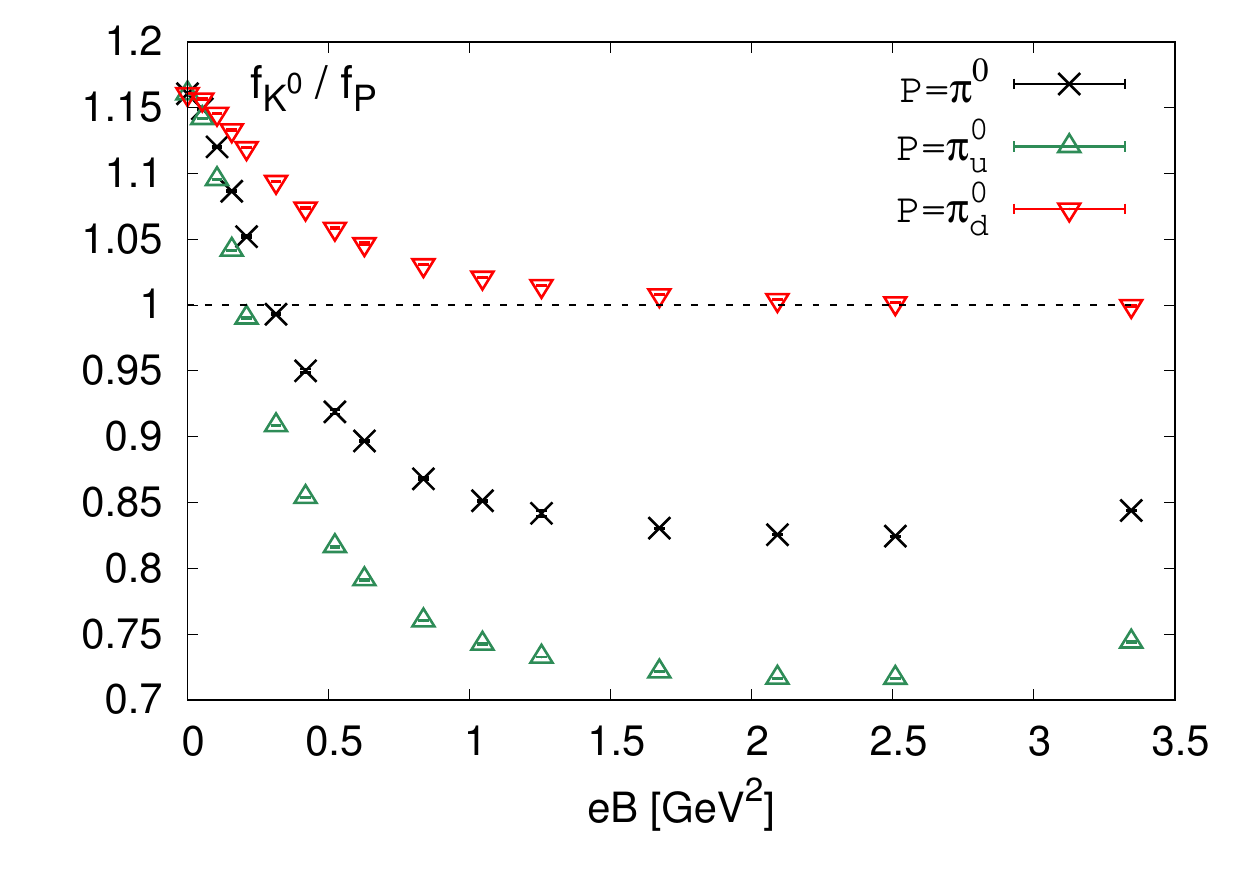} \\
	\caption{Left: $eB$ dependence of decay constants of $\pi_u^0$, $\pi_d^0$, $\pi^0$, and $K^0$. Right: Ratio of neutral kaon decay constant to other three decay constants as a function of $eB$.}
	\label{fig:dc}
\end{figure}
We start by showing decay constants of the neutral pion and kaon in Fig.~\ref{fig:dc}. At $eB=0$, we obtained the pion decay constant $f_\pi$ = 96.93(2) MeV and kaon decay constant $f_K=112.50(2)$ MeV, resulting in $f_K/f_\pi=1.1606(3)$. The errors quoted here are purely statistical ones. These three results are rather close to those obtained at the physical-mass point in the continuum limit as quoted in the latest FLAG review, i.e., $f_\pi$=92.1(6) MeV, $f_K$=110.1(5) MeV, and $f_K/f_\pi$=1.1917(37)~\cite{Aoki:2019cca}.\footnote{Note that there is a factor of $\sqrt{2}$ difference in convention for the decay constant in our paper and Ref.~\cite{Aoki:2019cca}. The numbers shown here from Ref.~\cite{Aoki:2019cca} are already divided by $\sqrt{2}$ for comparison to our results.} As seen from the left plot of Fig.~\ref{fig:dc}, all the decay constants increase with $eB$. The decay constant from the up quark flavor component of the neutral pion $f_{\pi^0_u}$ increases most rapidly with respect to $eB$ while $f_{K^0}$ is the least.
The neutral pion decay constant $f_{\pi^0}$ lies in between $f_{\pi^0_u}$ and $f_{\pi^0_d}$, which was extracted from the average of the correlation function $G_{\pi^0_u}$ and $G_{\pi^0_d}$ in the same way as we did for $M_{\pi^0}$ in Sec.~\ref{sec:res_mass}.
 It is also interesting to see that $f_{K^0}$ seems to degenerate with the decay constant of the down quark flavor component of the neutral pion $f_{\pi^0_d}$ in the large magnetic field. 
We further show the ratio of $f_{K^0}$ to the other three decay constants in the right plot of Fig.~\ref{fig:dc}. We see that the ratio $f_{K^0}/f_{\pi^0_d}$ first decreases with increasing $eB$ and saturates to unity at $eB\gtrsim 1.5 $GeV$^2$, while both $f_{K^0}/f_{\pi^0_u}$ and $f_{K^0}/f_{\pi^0}$ decrease faster in $eB$ as compared to $f_{K^0}/f_{\pi^0_d}$, and go below unity at $eB\gtrsim 0.2$ GeV$^2$. In the large magnetic field,  $f_{K^0}/f_{\pi^0_u}$ and $f_{K^0}/f_{\pi^0}$  seem to saturate at $\sim0.7$ and $\sim 0.85$, respectively. While in the range of $eB\in(1.5,2.5)$ GeV$^2$, they slightly increase by less than 5\% at our largest value of $eB$.

\begin{figure}[tbph]
	\centering
	\includegraphics[width=0.47\textwidth]{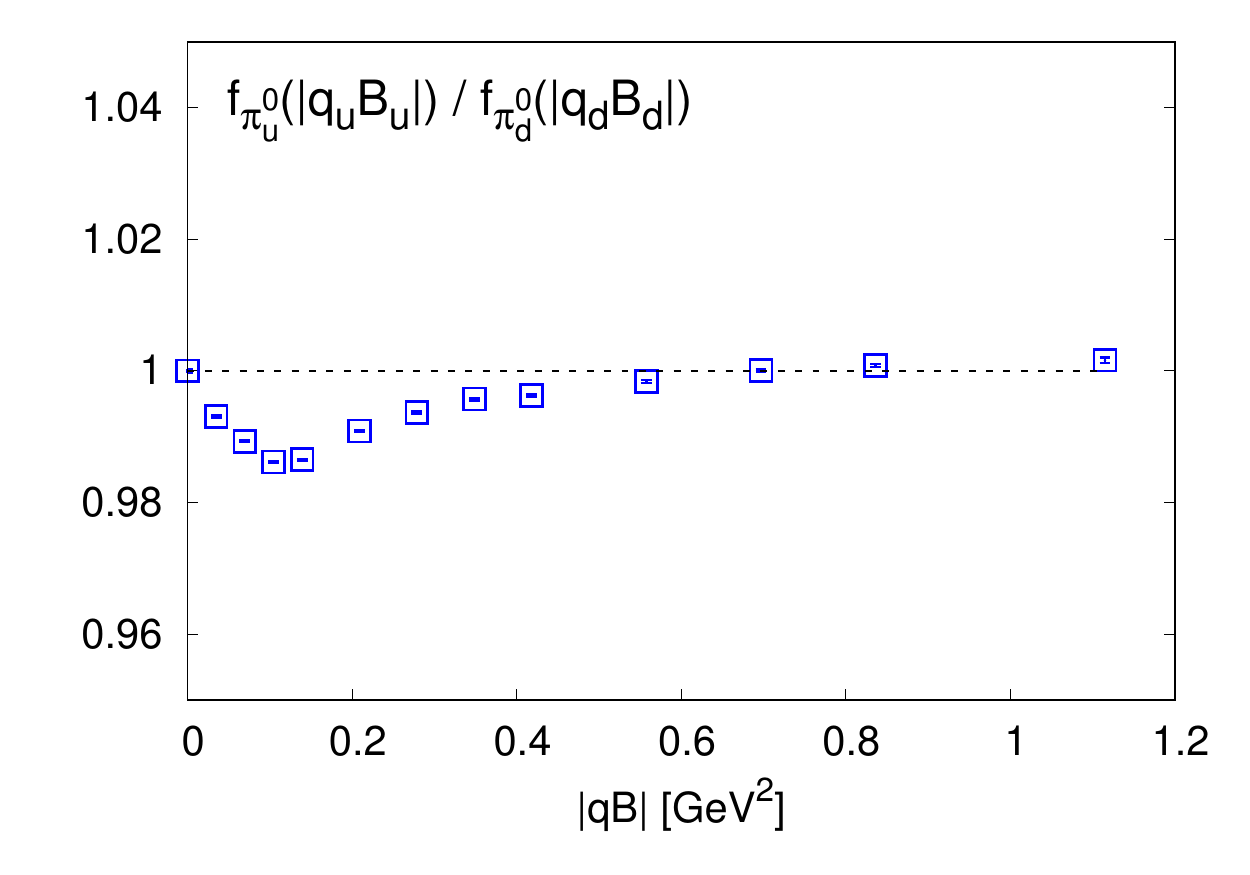}
	\caption{Ratio of $f_{\pi^0_u}$ to $f_{\pi^0_d}$ at the same value of $|qB|=|q_uB_u|=|q_dB_d|$ as a function of $|qB|$.  
	} 
	\label{fig:dc_qB}
\end{figure}

Because of the $qB$ scaling behavior of $M_{\pi^0_u} (M_{\pi^0_d})$ and $\Sigma_u$($\Sigma_d$) shown in Fig.~\ref{fig:neutral_mass} and Fig.~\ref{fig:pbp_ratio}, we also wonder about the case for the up and down quark flavor components of the neutral pion decay constant. 
We show the ratio $f_{\pi^0_u}(|q_uB_u|)/f_{\pi^0_d}(|q_dB_d|)$ as a function of $|qB|$ in Fig.~\ref{fig:dc_qB}. It can be clearly seen that the ratio is very close to 1, and the deviation is always less than 2\% in our current window of magnetic fields. Hence, the $qB$ scaling behavior is also found in the case of the neutral pion decay constant. This is a natural consequence of the $qB$ scaling of correlation functions shown in Fig.~\ref{fig:corr_qB} according to Eq.~\ref{eq:fpi_det}.

\begin{figure}[!tbph]
	\centering
	\includegraphics[width=0.47\textwidth]{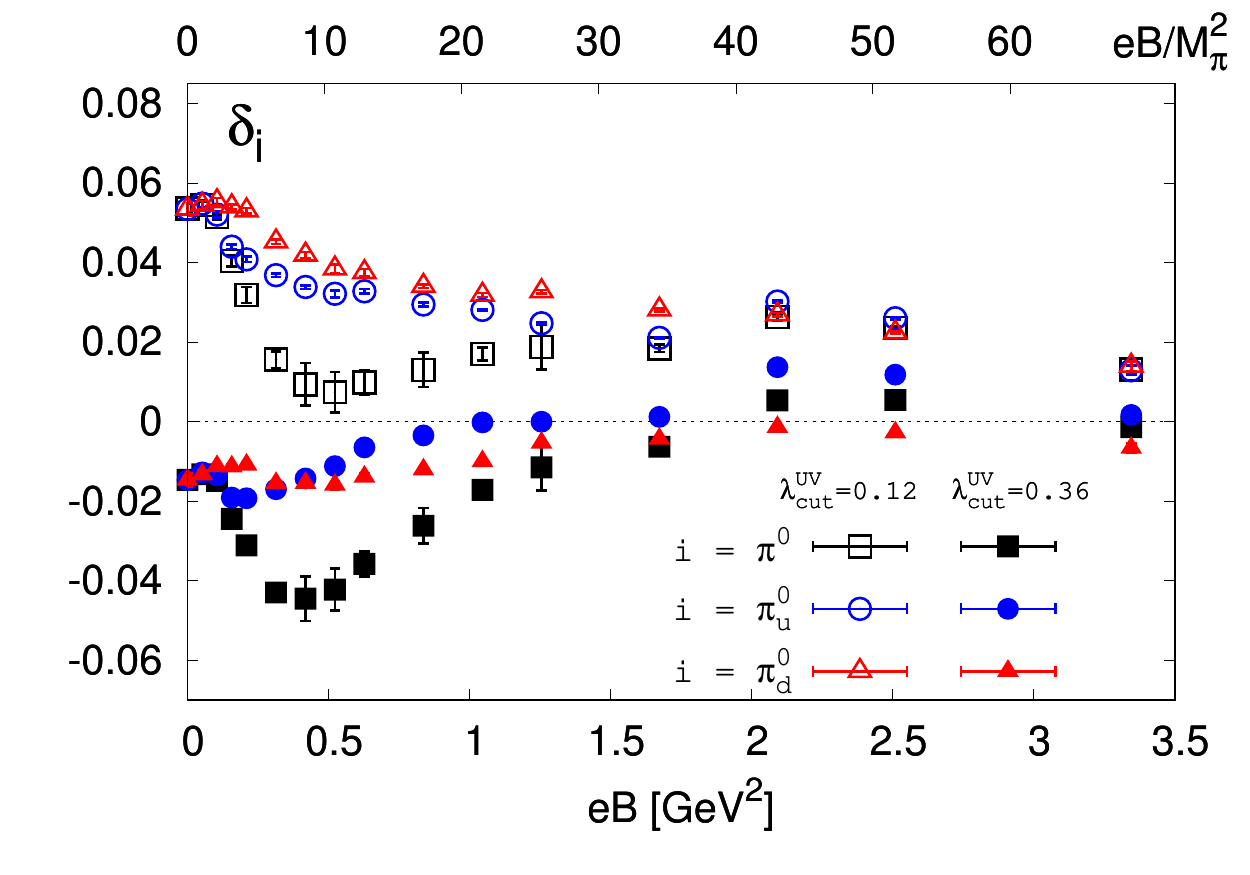}
	\includegraphics[width=0.47\textwidth]{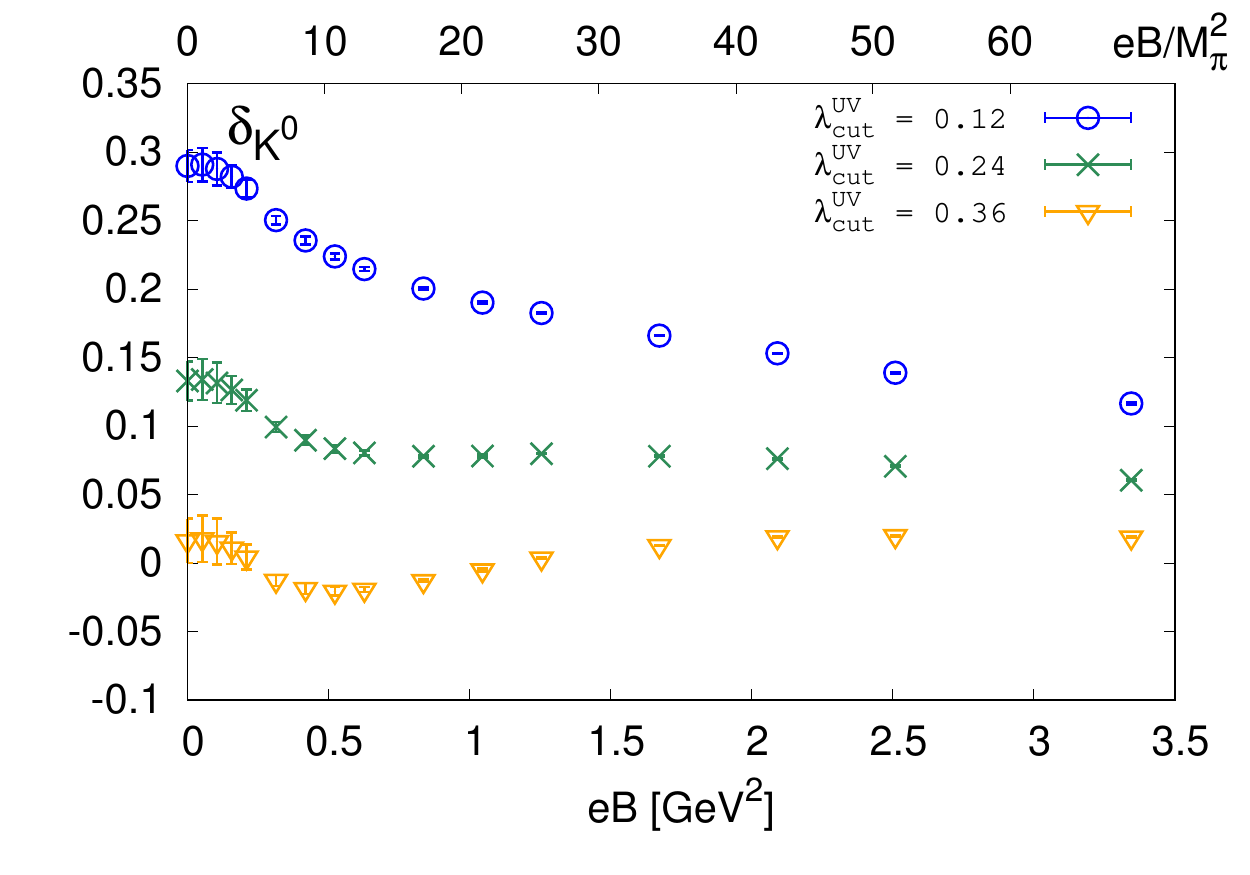}
	\caption{The corrections $\delta_{i=\pi^0,\pi^0_u,\pi^0_d}$ for the two-flavor GMOR relation (left) and $\delta_{K^0}$ for the three-flavor GMOR relation (right).}
	\label{fig:GMOR}
\end{figure}

Since we have obtained the masses and decay constants of neutral pseudoscalar mesons, light and strange quark chiral condensates, we are now ready to check the validity of the two-flavor and the three-flavor GMOR relations. We show the corrections to the two-flavor and three-flavor GMOR relations in a wide window of $eB$ from 0 to $\sim3.35$ GeV$^2$ in the left and right plots of Fig.~\ref{fig:GMOR}, respectively.\footnote{We remark here that these corrections obviously depend on the lattice cutoff effects, magnetic effects, and quark masses as the results shown in the current work are obtained from lattice QCD simulations at nonzero magnetic fields with a single lattice spacing and a single value of the pion mass without continuum and chiral extrapolations.} In the left plot of Fig.~\ref{fig:GMOR}, we show the correction $\delta_{\pi^0_{u,d}}$ (cf. Eq.~\ref{eq:GMOR_u} and Eq.~\ref{eq:GMOR_d}) for $u$ and $d$ quark flavor components separately. To get a UV-free chiral condensate as discussed in Sec.~\ref{sec:setup_UV}, we subtract the UV-divergence part $\pbp_l^{\rm UV}$ with $\lambda_{cut}^{\rm UV}=0.12$ and 0.36 from the chiral condensates $\pbp_l$. The results obtained using $\lambda_{cut}^{\rm UV}=0.12$ and 0.36 are shown as open and filled points, respectively. At $eB=0$, the correction to the GMOR relation is about 6\% at most. 
As the GMOR relation strictly holds in the chiral limit of quarks at $eB=0$ from the leading order chiral perturbation theory (cf. Eq.~\ref{eq:GMOR} without $\delta_\pi$), the next-to-leading order chiral corrections to the two-flavor GMOR relation at the physical pion mass $M_\pi$ is (6.2 $\pm$ 1.6)\%~\cite{Jamin:2002ev,Bordes:2010wy}. Although pion mass is 220 MeV at $eB=0$ in our study, the corrections to the GMOR relation at $eB$=0 shown in Fig.~\ref{fig:GMOR} are in the same ballpark. At large nonzero magnetic fields, we see that the correction either decreases or increases towards 0. And the value of $|\delta_{\pi^0_{u,d}} |$ becomes smaller at $eB \gtrsim 1.5~{\rm GeV}^2$, compared to the case at $eB = 0$. This is to say that the deviation of $\delta_{\pi^0_{u,d}}$ from zero is at most 6\% in the whole range of magnetic field strength studied, although the correction depends on $\lambda_{cut}^{\rm UV}$. We thus conclude that in our study, the GMOR relation for the up and down quark flavor components of $\pi^0$ holds with an accuracy of $\sim6\%$ in $eB\in[0,3.35)$ GeV$^2$. While for $\pi^0$, the correction $\delta_{\pi^0}$ was obtained by averaging up and down quark flavor components of the correlation function without the disconnected part. We also see that the GMOR relation for neutral pion $\pi^0$ (cf. Eq.~\ref{eq:GMOR}) holds quite well in the range of $eB$ from 0 to 3.35 GeV$^2$.

We then have a look at the correction $\delta_{K^0}$ to the three-flavor GMOR relation (cf. Eq.~\ref{eq:GMOR_K}) shown in the right plot of Fig.~\ref{fig:GMOR}. The UV divergence of the strange quark chiral condensate is taken care of in the same way as for the light quark chiral condensate. At $eB=0$, the chiral correction $\delta_K$ is much larger than $\delta_\pi$ due to an enhancement factor $M_K^2/M_\pi^2$~\cite{Gasser:1984gg}, and $\delta_K$ is about $\sim32$\% using the values of $M_\pi$ and $M_K$ in our setup . As seen from the right plot of Fig.~\ref{fig:GMOR} at $eB=0$, $\delta_K$ has the largest value of $\sim 30$\% with $\lambda_{cut}^{\rm UV}$=0.12 and the smallest value of $\sim 2$\% with $\lambda_{cut}^{\rm UV}$=0.36 (cf. discussions at the end of Sec.~\ref{sec:setup_UV}). In the latter case, the UV-divergence part of the chiral condensate is likely  underestimated. As $eB$ increases, $|\delta_{K^0}|$ becomes smaller for the case with $\lambda_{cut}^{\rm UV}=0.12$ and 0.24 and stays almost the same with $\lambda_{cut}^{\rm UV}=0.36$ .

\section{Conclusions}
\label{sec:conclusions}
In this paper, we investigated the masses and magnetic polarizabilities of light and strange pseudoscalar mesons, light quark chiral condensates, as well as neutral pion and kaon decay constants in the presence of background magnetic fields with $eB\lesssim 3.35$ GeV$^2$ in $N_f=2+1$ lattice QCD at zero temperature. The simulation was performed using HISQ fermions with $m_\pi\approx$ 220 MeV on 32$^3\times$96 and 40$^3\times$96 lattices at a single lattice cutoff $a=0.117$ fm. Our main results include the $eB$ dependence of pseudoscalar meson masses, the $qB$ scaling of various chiral observables, and the $eB$ dependence of the corrections to 2- and three-flavor GMOR relations. We find the $qB$ scaling behavior of $M_{\pi^0_u} (M_{\pi^0_d})$, $\Sigma_u$($\Sigma_d$) and $f_{\pi^0_u}$ ($f_{\pi^0_d}$) as well as $G_{\pi^0_u}(\tau) (G_{\pi^0_d}(\tau))$ in the range of $eB\in[0.05,3.35)$ GeV$^2$. 
Although the $qB$ scaling should be exact in quenched QCD and could be spoiled by dynamical quarks, it is found that effects of dynamical quarks are negligible in the case of $M_\pi(eB=0)=220$ MeV from our current study. The $qB$ scaling can also be deduced from lattice studies of $N_f=2+1$ QCD with $M_\pi=140$ MeV~\cite{Bali:2012zg}.

With a complete Dirac eigenvalue spectrum, we are able to estimate the UV contribution to the quark chiral condensate. This makes it possible for us to study the two-flavor and three-flavor GMOR relations. We found that the corrections to the two-flavor and three-flavor GMOR relations are about 6\% and 30\% at $eB=0$, respectively, in our setup. The correction to the two-flavor GMOR relation for the neutral pion is less than 6\% in the whole window of magnetic fields we studied and becomes less than 2\% at our strongest magnetic field. Thus, the two-flavor GMOR relation holds true with an accuracy of 6\% at $eB\in[0,3.35)$ GeV$^2$. The validity of the GMOR relation thus suggests that the mechanism of ``soft" breaking of chiral symmetry by light quark masses in QCD is not changed by the magnetic field~\cite{Agasian:2001ym}. 
It is known that at $eB=0$ the lighter a Goldstone (neutral) pion mass is, the easier it is to restore the chiral symmetry at a lower temperature~\cite{Ding:2019prx,Ding:2020rtq,Bazavov:2017xul, Kotov:2021rah, Aarts:2020vyb,Bhattacharya:2014ara,Umeda:2016qdo}. This makes the connection between the reduction of $T_{pc}$ and neutral pion mass in the nonzero magnetic field more clear as the mass of neutral pion, still being a Goldstone boson at $eB\ne0$, decreases as $eB$ grows. 

The GMOR relation also naturally reconciles the reduction of $T_{pc}$ and the magnetic catalysis at zero temperature thanks to the monotonous increasing decay constants in the magnetic field. Since both decay constants and pion masses are encoded in the pion correlation functions, one can further find that the reconciliation of magnetic catalysis and reduction of pion mass intrinsically lies in the Ward identity $\langle\bar{\psi}\psi\rangle_{u,d}=m_l\chi_{\pi^0_{u,d}}$ as shown in the left plot of Fig.~\ref{fig:WIandMixing}. This is because $\chi_{\pi^0_{u,d}}$ is the sum of pion correlation functions which decrease exponentially as $\exp(-M_{\pi^0_{u,d}}\tau)$ at large distances $\tau$. Thus, a smaller value of $M_{\pi^0_{u,d}}$ most likely leads to a larger value of $\pbp_{u,d}$. At a temperature proximate to $T_{pc}$ where the IMC is observed, one may expect a nonmonotonous behavior in the pion screening mass.

It is also interesting to make a comparison with the case of the vanishing magnetic field. When $eB=0$, both sides of the GMOR relation have similar behavior in terms of breaking fields, i.e., the quark mass. This is to say that all the chiral observables, chiral condensates, pion decay constants as well as the pion mass, decrease with lighter quark mass at $eB=0$. On the other hand, when the quark mass $m_q$ changes, the Ward identity $\pbp=m_q\chi_\pi$ always holds due to the intricate play between $m_l$ and $\chi_\pi$ at $eB=0$. These details are obviously different from the case in the nonzero magnetic field, where $\pbp$ and $f_{\pi^0}$ become larger and $M_{\pi^0}$ becomes smaller as $eB$ grows. However, in both cases of $eB=0$ and $eB\neq0$ the mass of the Goldstone pion is the key to understanding the reduction of $T_{pc}$, which represents an overall effect in both spontaneously and explicit breaking of chiral symmetry according to the GMOR relation. In both cases, $M_{\pi^0}$ decreases either as $m_l$ decreases or as $eB$ grows.

Concerning the meson spectrum, we found that the mass spectrum of lighter mesons are more affected by the magnetic field  for both charged and neutral pseudoscalar mesons. For the neutral pseudoscalar mesons, their masses monotonously decrease as the magnetic field strength grows and then saturate at nonzero values of $eB$ up to $\sim3.35$ GeV$^2$. The nonzero values of (connected) neutral pion mass thus disfavor an occurrence of a superconducting phase in the current window of magnetic fields.  For the charged pion and kaon, their masses show a nonmonotonous behavior in $eB$. They first increase in $eB$ following the LLL approximation, then slow down the increasing breaking away from the LLL approximation, and finally show a turning point with subsequent decreasing behavior in $eB$.  The novel decreasing behavior of the charged pion and kaon mass in $eB$ could be due to the dynamical quark effects and large magnetic field strength we simulated.
While the neutral pion cannot be considered as a pointlike particle even at our smallest value of $eB\sim0.05$ GeV$^2$$\sim M_\pi^2(B=0)$, the charged pion remains pointlike until $eB$ at about 6$M_\pi^2(B=0)$. The obtained magnetic polarizabilities of charged and neutral pions from the $eB$ dependence of their masses are at odds with the results from quenched QCD as well as the experimental measurements where the magnetic polarizability is assumed to be the additive inverse of the electric polarizability. Together with the decreasing behavior of $M_{\pi^-}$ at large $eB$, they leave plenty of room for studies on the possible effects from dynamical quarks as well as discretization effects on the lattice in the weak and strong magnetic fields.

On the other hand, the pion and kaon decay constants and their ratio are obtained as $f_\pi$ = 96.93(2) MeV, $f_K=112.50(2)$ MeV, and $f_K/f_\pi=1.1606(3)$ at $eB=0$ in our study. These results deviate by 5\% from the state-of-the-art lattice QCD results obtained at the physical-mass point in the continuum limit. As $eB$ increases, both the neutral and kaon decay constants increase, and the ratios between $f_{K^0}$ and the other three decay constants ($f_{\pi^0},~f_{\pi^0_u},~f_{\pi^0_d}$) monotonously decrease and then saturate at nonzero values at large $eB$. Since we only deal with the decay constants of neutral pseudoscalar mesons, which are related to the axial vector current parallel to the magnetic field at zero momentum, it would be interesting to study the new decay constants at nonzero momentum and those related to the vector current as well in the future. Because of the single lattice cutoff used in our computations, it would be interesting to perform computations towards the continuum limit to further confirm our findings presented in the current paper.


\section*{Acknowledgements}
We thank Swagato Mukherjee for the early involvement of the work and enlightening discussions, and thank Gunnar Bali, Toru Kojo, Jinfeng Liao, Zhaofeng Liu, and Pengfei Zhuang for interesting discussions. This work was supported by the National Natural Science Foundation of China under Grants No. 11535012, No. 11775096, and No. 11947237, and the Rikagaku Kenky$\mathrm{\bar{u}}$jo (RIKEN) Special Postdoctoral Researcher program. The numerical simulations have been performed on the GPU cluster in the Nuclear Science Computing Center at Central China Normal University (NSC$^3$), Wuhan, China.

\appendix
\section{Generalized Ward-Takahashi identities in a background magnetic field}
\label{sec:app_WI}
\numberwithin{equation}{section}
\setcounter{equation}{0}

The axial and vector Ward-Takahashi identities in the continuum QCD in the nonzero magnetic field have already been derived in, e.g., Ref.~\cite{Bali:2017ian}. In this appendix, we will further derive the Ward-Takahashi identity related to the integrated pseudoscalar operator correlation functions with chiral transformation in the presence of a magnetic field interacting with two degenerate light quark flavors in the continuum QCD. The main results are shown in Eqs.~\ref{eq:app_WI_PS1},~\ref{eq:app_WI_PS2}, and~\ref{eq:app_WI_PS3}, which are discussed in Fig.~\ref{fig:WIandMixing} in Sec.~\ref{sec:setup_mixing}.

We start by showing the Ward-Takahashi identity at a vanishing magnetic field with our conventions.
The expectation value of an observable $O$ in QCD is given by
\begin{align}
\av{O} = 
\frac{1}{Z} \int \mathcal{D} G_\mu \mathcal{D} \bar{\psi} \mathcal{D} \psi \; \e^{-S_\text{QCD}} O,
\end{align}
where $O=O(y_1,\cdots, y_n)$ is a position dependent operator, $\av{1}=1$, and
$S_\text{QCD} $ is the Euclidean action,
\begin{align}
S_\text{QCD} = \int d^4 x \Big(
\frac{1}{2}\tr G_{\mu\nu}G_{\mu\nu} + \bar{\psi}(\fsla{D} + \massmat )\psi
\Big),
\end{align}
and $G_{\mu\nu} = \partial_\mu G_{\nu} - \partial_\nu G_{\mu} - \im g [G_{\mu}, G_{\nu}]$
where $G_\mu$ represents the gluon field. $\massmat = \diag(m_u,m_d,\cdots)$ is a mass matrix.
$\psi = (u,d,\cdots)^\top$ is a flavor multiplet of quarks, whose spinor and color indexes are suppressed.
$D_\mu = \partial_\mu - \im g G_{\mu}$ is the covariant derivative at the zero magnetic field.

The Ward-Takahashi identity \cite{Ward:1950xp, Takahashi:1957xn} for a local infinitesimal transformation $\bar{\psi}'(x) \to  \bar{\psi}(x) + \alpha(x) \bar{\psi}(x)\hat{\lambda} , \psi'(x) \to \psi(x) + \alpha(x) \lambda \psi(x)$ is written as,
\begin{align}
\av{ O \frac{\delta \log\mathcal{J} }{ \delta \alpha(x)} }
-\av{ O \frac{\delta S_\text{QCD} }{\delta \alpha(x)} }
+ \av{ \frac{\delta O}{\delta \alpha(x)}  }   = 0. \label{eq:WT_id_generic}
\end{align}
Here, $\mathcal{J}$ is the Jacobian of the transformation, which represents the anomaly and has the following form at $eB=0$~\cite{Fujikawa:2004book}:
\begin{equation}
\mathcal{J}=\exp \left[-2 i \int d^{4} x \alpha(x) \frac{g^{2}}{32 \pi^{2}} \operatorname{tr}\left[\epsilon^{\mu \nu \alpha \beta} G_{\mu \nu} G_{\alpha \beta} \right]\right]\,.
\end{equation}
Here $\epsilon^{\mu \nu \alpha \beta}$ is the completely antisymmetric tensor normalized by the convention $\epsilon^{1234}$=1 in the Euclidean metric. Note that hereafter the super(sub) index $\alpha$ is different from the infinitesimal quantity $\alpha(x)$. $\lambda$ and $\hat{\lambda}$ are products of matrices in Dirac and flavor spaces. The original relation between renormalization factors in QED was derived in \cite{Ward:1950xp},
and it was generalized as a nonperturbative relation from the symmetry argument \cite{Takahashi:1957xn}.

In the case of nonzero magnetic fields, 
the covariant derivative with the magnetic field thus becomes
\begin{align}
D_\mu \to \tilde{D}_{\mu}=\partial_{\mu}-\mathrm{i} g G_{\mu}-\mathrm{i}\, e A_{\mu} Q^{3},
\end{align}
where $Q^3$ is the charge matrix
\begin{align}
Q^{3}=\frac{1}{6} \sigma^0+\frac{1}{2} \sigma^{3}=\frac{1}{6} \mathbf{1}+t^{3},
\end{align}
and $\sigma^i$ is Pauli matrices with $i=1,2,3$ while $t^i\equiv\sigma^i/2$, and $\sigma^0 =\mathbf{1}$. And $Q^{3}$ has the following commutation relation with $t^i\equiv\sigma^i/2$:
\begin{align}
Q^{3} t^{i}=t^{i} Q^{3}-\im \, T^{12}_i,
\end{align}
where $T^{12}_i= \left(\delta_{i2} t^{1}-\delta_{i 1} t^{2}\right)$. The Jacobian at $eB\neq0$ then becomes
\begin{equation}
\mathcal{J}=\exp \left[-2 i \int d^{4} x \alpha(x) \frac{1}{32 \pi^{2}} \operatorname{tr}\left[\epsilon^{\mu \nu \alpha \beta} (g^{2}G_{\mu \nu} G_{\alpha \beta} + e^2F_{\mu \nu} F_{\alpha \beta})\right]\right],\
\end{equation}
where $F_{\mu \nu}=\partial_\mu A_\nu -\partial_\nu A_\mu$.

Now we calculate the form of Ward identity under chiral rotation. 
Let us introduce a scalar function $\alpha^i(x)$, which represents an infinitesimal local transformation.
The local chiral rotation is then,
\begin{align}
\begin{cases}
\bar{\psi}(x) \to& \bar{\psi}'(x) = \bar{\psi}(x) + \alpha^i(x)  \bar{\psi}(x) t^i \gamma_5 ,\\
\psi(x) \to& \psi'(x) = \psi(x) +  \alpha^i(x) t^i \gamma_5 \psi(x)\, .
\end{cases}
\end{align}

In the following derivation, we assume a flavor symmetry $\massmat = m{ \boldsymbol 1}$. We now derive the variation of the QCD action under the chiral transformation at $eB\neq0$ as follows.

We focus on the traceless part, namely, we only take $i= 1,2,3$.
With the the help of Leibniz rule, $\delta S_\text{QCD}$ transformed as
\begin{align}
\delta S_\text{QCD} = \bar{\psi}(x)  \gamma_\mu  \big(  \partial_\mu \alpha^i(x)  \big) t^i \gamma_5 \psi(x) 
+2m \alpha^i(x)  \bar{\psi}(x) t^i \gamma_5 \psi(x)
-e\alpha^i(x) A_{\mu} \bar{\psi}(x)   \gamma_\mu  \gamma_5   T^{12}_i  \psi(x) .
\end{align}

In the last term, we use the Landau gauge for the magnetic field $A_\mu$ pointing along the $z$ ($x_3$) direction in the infinite volume, $A_\mu(x) = (A_1,A_2,A_3,A_4) = (0,B x_1 ,0,0)$.
Using integrations by parts , the variational part from the action is
\begin{align}
\frac{\delta S_\text{QCD} }{\delta  \alpha^i(x)}
&=
-  \partial_\mu \big(  \bar{\psi}(x)  \gamma_\mu  t^i \gamma_5 \psi(x) \big) +2 m   \bar{\psi}(x) t^i \gamma_5 \psi(x)
+ \Delta_i(x,eB),
\end{align}
where 
\begin{align}
\Delta_i(x,eB) = -eA_{\mu}(x) \bar{\psi}(x)   \gamma_\mu  \gamma_5  T^{12}_i \psi(x)\, 
= -e B x_1 \bar{\psi}(x) \gamma_2  \gamma_5  (\delta_{i2} t^{1}-\delta_{i1} t^{2}) \psi(x). 
\label{eq:app_WI_Delta}
\end{align}
Since the Dirac fields transform as $\bar{\psi} (t,\vec{x}) \to \bar{\psi} (t,-\vec{x}) \gamma^0$,
${\psi} (t,\vec{x}) \to \gamma^0 {\psi} (t,-\vec{x}) $ under the parity, while the bilinear term  transforms as the odd parity,
$x_1 \bar{\psi} \gamma_2 \gamma_5 \psi(t,\vec{x})
\to -x_1 \bar{\psi} \gamma_2 \gamma_5 {\psi} (t,-\vec{x})
$. Thus $\Delta_i(x,eB) $ is an odd function of $x_1$. 

Next, we derive the variation of the pseudoscalar operator under chiral rotation. The pseudoscalar operator for the SU(2) case is
$P^i(y) = \bar{\psi}(y) \gamma_5 t^i \psi(y)$,
where $i=1,2,3$. 
The variation reads off
\begin{align}
\frac{\delta P^i(y) }{\delta \alpha^j(x) }
= \frac{\delta \left(\bar{\psi}(y) \gamma_5 t^i \psi(y)+  \frac{1}{2}\alpha^i(y)  \bar{\psi}(y) \psi(y) \right) }{\delta \alpha^j(x) }
=\frac{1}{2}\,\bar{\psi}(y) \psi(y) \delta(x-y) \delta^{ij}\,.
\end{align}

With the variations derived above, now we are ready to derive the Ward-Takahashi identities for the nonanomalous case,
\begin{align}
\av{ O \frac{\delta S_\text{QCD} }{\delta \alpha^j(x)} } =  \av{ \frac{\delta O}{\delta \alpha^j(x)}  } ,
\label{eq:app_WI_non}
\end{align}
for chiral rotations with pseudoscalar operator as follows in the nonzero magnetic field. 
We choose an operator as $O(y)=P^{j}(y)= \bar{\psi}(y) \gamma_{5} t^{j} \psi(y)$.
By integrating over $x$ on the left-hand side (LHS), the first term in the right-hand side (RHS) vanishes
\begin{align}
\label{eq:app_WI_PS_tem}
\int d^4 x \,\text{LHS}  &=  \int d^4 x \, \left(- \partial^x_\mu \av{  P^{j}(y)   \big(  \bar{\psi}(x)  \gamma_\mu  t^i\gamma_5 \psi(x) \big) }
+2 m \av{  P^{j}(y)  \bar{\psi}(x) t^i\gamma_5 \psi(x) }
+\av{  P^{j}(y) \Delta_i(x,eB) } \right) \\
&= 2\, m \int d^4 x \av{  P^{j}(y) P^i(x) }
+ \int d^4 x \av{  P^{j}(y) \Delta_i(x,eB) }.
\end{align}
Integrating over $x$ in the right-hand side of Eq.\ref{eq:app_WI_non} we arrive at
\begin{align}
\int d^4 x \,\text{RHS} = 
\int d^4 x \, \av{ \frac{\delta P^j(y) }{\delta \alpha^i(x) } }
=
\frac{1}{2}\,\av{ \bar{\psi}(y) \psi(y) }  \delta^{ij}.
\end{align}
Thus, the identity becomes
\begin{align}
4 m \int d^4 x \av{  P^{j}(y) P^i(x) } + 2 \int d^4 x \av{  P^{j}(y) \Delta_i(x,eB) }
&=
\av{ \bar{\psi}(y) \psi(y) }  \delta^{ij}.
\end{align}
After integrating over $y$ and dividing by four-volume $V$,
\begin{align}
4m \frac{1}{V}\int d^4y d^4 x \av{  P^{i}(y) P^j(x) } + \frac{2}{V}\int d^4 y \int d^4 x \av{  P^{j}(y) \Delta_i(x,eB) }
&=
\frac{1}{V}\int d^4y \av{ \bar{\psi}(y) \psi(y) }  \delta^{ij}.
\end{align}
In our convention the right-hand side of the above identity gives a two-flavor chiral condensate, i.e., $\pbp_u + \pbp_d$ with $i=j$. At nonzero magnetic field the neutral pion operator is $(\alpha \bar{u}\gamma_5u - \beta \bar{d}\gamma_5d )$ with $\alpha^2+\beta^2=1$, thus in the case of $i=j=3$ and $\alpha=\beta=1/\sqrt{2}$, the left-hand side of the above identity is proportional to the correlation function of the neutral pion. Note that when $i=3$, the magnetic field related term $\Delta_{i}(x,eB)$ (cf. Eq.~\ref{eq:app_WI_Delta}) vanishes. The above identity thus becomes
\begin{equation}
2m_l\,\tilde{\chi}_{\pi^0} = \pbp_u + \pbp_d\,.
\label{eq:app_WI_PS1}
\end{equation}
Although the above relation was derived assuming the flavor symmetry $\massmat = m{ \boldsymbol 1}$ ($m_l=m_u=m_d$), it can also be extended to the case of $m_u\neq m_d$, i.e., in the above relation $2m_l$ needs to be replaced by $m_u+m_d$ and there exist contributions from disconnected diagrams to $\tilde{\chi}_{\pi^0}$ even at $eB=0$. Following same procedures, the above relation can also be extended to the $K^0$ meson with corresponding operator $\bar{d}\gamma_5s$ and the fictitious $\eta_s^{0}$ meson with $\bar{s}\gamma_5s$,
\begin{eqnarray}
\label{eq:app_WI_PS2}
m_s \tilde{\chi}_{\eta^0_s} &=& \pbp_s +\Delta^s_{\mathcal{J}}\,,\\
(m_d+m_s) \chi_{K^0} &=& \pbp_d + \pbp_s\,.
\label{eq:app_WI_PS3}
\end{eqnarray}
Here $\tilde{\chi}_{\pi^0}$, $\chi_{K^0}$, and $\tilde{\chi}_{\eta^0_s}$ are the space-time sum of the two-point correlation functions for neutral pion, neutral kaon, and $\eta_s^0$ with $m_l=m_u=m_d$ and the corresponding operators $P=1/\sqrt{2}(\bar{u}\gamma_5u - \bar{d}\gamma_5d)$, $\bar{d}\gamma_5s$, and $\bar{s}{s}$, respectively. Hereafter the superscript ``$\tilde{\chi}$"  denotes that correlators in the isosinglet channel includes contributions from both connected and disconnected diagrams. The $\Delta^s_{\mathcal{J}}$-term, arising from the flavor singlet transformation involving the first term in Eq.~\ref{eq:WT_id_generic}, has the following form:
\begin{equation}
\Delta^s_{\mathcal{J}}=\frac{1}{V}\int d^4y\int d^4x\av{(\bar{s}(y)\gamma_5 s(y))\left(\frac{-i}{16\pi^2}\epsilon^{\mu \nu \alpha \beta} \left[g^2G_{\mu \nu}(x) G_{\alpha \beta}(x)+\Delta_{\mathcal{J},B}\right]\right)}\,,
\end{equation}
with $\Delta_{\mathcal{J},B}=e^2F_{\mu \nu}(x) F_{\alpha \beta}(x)$. $\Delta_{\mathcal{J},B}$ vanishes in either zero magnetic field or zero electric field. In our current setup $\Delta_{\mathcal{J},B}=0$ as $A_\mu(x)= (0,B x_1 ,0,0)$.
Similarly one can also obtain the following two relations:
\begin{align}
\label{eq:WIpiu}
 m_u \tilde{\chi}_{\pi^0_u} &= \pbp_u +\Delta^u_{\mathcal{J}}\,,\\
 m_d \tilde{\chi}_{\pi^0_d} &= \pbp_d +\Delta^d_{\mathcal{J}}\,,
  \label{eq:WIpid}
\end{align}
with $P=\bar{u}\gamma_5u$ and $\bar{d}\gamma_5d$, respectively.
$\Delta^u_{\mathcal{J}}$ and $\Delta^d_{\mathcal{J}}$ have the following forms:
\begin{align}
\Delta^u_{\mathcal{J}} &= \frac{1}{V}\int d^4y\int d^4x\av{(\bar{u}(y)\gamma_5 u(y))\left(\frac{-i}{16\pi^2}\epsilon^{\mu \nu \alpha \beta} \left[g^2G_{\mu \nu}(x) G_{\alpha \beta}(x)+\Delta_{\mathcal{J},B}\right]\right)}\,,\\
\Delta^d_{\mathcal{J}} &= \frac{1}{V}\int d^4y\int d^4x\av{(\bar{d}(y)\gamma_5 d(y))\left(\frac{i}{16\pi^2}\epsilon^{\mu \nu \alpha \beta} \left[g^2G_{\mu \nu}(x) G_{\alpha \beta}(x)+\Delta_{\mathcal{J},B}\right]\right)}\,.
\end{align}
It is obvious that the sum of Eq.~\ref{eq:WIpiu} and Eq.~\ref{eq:WIpid} at $eB=0$ with $m_u=m_d$ recovers Eq.~\ref{eq:app_WI_PS1}.

The identity~\ref{eq:app_WI_PS1} at the zero magnetic field was also obtained using a diagrammatic method (cf. Eq. 6.4 in Ref.~\cite{Kilcup:1986dg}). Since the magnetic field represented by the U(1) field can be factored out from the gauge field, the extension to nonzero magnetic fields remains the same, which can be simply observed from the diagrammatic method in Ref.~\cite{Kilcup:1986dg}. As seen from Fig.~\ref{fig:WIandMixing} and the discussions in Sec.~\ref{sec:setup_mixing}, the identities~\ref{eq:app_WI_PS1},~\ref{eq:app_WI_PS2}, and~\ref{eq:app_WI_PS3} hold well in the staggered discretization scheme.

\section{GMOR relation for up and down quark components of neutral pion at $eB=0$}
\label{A:GMOR}
In this appendix we show the derivation of the GMOR relation for pure up and down quark components of neutral pion at the vanishing magnetic field based on the Ward identity, Eq.~\ref{eq:WT_id_generic}.

According to Goldstone's theorem, the pions are created from the vacuum by a chiral SU(2) current,
\begin{align}
\av{0|J_\mu^{5i} | \pi^j(p)} = ip_\mu f_\pi e^{-ipx} \delta^{ij},
\label{eq:app_gsPion}
\end{align}
where the coefficient $f_\pi$ is the pion decay constant. On the other hand, the divergence of the axial vector current relates the pseudoscalar field through the axial Ward identity. Together with the partially conserved axial vector current relation, we have
\begin{equation}
M^2_\pi f_\pi \av{0|\phi^i|\pi^j}= \av{0|\partial_\mu J_\mu^{5i}|\pi^j} = 2m\av{0|P^i|\pi^j},
\label{eq:app_fpi}
\end{equation}
where $\phi^i$ is one of renormalized physical pion field components, $M_\pi$ is the pion mass, and $m$ is the quark mass.

To decompose the neutral pion field to into up quark component, we choose an operator $O_u=\bar{\psi}\Gamma_u\psi$ with
\begin{equation}\label{trans}
\Gamma_u=\lambda_u=\hat{\lambda}_u\ =\ \gamma_5\frac{1}{2}(\sigma^1+\mathrm{i}\sigma^2)\frac{1}{2}(\sigma^1-\mathrm{i}\sigma^2) =
\gamma_5 \left(\begin{array}{ll} 
1 & 0 \\
0 & 0
\end{array}\right) \equiv \gamma_5\sigma^u.
\end{equation}
The corresponding transformation is $\bar{\psi}'(x) \to  \bar{\psi}(x) + \bar{\psi}(x)\alpha(x) \hat{\lambda}_u,~ \psi'(x) \to \psi(x) + \alpha(x) \lambda_u\psi(x)$. Now we consider the up quark component of axial and pseudoscalar currents, and Eqs.~\ref{eq:app_gsPion} and~\ref{eq:app_fpi} thus become
\begin{align}
\begin{split}
\av{0|J_\mu^{50} | \pi^0(p)} &= ip_\mu f_{\pi^0_u} e^{-ipx} \,,\\
M^2_{\pi^0_u} f_{\pi^0_u} \av{ 0|\phi^0_u|\pi^0}&=\av{0|\partial_\mu J_{\mu}^{50} | \pi^0} = 2m_u\av{ 0 | P^u| \pi^0},
\end{split}
\label{eq:PCACu}
\end{align}
where $P^u=\bar{u}\gamma_5 u$ is the operator that leads to a ground state mass of $M_{\pi^0}$, and $J_\mu^{50}$ is defined such that $\partial_\mu J_\mu^{50}=\partial_\mu(\bar{\psi}\gamma_\mu\gamma_5\sigma^{u}\psi) + \frac{i}{16\pi^2}\epsilon^{\mu \nu \alpha \beta} g^2G_{\mu \nu}G_{\alpha \beta}$.
Thus, the Ward identity, Eq.~\ref{eq:WT_id_generic}, reads off 
\begin{equation} 
\frac{1}{V}m_u \int d^4 y d^4 x \av{  P^u(y) P^u(x) } =\frac{1}{V} \int d^4 y \av{ \bar{\psi}(y)\psi(y) }_u + \Delta_{\mathcal{J}}^{u}\,.
\end{equation}
Inserting a complete set of states on the left-hand side of the above equation and using Eq.~\ref{eq:PCACu}
\begin{equation}
\text{LHS}=\frac{f^2_{\pi^0_u} M^4_{\pi^0_u}}{4m_u}\int d^4yd^4x \av{\phi^0_u(y)\phi^0_u(x)} = \frac{f^2_{\pi^0_u} M^2_{\pi^0_u}}{2m_u}.
\end{equation}
This then leads to the GMOR relation involving only up quark chiral condensate
\begin{equation}
 2m_u ~\left( \pbp_u + \Delta_\mathcal{J}^u \right)= f_{\pi^0_u}^2 M_{\pi^0_u}^2.
\label{eq:app_GMORpi0u}
\end{equation}
Similar procedures can be followed to obtain the GMOR relation involving only down quark chiral condensate
\begin{equation}
 2m_d ~\left( \pbp_d + \Delta_\mathcal{J}^d \right) = f_{\pi^0_d}^2 M_{\pi^0_d}^2.
\label{eq:app_GMORpi0d}
\end{equation}

\section{Simulations in the HISQ discretization scheme}
\label{sec:app_HISQ}
\numberwithin{equation}{section}
\setcounter{equation}{0}
In this appendix, we describe the implementation of the magnetic field in the lattice QCD simulations using the HISQ action, in particular, the procedure to compute the fermion force. The HISQ action is constructed by the Kogut-Susskind one-link action $D_\text{KS}$
and Naik improvement term $ D_\text{Naik}$ with smeared links,
\begin{align}
D_\text{HISQ}[X(U),W(U)] \equiv c_{1000} D_\text{KS}[X(U)] + c_{3000} D_\text{Naik}[W(U)],
\end{align}
where all coefficients are chosen at the vanishing magnetic field~\cite{Bazavov:2010ru} and
\begin{align}
D_\text{KS}[X(U)] &= \sum_\mu \eta_\mu(n)\left[X_{\mu}(n) \delta_{n+\mu,n'} -X^\dagger_{\mu}(n-\hat\mu)
\delta_{n-\mu,n'}
\right],
\end{align}
and
\begin{align}
D_\text{Naik}[W(U)] &=\sum_\mu \eta_\mu(n)\big[
W_{\mu}(n) W_{\mu}(n+\hat\mu) W_{\mu}(n+2\hat\mu) \delta_{n+3\mu,n'}  -
W^\dagger_{\mu}(n-\hat\mu) W^\dagger_{\mu}(n-2\hat\mu) W^\dagger_{\mu}(n-3\hat\mu)
\delta_{n-3\mu,n'}
\big],
\end{align}
where $\eta_\mu(n)$ is the staggered phase, $X$ denotes level-two fat-7 smeared links, and $W$ denotes reunitarized links, which are constructed by thin links $U$.

The magnetic field on the lattice is represented by \uone ~links in the Landau gauge for the electric charge $q$,
\begin{align}
u_x(n_x,n_y,n_z,n_t)&=
\begin{cases}
\exp[-iq \hat{B} N_x n_y], \;\;&(n_x= N_x-1)\\
1 \;\;&(\text{otherwise})\\
\end{cases}\notag\\
u_y(n_x,n_y,n_z,n_t)&=\exp[iq \hat{B}  n_x], \notag \\
u_z(n_x,n_y,n_z,n_t)&=u_t(n_x,n_y,n_z,n_t)=1,\notag
\end{align}
where $\hat B=a^2 B$ with the lattice spacing $a$. In the HISQ action, the magnetic field can be 
realized by just replacing all smeared links as
\begin{align}
D_\text{HISQ}[X,W] 
\to
D_\text{HISQ}[uX,uW] \label{eq:mag_HISQ}\,,
\end{align}
while keeping the bare links in the gauge action because gluons do not carry electric charges.
Note that we multiply the magnetic field to the smeared links instead of bare links, which could be different from the case in the implementation of imaginary chemical potential as the magnetic field variable $u$ depends on the coordinate.
We suppress the Naik term contribution for simplicity below, but
the extension is straightforward.

We employ the rational hybrid Monte Carlo algorithm to generate gauge configurations. The fermion force is defined as
\begin{align}
F_\mu(n) =  \left[ U_\mu(n) \frac{\partial S_\text{f}[X]}{ \partial U_\mu(n)} \right]_\text{TA} \;\;\;\;(\text{no sum})\,,
\end{align}
where $S_\text{f}$ is a rationally approximated pseudofermion action, and TA means removing trace and anti-Hermitianizing operations. In the presence of magnetic fields, the force is modified as
\begin{align}
F_\mu(n) =  \left[ U_\mu(n) \frac{\partial S_\text{f}[uX]}{ \partial U_\mu(n)} \right]_\text{TA} \;\;\;\;(\text{no sum})\, .
\end{align}
For the case of a zero magnetic field, the derivative with respect to bare links can be
calculated with the chain rule, which can be symbolically expressed as
\begin{align}
\frac{\partial S_\text{f}[X]}{\partial U}=\frac{\partial S_\text{f}[X]}{\partial X} \frac{\partial X}{\partial W} \frac{\partial W}{\partial V} \frac{\partial V}{\partial U},
\label{eq:app_derivative}
\end{align}
where $V$ is the level-one fat-7 link.
The first term on the right-hand side ${\partial S_\text{f}[X]}/{\partial X}$ is formally the same function form as the standard staggered force except that smeared links are used instead of thin links.
This term can be symbolically written as
\begin{align}
\frac{\partial S_\text{f}[X]}{\partial X} \sim \sum_k \psi^\dagger_k \otimes \Psi_k + \cdots
\end{align}
where $\Psi_k = (D^\dagger_\text{HISQ}[X] D_\text{HISQ}[X] +\beta_k)^{-1} \phi$,
$\psi_k = D_\text{HISQ}[X] \Psi_k$
and $\phi$ is a pseudofermion field and $\beta_k$ represents a rational coefficient of order $k$.
For the case of nonzero magnetic fields, Eq.~\ref{eq:app_derivative} becomes
\begin{align}
\frac{\partial S_\text{f}[uX]}{\partial U}=\frac{\partial S_\text{f}[uX]}{\partial X} \frac{\partial X}{\partial W} \frac{\partial W}{\partial V} \frac{\partial V}{\partial U}.
\end{align}
Since $X, W$, and $V$ are dummy variables in the chain rule, the above term can be rewritten as 
\begin{align}
\frac{\partial S_\text{f}[uX]}{\partial U}
=\frac{\partial S_\text{f}[\tilde{X} ]}{\partial \tilde{X}} \frac{\partial \tilde{X}}{\partial {W}} \frac{\partial {W}}{\partial {V}} \frac{\partial {V}}{\partial {U}},
\end{align}
where variables with tilde represent \uone ~rotated variables $\tilde X = uX$.
Thus, the functional form of the first term on the right-hand side of the equal sign is the same as that at a zero magnetic field (cf. Eq.~\ref{eq:app_derivative}).
Moreover, in the second term, the magnetic field is just factored out as ${\partial \tilde{X}}/{\partial {W}}= u {\partial {X}}/{\partial {W}}$ due to the structure of fat-7 links. Finally, in the presence of magnetic fields, the fermion force is
\begin{align}
F^\text{mag-HISQ}_\mu(n) =  \left[ u_\mu(n) U_\mu(n) \mathcal{F} \right]_\text{TA} \;\;\;\;(\text{no sum})
\end{align}
with
\begin{align}
\mathcal{F}= \frac{\partial S_\text{f}}{\partial {X}}\Big|_{X\to uX}\frac{\partial {X}}{\partial {W}} \frac{\partial {W}}{\partial {V}} \frac{\partial {V}}{\partial {U}},
\end{align}
where $\frac{\partial S_\text{f}}{\partial {X}}\big|_{X\to uX} $ means that the staggered force term has an argument $uX$ instead of $X$.
Thus, the force calculation in the molecular dynamics is summarized as follows:
\begin{enumerate}
	\item Prepare smeared links $X, W$ and $V$ from $U$.
	\item Multiply the \uone~variable $u$ on the smeared links $X$.
	\item Calculate a part of fermion force,
	$\Psi_k = (D^\dagger_\text{HISQ}[uX] D_\text{HISQ}[uX] +\beta_k)^{-1} \phi$
	and $\psi_k = D_\text{HISQ}[X] \Psi_k$ for all $k$ for the rational approximation.
	\item Remove \uone~variable $u$ from the smeared links $uX$.
	\item Construct $\mathcal{F}$.
	\item Finalize the force $F^\text{mag-HISQ}_\mu(n)$.
\end{enumerate}

\bibliography{Refs}
%
%
%
\end{document}